\documentclass{aa}  

\usepackage{natbib}
\bibpunct{(}{)}{;}{a}{}{,} 
\usepackage{graphicx}
\usepackage{txfonts}
\usepackage{twoopt}
\usepackage[breaklinks=true,colorlinks=true,allcolors=blue]{hyperref}
\usepackage{pifont}
\usepackage{array}
\usepackage{adjustbox}
\usepackage{enumitem}
\usepackage{xcolor}
\usepackage{placeins}

\makeatletter
  \newcommandtwoopt{\citeads}[3][][]{\href{http://adsabs.harvard.edu/abs/#3}%
    {\def\hyper@linkstart##1##2{}%
     \let\hyper@linkend\@empty\citealp[#1][#2]{#3}}}
  \newcommandtwoopt{\citepads}[3][][]{\href{http://adsabs.harvard.edu/abs/#3}%
    {\def\hyper@linkstart##1##2{}%
     \let\hyper@linkend\@empty\citep[#1][#2]{#3}}}
  \newcommandtwoopt{\citetads}[3][][]{\href{http://adsabs.harvard.edu/abs/#3}%
    {\def\hyper@linkstart##1##2{}%
     \let\hyper@linkend\@empty\citet[#1][#2]{#3}}}
  \newcommandtwoopt{\citeyearads}[3][][]%
    {\href{http://adsabs.harvard.edu/abs/#3}
    {\def\hyper@linkstart##1##2{}%
     \let\hyper@linkend\@empty\citeyear[#1][#2]{#3}}}
\makeatother

\newcommand{\cmark}{\ding{51}}%
\newcommand{\xmark}{\ding{55}}%

\def\hi{H\,{\sc i}}
\newcommand\kms{km~s$^{-1}$}
\newcommand\msun{$M_\odot$}
\newcommand\mhi{$M_{\text{\hi}}$}

\def\arcsec{$^{\prime\prime}$}
\def\dg{$^{\circ}$}
\def\w50{$W_{50}$}

\def\mstar{$M_{\star}$}

\def\be{\begin{equation}}
\def\ee{\end{equation}}

\def\vsys{$V_{\rm sys}$}
\def\snrgp{$(S/N)_{\rm ch}$}
\def\mbar{$M_{\rm b}$}
\def\vrot{$V_{\rm rot}$}
\def\disp{$\sigma_{\rm V}$}
\def\barolo{$^{\rm 3D}$Barolo}
\def\PAhi{$PA_{\rm \text{\hi}}$}
\def\PAop{$PA_{\rm op}$}
\def\PAkin{$PA_{\rm kin}$}
\def\mmuefr{$\langle \mu \rangle_{e,r}$}
\def\mmuefX{$\langle \mu \rangle_{e,X}$}

\def\paironeleft{AHCJ0203+3714-1}
\def\paironeright{AHCJ0203+3714-2}
\def\pairtwoleft{UGC 8605}
\def\pairtworight{UGC 8602}
\def\pairthrleft{CGCG 290-011}
\def\pairthrright{UGC 5480}
\def\bSNRone{UGC 8503}
\def\svel{AGC 239039}
\def\lsmerge{AGC 234932}
\def\fon{AGC 239112}
\def\inter{UGC 5541}
\def\bSNRtwo{AHCJ1308+5437}
\def\tstd{AHCJ1359+3726}
\def\nce{AHCJ2207+4008}
\def\beaut{AHCJ2207+4143}
\def\stick{AHCJ2232+3938}
\def\pavel{AHCJ2239+3832}
\def\reg{AHCJ2216+4024}
\def\paver{AHCJ2218+4059}
\def\mess{UGC 12027}
\def\wtm{UGC 8363}
\def\dwtm{UGCA 363}
\def\misal{UGC 12005}
\def\weir{AHCJ2249+3948}

\usepackage{subcaption}

\begin{document} 
\setlength{\extrarowheight}{5pt}

   \title{Photometry and kinematics of dwarf galaxies from the Apertif \hi\ survey}

   \subtitle{}

   \author{
   B. Šiljeg\inst{1,2}
   \and
   E. A. K. Adams\inst{1,2}
   \and
   F. Fraternali\inst{2}
   \and
   K. M. Hess\inst{3,1,4}
   \and
   T. A. Oosterloo\inst{1,2}
   \and
   A. Marasco\inst{5}
   \and
   B. Adebahr\inst{6}
   \and
   H. Dénes\inst{1,11}
   \and
   J. Garrido\inst{4}
   \and
   D. M. Lucero\inst{7}
   \and
   P. E. Mancera Piña\inst{8}
   \and
   V. A. Moss\inst{9}
   \and
   M. Parra-Royón\inst{4}
   \and
   A. A. Ponomareva\inst{10}
   \and
   S. Sánchez-Expósito\inst{4}
   \and
   J. M. van der Hulst\inst{2}}

    \institute{
    ASTRON, the Netherlands Institute for Radio Astronomy, Oude Hoogeveenseweg 4, 7991 PD Dwingeloo, The Netherlands
    \and
    Kapteyn Astronomical Institute, University of Groningen, P.O. Box 800, 9700 AV, Groningen, The Netherlands
    \and
    Department of Space, Earth and Environment, Chalmers University of Technology, Onsala Space Observatory, 43992 Onsala, Sweden
    \and
    Instituto de Astrof\'{i}sica de Andaluc\'{i}a (CSIC), Glorieta de la Astronom\'{i}a s/n, 18008 Granada, Spain
    \and
    INAF - Padova Astronomical Observatory, Vicolo dell’Osservatorio 5, I-35122 Padova, Italy
    \and
    Ruhr University Bochum, Faculty of Physics and Astronomy, Astronomical Institute (AIRUB), 44780 Bochum, Germany
    \and
    Department of Physics, Virginia Polytechnic and State University, Blacksburg, VA 24061-0435, USA
    \and
    Leiden Observatory, Leiden University, PO Box 9513, NL-2300 RA Leiden, The Netherlands
    \and
    CSIRO Space and Astronomy, PO Box 76, Epping, NSW 1710, Australia
    \and
    Oxford Astrophysics, University of Oxford, Denys Wilkinson Building, Keble Road, Oxford OX1 3RH, UK
    \and 
    School of Physical Sciences and Nanotechnology, Yachay Tech University, Hacienda San José S/N, 100119, Urcuquí, Ecuador}

   \date{}

 
  \abstract
   {Understanding the dwarf galaxy population in low density environments (in the field) is crucial for testing the current $\Lambda$CDM cosmological model. The increase in diversity towards low-mass galaxies is seen as an increase in the scatter of scaling relations such as the stellar mass-size and the baryonic Tully-Fisher relation (BTFR), and is also demonstrated by recent in-depth studies of an extreme subclass of dwarf galaxies of low surface brightness but large physical sizes called ultra-diffuse galaxies (UDGs).}
   {We aim to select dwarf galaxies independent of their stellar content, and make a detailed study of their gas and stellar properties. We select galaxies from the APERture Tile In Focus (Apertif) \hi\ survey, and apply a constraint on their $i-$band absolute magnitude to exclude high-mass systems. The sample consists of 24 galaxies, 22 of which are resolved in \hi\ by at least 3 beams, and span \hi\ mass ranges of 8.6 $\lesssim \log$(\mhi/\msun) $\lesssim$ 9.7 and stellar mass range of
   8.0 $\lesssim \log$(\mstar/\msun) $\lesssim$ 9.7 (with only three galaxies having log (\mstar/\msun)>9).}
   {We determined the geometrical parameters of the \hi\ and stellar discs, built kinematic models from the \hi\ data using \barolo, and extracted surface brightness profiles in \textit{g-, r-} and \textit{i-}band from the Pan-STARRS 1 photometric survey. We used these measurements to place our galaxies on the stellar mass-size relation and the BTFR, and we compared them with other samples from the literature.} 
   {We find that at fixed stellar mass, our \hi\ selected dwarfs have larger optical effective radii than isolated, optically-selected dwarfs from the literature, and we found misalignments between the optical and \hi\ morphologies for some of our sample.
   For most of our galaxies, we used the \hi\ morphology to determine their kinematics, and we stress that deep optical observations are needed to trace the underlying stellar discs.
   Standard dwarfs in our sample follow the same BTFR of high-mass galaxies, whereas UDGs are slightly offset toward lower rotational velocities, in qualitative agreement with results from previous studies. Finally, our sample features a fraction (25\%) of dwarf galaxies in pairs that is significantly larger with respect to previous estimates based on optical spectroscopic data.}
   {}

   \keywords{Galaxies: dwarf -- Galaxies: fundamental parameters -- Galaxies: ISM -- Galaxies: kinematics and dynamics -- Galaxies: photometry}

    \titlerunning{Dwarf galaxies in the Apertif \hi\ survey}
    \authorrunning{B. Šiljeg et al.}
   \maketitle

%

\section{Introduction}
\label{sec:introduction}

    Low-mass galaxies with stellar masses $\lesssim 10^9$\msun\ provide a crucial testing ground for the currently favored cosmological model, with dark energy plus cold dark matter ($\Lambda$CDM), through detailed comparison of observations and cosmological hydrodynamical simulations. Due to their shallow gravitational potential wells, dwarf galaxies are more sensitive to different baryonic effects, such as stellar feedback (e.g. stellar winds, supernovae, photoionisation) and gas turbulence, and can hence inform on the current implementation of sub-grid physics in simulations \citepads[e.g.][]{2017ARA&A..55...59N}. Properly accounting for these effects is important for studying the underlying connection between the baryonic and dark matter, and thereby testing predictions of the $\Lambda$CDM model. The significance of the dwarf galaxy population is already seen in existing discrepancies, such as the large diversity in shapes of rotation curves of observed dwarfs \citepads[e.g.][]{2008AJ....136.2648D, 2015AJ....149..180O,2015MNRAS.452.3650O}, or the too-big-to-fail problem \citepads{2012MNRAS.425.2817F,2015A&A...574A.113P, 2016A&A...591A..58P} (for a more complete picture, we direct the reader to review articles \citetads{2017ARA&A..55..343B}, \citetads{2018PhR...730....1T} and \citetads{2022NatAs...6..897S}).

    Galaxies in relatively isolated environments (field galaxies) make the best targets for tackling the above problems as their properties are a direct outcome of the initial conditions, dictated by cosmology, and their internal evolution, with minimal environmental impact. Field low-mass galaxies tend to be star-forming and gas-rich with gas fractions growing towards lower stellar masses \citepads[e.g.][]{2012ApJ...756..113H,2018MNRAS.476..875C}. This makes neutral hydrogen (\hi) a critical (and often dominant) baryonic component of these systems, and hence \hi\ observations an extremely useful tool for studying these systems. Furthermore, \hi\ observations obtained with interferometers can provide spatially-resolved kinematic information, thereby allowing for the modeling of rotation curves and the full insight into the underlying gravitational potential (baryons plus dark matter).

    Recent studies on a specific sub-class of low-mass galaxies with low surface brightnesses and large physical sizes, the so-called ultra-diffuse galaxies (UDGs), have pointed towards additional challenges for current models which are struggling to reproduce their observed properties \citepads[e.g.][]{2017MNRAS.466L...1D,2022ApJ...936..166K}. The UDGs are defined as having mean effective surface brightness \mmuefX\ > 24 mag arcsec$^{-2}$ and effective radii $R_{e,X} > 1.5$ kpc \citepads{2016A&A...590A..20V}, where $X$ is an optical photometric band (usually $g$ or $r$). They were initially identified in large amounts in cluster environments \citepads{2015ApJ...798L..45V,2015ApJ...809L..21M,2016A&A...590A..20V,2017A&A...608A.142V,2019MNRAS.485.1036M,2022A&A...665A.105L}, but have since been progressively detected also in the field \citepads{2017ApJ...842..133L,2017MNRAS.468..703R,2023ApJS..267...27Z}. While the cluster population is gas-poor, as expected for cluster environments, field UDGs tend to be gas-rich, with gas-fractions going up to very high values ($\langle$\mhi/\mstar$\rangle \simeq 35$ for extremely optically faint systems, \citetads{2017ApJ...842..133L,2019MNRAS.490..566J,2022A&A...659A..14P}). However, it is not yet clear if the two populations have the same origin. The current formation scenarios for the field population (and possibly also the cluster population) include strong stellar outflows \citepads[e.g.][]{2017MNRAS.466L...1D} and high-spin parameters of host dark matter halos \citepads[e.g.][]{2017MNRAS.470.4231R}, while the cluster population has additional formation mechanisms based on tidal interactions and collisions \citepads[e.g.][]{2019MNRAS.490.5182L}.
    
    Generally, observed dwarf galaxies exhibit large diversity in their properties when compared to higher mass systems. One such example is the diversity of physical sizes (measured as effective radii $R_e$) present as a larger scatter in the observed stellar mass-size relation for dwarf galaxies than for higher-mass galaxies \citepads[e.g.][]{2016MNRAS.462.1470L}. The increase in the scatter at low masses seems to be also present in the baryonic Tully-Fisher relation (BTFR) \citepads[see e.g.][]{2009A&A...505..577T,2017MNRAS.464.2419S,2017MNRAS.466.4159I,2022ApJ...940....8M}, which connects the baryonic mass to circular velocity and is a very tight scaling relation for high mass late-type galaxies \citepads[e.g.][]{2016ApJ...816L..14L,2019MNRAS.484.3267L}. It is not yet clear if the observed scatter at low-masses is intrinsic, or if some of the assumptions (e.g. that the kinematics of \hi\ correctly traces the circular speed of the underlying dark matter halo) do not hold at these scales \citepads{2017A&A...607A..13V,2023MNRAS.522.3318D}. The \hi-rich UDGs further complicate this question as they seem to be systematically offset from the BTFR towards low circular velocities as revealed by a spatially-resolved \hi\ study of \citetads{2019ApJ...883L..33M,2020MNRAS.495.3636M}, but is also supported by unresolved studies \citepads{2017ApJ...842..133L,2020ApJ...902...39K,2023ApJ...947L...9H}. 
    
    High gas fractions of field dwarf galaxies make \hi\ surveys extremely useful for finding these systems. This is especially true for the low surface brightness dwarfs, which are exceptionally hard to find in optical surveys. The Arecibo Legacy Fast ALFA (ALFALFA) \hi\  survey \citepads{2005AJ....130.2598G}, a single dish untargeted \hi\ survey conducted with the Arecibo telescope, has already demonstrated the potential of \hi\ studies in finding galaxies overlooked in optical catalogs \citepads[e.g.][]{2015AJ....149...72C,2015ApJ...801...96J,2017ApJ...842..133L}. As one might expect, \hi\ selected galaxies have systematic differences with respect to the optically selected ones, e.g. they tend to be star-forming and bluer \citepads[e.g.][]{2007PhDT.......445K,2012ApJ...756..113H,2012AJ....143..133H,2020AJ....160..271D}, as expected from the presence of \hi\ as fuel for star formation. 

    As mentioned, for a robust determination of kinematic and dynamical properties of galaxies, resolved \hi\ observations have proven crucial as they allow us to determine how regular the system is and, for those dominated by rotation, extract a rotation curve that can be used for dynamical modeling \citepads[e.g.][]{1985ApJ...295..305V,2008AJ....136.2648D,2016MNRAS.462.3628R,2022MNRAS.514.3329M}. Previous \hi\ resolved studies of low-mass galaxies have often been a result of a followup interferometric observations of preselected samples, selected from either untargeted single-dish \hi\ surveys \citepads[e.g.][]{2011ApJ...739L..22C,2021ApJ...909...19G} or optical surveys \citepads[e.g.][]{2012AJ....144..134H}. Being constrained by preselection of a limited number of candidate sources, these studies have been focusing either on very local targets \citepads[e.g.][]{2002A&A...390..829S,2012AJ....144..134H} or on specific types of galaxies, e.g. lowest \hi\ masses \citepads{2011ApJ...739L..22C} or UDGs \citepads{2021ApJ...909...19G}. Due to these constraints, the intermediate regime between standard dwarf galaxies and extreme UDGs has not been explored in detail. A previous untargeted \hi\ survey conducted with the Westerbork Synthesis Radio Telescope (WSRT) \citepads{2007PhDT.......445K} had a good spatial resolution ($\sim$ 30\arcsec) for resolving very closeby dwarfs. However, this survey focussed on a specific sky region towards the Canes Venatici groups of galaxies, and with a limited total sky area (30\dg $\times$ 30\dg) when compared to single dish surveys (e.g. ALFALFA footprint of 6900 deg$^2$), thereby potentially providing a less representative sample of the total dwarf galaxy population.
    
    In addition to kinematic modeling, resolved \hi\ data allow the comparison between stellar and \hi\ geometries. While these are generally consistent for most galaxies, UDGs have been shown to often have misaligned \hi\ and stellar morphologies \citepads{2021ApJ...909...19G}. Exploration of these misalignments is especially important for the correct determination of rotational velocity which strongly depends on the disk inclination. Most previous studies have been assuming stellar-to-gas disk alignment and have considered only one component for the measurement of disk geometry, with some measurements done on \hi\ total intensity maps \citepads{2020MNRAS.495.3636M} and others using optical images \citepads{2020ApJ...902...39K,2023ApJ...947L...9H}. 

    In this work, we draw a sample of low-mass galaxies from an untargeted \hi\ survey undertaken with the phased-array feed, APERture Tile In Focus (Apertif) \citepads{2022A&A...658A.146V,2022A&A...667A..38A}, for the Westerbork Synthesis Radio Telescope (WSRT). With a large sky coverage enabled by Apertif and the good spatial resolution of the WSRT, this allows us to both find the field population of low-mass galaxies and to conduct the kinematic modeling of the galaxies to obtain their rotation curves. It also allows us to compare the stellar and \hi\ geometries of low-mass galaxies. With this, we will be able to position our sample in the stellar mass-size relation and the BTFR, working towards linking the standard dwarf population to the UDG population.
    
    Additionally, the Apertif dataset offers a unique opportunity to explore the frequency of dwarf galaxy pairs or multiples using \hi\ observations, as they are easily resolved with Apertif. Dwarf galaxy interactions are thought to play a role in the evolution of dwarf galaxies, e.g. by igniting or temporarily suppressing the star formation process \citepads[e.g.][]{2015ApJ...805....2S,2024ApJ...963...37K}. However, it is not yet clear how often these interactions take place, and what role they have in the evolution of the overall population of dwarf galaxies. Previous studies in this domain have exclusively used optical spectroscopic data \citepads[e.g.][]{2013MNRAS.428..573S,2018MNRAS.480.3376B}. The importance of including \hi\ in such studies can already be seen by the smaller fraction of low-mass galaxies present in spectroscopic surveys when compared to \hi\ surveys \citepads{2012AJ....143..133H}.

    The paper is organized as follows. In Sect. \ref{sec:data}, we describe the Apertif \hi\ data, PanSTARRS 1 (PS1) data that we use for the optical counterparts, and the source selection. Sect. \ref{sec:analysis} gives an overview of the kinematic analysis done on the \hi\ data and the photometry measurements applied on the PS1 data. In Sect. \ref{sec:results}, we present our results. We compare global properties of our sample with other \hi\ selected samples in the literature, we compare geometries of the \hi\ and stellar disks in our sample, and comment on properties of UDGs in our sample. We place our galaxies on the stellar mass-size relation and the BTFR in Section \ref{sec:results_scaling_relations}. In Sect. \ref{sec:discussion}, we discuss the frequency of pairs in our sample and possible biases in our selection procedure; and we further discuss the properties of our \hi\ selected sample compared to optically selected ones. Finally, in Sect. \ref{sec:conclusion}, we state our conclusions.

\section{Data}
\label{sec:data}

\subsection{Apertif \hi\ data}
\label{sec:Apertif_data}
    Apertif \citepads{2022A&A...658A.146V} is a phased-array feed receiver system designed for the Westerbork Synthesis Radio Telescope (WSRT). It produced 40 instantaneous compound beams on the sky, thereby increasing the field of view of the telescope up to 8 deg$^2$ making it a natural instrument for a wide area survey.

    The Apertif imaging survey \citepads{2022A&A...667A..38A} operated between the 1$^{\rm st}$ of July 2019 and 28$^{\rm th}$ of February 2022. It observed selected regions of the sky above a declination of 30\dg, simultaneously conducting neutral hydrogen (HI) spectral line, continuum and polarisation imaging surveys. Each individual Apertif observation is 11.5 hours. The \hi\ cubes are produced over the topocentric frequency range 1292.5 - 1429.3 MHz \citepads{2022A&A...667A..38A} and have spatial resolution of $ 15$\arcsec$ \times 15$\arcsec$ / \sin \delta $. The Apertif imaging surveys have two tiers: the Apertif Wide-area Extragalactic Survey (AWES) aimed at covering large area of the sky with $\sim$ 2200 deg$^2$ of coverage and one or two observations per field; and the Apertif Medium-deep Extragalactic Survey (AMES) aims to go deeper targeting specific smaller area of the sky with up to 10 observations per field and the total sky coverage of $\sim 130$ deg$^2$ \citepads[see][]{2022A&A...667A..38A}.

    We used a preliminary \hi\ source list made by running the Source Finding Application (SoFiA) \citepads{2015MNRAS.448.1922S,2021MNRAS.506.3962W} on individual \hi\ cubes from single observations taken from August through December 2019 (Hess, priv. comm.). These cubes are separate for individual compound beams and have a typical noise of around 1.6 mJy$\,$beam$^{-1}$ over 36.6 kHz ($\sim$8 \kms) \citepads{2022A&A...667A..38A}. The SoFiA algorithm conducts source finding by using multiple resolution kernels and searches for detections across different kernels based on the $S/N$. For the production of the source list used in this work, SoFiA was run using 3 spatial kernels: 15\arcsec, 23.4\arcsec\ and 39\arcsec; and 3 spectral kernels: 7.7 \kms, 23.2 \kms\ and 54.1 \kms (Hess et al., in prep.). The mask of the source was then constructed using all pixels that had $S/N$ > 3.8 in one or more combinations of spatial and spectral smoothing kernels.
    In the next step, SoFiA corrects for false positive detections by comparing positive and negative detections, with reliability threshold of 0.85 in our case. For sources deemed real, the pipeline proceeds to calculate the \hi\ properties of the detection by producing moment maps and global spectral profiles. Global spectral profiles are constructed using a 2D projection of the initial 3D mask of the source and integrating corresponding values in each channel, both inside and outside the original 3D mask. 
    We note that systemic velocities in the preliminary source list are given in topocentric reference frame. The barycentric correction would be $\lesssim$ 30 \kms, i.e. 1.5\% at velocities used in this work (>2000 \kms). Preliminary distances are calculated from systemic velocities assuming Hubble flow with $H_0=70$ \kms\ Mpc$^{-1}$.
    
    We use the preliminary source list only for the selection of the sample (described in Sect. \ref{sec:source_selection}), while the \hi\ cubes used for the kinematic modeling come from a subsequent source finding (currently in progress) that was run on data that have been co-added from all available observations and mosaicked between compound beams within individual fields (full description of the improved processing is in Hess et al., in prep). Therefore, the data products used in this work will have better signal-to-noise ratios (S/N) than predicted by the preliminary source list. 
    
    The properties of the cubes used in this work are given in Table \ref{tab:technical_params}.
    We note that the spectral resolution degrades during the co-adding step due to shifting to a common velocity frame. Noise levels also vary due to differing numbers of observations per field, independent mosaicking of individual fields producing non-uniform noise on the large scale, and instrumental effects (e.g. continuum subtraction) within the Apertif cubes themselves.

\begin{table*}
\centering
\caption{Properties of \hi\ cubes for galaxies in the sample.}
\begin{tabular}{lllllll}
\hline \hline
 & Apertif name & $N_{\rm obs}$ & Noise & BMAJ $\times$ BMIN & BPA & spec. res. \\
 & & & [mJy/beam] & [arcsec $\times$ arcsec] & [\dg] & [km/s] \\
\hline
1\&2 & AHCJ020345.3+371444 & 10 & 0.78 & $27.0 \times 13.8$ & 2.6 & 8.05 \\
3\&4 & AHCJ133650.2+320534 & 1 & 1.09 & $28.5 \times 14.0$ & -2.6 & 7.73 \\
5\&6 & AHCJ101017.8+582856 & 2 & 1.11 & $20.9 \times 15.9$ & 0.6 & 7.99 \\
7 & AHCJ133045.2+324548 & 1 & 3.0 & $28.6 \times 14.1$ & 0.4 & 7.73 \\
8 & AHCJ133042.2+294735 & 1 & 1.4 & $31.2 \times 14.4$ & -1.4 & 7.73 \\
9 & AHCJ133507.0+313118 & 1 & 1.05 & $28.5 \times 14.0$ & -2.6 & 7.73 \\
10 & AHCJ133704.8+315336 & 1 & 1.14 & $28.5 \times 14.0$ & -2.6 & 7.73 \\
11 & AHCJ101655.0+582325 & 2 & 0.79 & $20.9 \times 15.9$ & 0.6 & 7.99 \\
12 & AHCJ130830.2+543756 & 2 & 1.55 & $19.4 \times 14.3$ & 0.4 & 8.69 \\
13 & AHCJ135938.5+372631 & 1 & 1.42 & $27.0 \times 14.2$ & -4.8 & 7.73 \\
14 & AHCJ220741.6+400853 & 2 & 1.17 & $37.4 \times 24.4$ & -1.2 & 8.42 \\
15 & AHCJ220743.8+414343 & 2 & 1.09 & $37.4 \times 24.4$ & -1.2 & 8.42 \\
16 & AHCJ223258.7+393853 & 2 & 1.03 & $34.3 \times 20.3$ & -0.1 & 10.38 \\
17 & AHCJ223902.0+383211 & 2 & 1.14 & $34.3 \times 20.3$ & -0.1 & 10.38 \\
18 & AHCJ221640.4+402424 & 2 & 0.85 & $26.4 \times 16.6$ & -1.1 & 8.42 \\
19 & AHCJ221800.3+405946 & 2 & 0.75 & $26.4 \times 16.6$ & -1.1 & 8.42 \\
20 & AHCJ222407.1+411511 & 2 & 0.93 & $26.4 \times 16.6$ & -1.1 & 8.42 \\
21 & AHCJ131846.6+274359 & 2 & 2.29 & $46.5 \times 20.4$ & -0.4 & 9.05 \\
22 & AHCJ133339.8+602315 & 2 & 1.86 & $20.7 \times 16.6$ & 4.4 & 10.97 \\
23 & AHCJ222230.6+360028 & 2 & 0.98 & $25.9 \times 14.5$ & -2.7 & 8.57 \\
24 & AHCJ224941.0+394852 & 2 & 0.93 & $24.3 \times 14.1$ & 0.1 & 7.75 \\
\hline
\label{tab:technical_params}
\end{tabular}
\tablefoot{The Apertif name is based on the position of the source corresponding to the right ascension and declination in J2000 as following: AHCJhhmmss.s+ddmmss. The first three rows correspond to detections containing a pair of galaxies, so we assign them two numbers in the leftmost column for easier correspondence between all tables in the paper. $N_{\rm obs}$ is the number of observations. Noise is the root mean square (rms) noise in the data cube. BMAJ and BMIN represent FWHM of major and minor axes of the synthesized beam, while BPA is the position angle of the beam. 
Spectral resolution is measured by summing in quadrature the channel width of Apertif observations with the standard deviation of relative shifts between spectral axes of observations taken at different times. Note that the channel width is the same for all cubes, and is equal to 7.73 \kms.}
\end{table*}

\subsection{Optical data}
\label{sec:optical_data}

    For the determination of the stellar properties of our galaxy sample, we use data from the Pan-STARRS 1 (PS1) survey \citepads{2016arXiv161205560C}, as the only optical survey to date encompassing the whole Apertif coverage. This is a broadband photometric survey made with the 1.8 meter telescope stationed at the Haleakala Observatories in Hawaii. The PS1 survey covers the whole sky region north from $\delta = -30$\dg\ which completely includes the Apertif coverage. In this work, we use $g$-, $r$- and $i$-bands PS1 images with median seeings of 1.31\arcsec, 1.19\arcsec and 1.11\arcsec. Following the procedure described in the appendix A from \citetads{2020A&A...644A..42R}, we measured the surface brightness depth of PS1 images at the 3$\sigma$ noise level over $10\text{\arcsec}\times 10\text{\arcsec}$ area obtaining $\mu (3\sigma_{10\text{\arcsec}\times 10\text{\arcsec}}) = (27.43, 27.21, 27.12)$ mag arcsec$^{-2}$ in $g$-, $r$- and $i$-bands, respectively. 
    In comparison, PS1 images are around 0.5 mag arcsec$^{-2}$ deeper than the Sloan Digital Sky Survey (SDSS) \citepads{2018ApJS..235...42A} and about 1 mag arcsec$^{-2}$ shallower than The Dark Energy Camera Legacy Survey (DECaLs) \citepads{2019AJ....157..168D}.
    
\subsection{Source selection}
\label{sec:source_selection}

    As we are interested in gas-rich low surface brightness, low stellar mass galaxies, we base our selection on the preliminary \hi\ source list from Apertif (see Sect. \ref{sec:Apertif_data}). This enables us to find galaxies that could otherwise go undetected in optical surveys. In addition, we use the PS1 source catalog as a final check in order to exclude high stellar mass galaxies with low gas mass fractions.
    
    The selection of sources from the preliminary \hi\ source list of Apertif was based on the following constraints:
    \begin{enumerate}[label=(\alph*)]
        \item \hi\ mass: \mhi\ < $10^{10}$ \msun,
        \item width of the global profile at 50\% of the peak emission: \w50\ < 150 \kms,
        \item average signal-to-noise ratio per channel in the global profile: \snrgp\ > 3,
        \item systemic velocity: \vsys\ > 2000 \kms,
        \item the \hi\ disk resolved by at least 3 Apertif beam elements.
    \end{enumerate}
    Additionally, we considered only sources for which all observations are fully processed and have been co-added so the improved data products are available. In the following, we describe the procedure in more detail.

    Conditions a) and b) are imposed in order to exclude high-mass galaxies. While the latter condition seems to also exclude high inclination galaxies, we point the reader to Section \ref{sec:discussion_HI_sample} where we discuss the possible biases of our selection procedure on the inclination.

    Condition c) excludes low $S/N$ detections. As we intend to kinematically model our sample using the total 3D information (see Sect. \ref{sec:analysis-kinematic_model}), we apply our condition on the average signal-to-noise ratio per channel in the global profile, defined as:
    \begin{equation}
        \text{\snrgp} = \frac{F_{\text{\hi}}\, [\rm{ Jy\, Hz}]}{\mathcal{N}_{\rm ch}\cdot \Delta \nu\,[\rm Hz]\cdot \rm{rms}\, [\rm Jy] }
        \label{eq:SNR_GP}
    \end{equation}
    where $F_{\text{\hi}}$ is the total detected flux, $\mathcal{N}_{\rm ch}$ the number of channels with detected emission, $\Delta \nu$ the channel width, and rms the root mean square noise measured in channels without emission in the global profile of the source (constructed as described in Sect. \ref{sec:Apertif_data}). We chose this definition because it gives an estimate of the average level of emission $\left(F_{\text{\hi}}\, [\rm{ Jy\, Hz}] / (\mathcal{N}_{\rm ch}\cdot \Delta \nu\,[\rm Hz] )\right)$ compared to the average level of integrated noise ($\rm{rms}\, [\rm Jy]$) per channel, providing a better estimate for success in producing reliable 3D kinematic models by ensuring sufficient signal in each channel.
    
    Condition d) aims to minimize the uncertainties introduced by peculiar motions on the estimated distances. 
    The threshold of 2000 \kms\ is chosen so that distance errors are not dominated by typical peculiar velocities of galaxies in the field (one to a few hundred \kms).

   Condition e) aims to ensure that the galaxy is spatially resolved enough for the construction of a reliable 3D kinematic model. Given that \hi\ diameter is not measured in the source finding with SoFiA, we estimate the physical sizes of sources in the preliminary source list using the \hi\ mass - \hi\ diameter relation from \citetads{2016MNRAS.460.2143W}:
    \begin{equation}
        \log \frac{d_{\rm \text{\hi}}}{\text{kpc}} = (0.506 \pm 0.003)\, \log \frac{M_{\rm \text{\hi}}}{\text{\msun}} - (3.293 \pm 0.009)
    \end{equation}
    where $d_{\rm \text{\hi}}$ is measured at a surface density of 1 \msun pc$^{-2}$. 
    Due to the unknown orientation of the galaxy, we take the geometric mean of the beam axes as an estimate of the beam diameter. We then include sources for which the estimated angular diameter is 3 or more times larger than the beam diameter. We note that this condition ultimately corresponds to an \hi\ flux cut, but given that we apply it to individual detections (with different beam sizes, see Table \ref{tab:technical_params}), it cannot be applied as a universal flux cut on the sample.

    After applying the above constraints on the \hi\ content and obtaining a list of candidates, we move on to the optical counterpart. In order to exclude optically bright and high stellar mass sources from our selection, we put a lower limit of $-18.5$ on the $i$-band absolute magnitude of candidate sources. This threshold roughly corresponds to stellar masses of $\lesssim 10^9$ \msun, based on a typical $g-r$ color of 0.2, following the relation from \citetads{2016AJ....152..177H}. For this, we search the PS1 source catalog inside the 20\arcsec\ radius from the \hi\ center position of each source. We exclude the contamination of foreground stars by following the procedure described in \citetads{2014MNRAS.437..748F}, i.e. we put a lower limit threshold of 0.05 mag on the difference between the measured Kron and point spread function (PSF) magnitudes in $i$-band for each detection. Finally, we transfer apparent magnitudes in $i$-band ($m_{\rm i}$) to absolute magnitudes ($M_{\rm i}$) using the Hubble flow distance estimate from the \hi\ source list and apply the cut of $M_{\rm i} > -18.5$. 
    
    After applying the selection criteria described above, we manually check the PS1 and \hi\ data of the selected candidates. In a few cases, SoFiA detected an \hi\ tail of a larger galaxy as a separate source. In addition, some pairs of galaxies were detected as a single source and the center position of the detection was placed between the two galaxies. These galaxies were in some cases outside of the 20\arcsec\ radius in which the PS1 catalog was searched for detections. For these cases, we manually searched the PS1 catalog again and excluded the sources in which the corresponding galaxies failed to pass the required selection criteria.

    Out of 1231 sources in the source list, 76 detections passed the \hi\ criteria, out of which 47 had fully processed data needed for the production of improved data products (data with co-added observations and mosaicked across compound beams). Out of these, 24 passed the optical criteria.

    Properties of \hi\ cubes of the obtained sample are provided in Table \ref{tab:technical_params}. Refined physical properties of the selected galaxies (see Sect. \ref{sec:dist_and_masses}) are listed in Table \ref{tab:general_properties}.
    Note that for galaxies in pairs (see Sect. \ref{sec:results}), the cube was separated in two to allow for measurements of properties of individual galaxies. The obtained \hi\ masses range between $8.6\lesssim \log \text{(\mhi/\msun)} \lesssim 9.7$, while the stellar masses range between $8.0\lesssim \log \text{(\mstar/\msun)} \lesssim 9.7$. While the stellar mass range seems to go to higher values than should be permitted by our selection, we note that only 3 galaxies in the sample have $\log \text{(\mstar/\msun)} > 9.0$.
    
    When available, we provide names from the literature obtained by searching the NASA/IPAC Extragalactic Database\footnote{\url{https://ned.ipac.caltech.edu/}} (NED). For 7 galaxies in our sample, there was no recorded entry in NED, and 4 previously known galaxies have no archival redshift measurement. All of the galaxies previously detected in \hi\ had only single-dish observations and were spatially unresolved. When referring to individual galaxies, we will use their literature names when they are shorter, and Apertif names otherwise (shortened as AHCJhhmm+ddmm).

\begin{table*}
\centering
\caption{Global properties of the sample.}
\begin{tabular}{llllllllll}
\hline \hline
& Apertif name & Literature name & D & \vsys & \w50 & $F_{\text{\hi}}$ & log & log  & log  \\
& AHC- &  & [Mpc] & [\kms] & [\kms] & [Jy \kms] &  (\mhi/\msun) & (\mstar/\msun) & ($M_{\rm b}$/\msun) \\
\hline
1 & J020345.3+371444 & - & 50 & ${4526} \pm {4}$ & 36 & 0.99 & ${8.77}_{-0.17}^{+0.12}$ & ${8.37}_{-0.26}^{+0.28}$ & ${9.01}_{-0.17}^{+0.12}$ \\
2 & J020345.3+371444 & a* & 50 & ${4517} \pm {4}$ & 36 & 3.39 & ${9.30}_{-0.17}^{+0.12}$ & ${8.78}_{-0.27}^{+0.28}$ & ${9.51}_{-0.17}^{+0.12}$ \\
3 & J133650.2+320534 & UGC 8605 & 48 & ${3004} \pm {4}$ & 55 & 2.46 & ${9.12}_{-0.17}^{+0.12}$ & ${8.76}_{-0.23}^{+0.24}$ & ${9.37}_{-0.16}^{+0.12}$ \\
4 & J133650.2+320534 & UGC 8602 & 48 & ${3066} \pm {4}$ & 77 & 1.78 & ${8.98}_{-0.17}^{+0.13}$ & ${8.69}_{-0.29}^{+0.32}$ & ${9.25}_{-0.21}^{+0.14}$ \\
5 & J101017.8+582856 & CGCG 290-011 & 35 & ${2134} \pm {4}$ & 70 & 1.25 & ${8.55}_{-0.18}^{+0.12}$ & ${8.61}_{-0.21}^{+0.20}$ & ${8.95}_{-0.17}^{+0.12}$ \\
6 & J101017.8+582856 & UGC 5480 & 35 & ${2158} \pm {4}$ & 110 & 3.53 & ${9.00}_{-0.17}^{+0.12}$ & ${8.84}_{-0.21}^{+0.21}$ & ${9.31}_{-0.16}^{+0.12}$ \\
7 & J133045.2+324548 & UGC 8503 & 67 & ${4680} \pm {4}$ & 56 & 3.81 & ${9.61}_{-0.12}^{+0.09}$ & ${9.28}_{-0.20}^{+0.20}$ & ${9.87}_{-0.12}^{+0.09}$ \\
8 & J133042.2+294735 & AGC 239039 & 47 & ${2949} \pm {4}$ & 86 & 2.02 & ${9.01}_{-0.18}^{+0.12}$ & ${8.56}_{-0.28}^{+0.31}$ & ${9.24}_{-0.18}^{+0.13}$ \\
9 & J133507.0+313118 & AGC 234932 & 70 & ${4953} \pm {4}$ & 49 & 1.38 & ${9.20}_{-0.17}^{+0.13}$ & ${8.53}_{-0.19}^{+0.18}$ & ${9.39}_{-0.15}^{+0.11}$ \\
10 & J133704.8+315336 & AGC 239112 & 48 & ${3022} \pm {4}$ & 48 & 1.28 & ${8.83}_{-0.18}^{+0.12}$ & ${8.04}_{-0.24}^{+0.25}$ & ${9.01}_{-0.16}^{+0.12}$ \\
11 & J101655.0+582325 & UGC 5541 & 37 & ${2265} \pm {4}$ & 139 & 5.57 & ${9.24}_{-0.18}^{+0.12}$ & ${8.66}_{-0.18}^{+0.17}$ & ${9.46}_{-0.15}^{+0.11}$ \\
12 & J130830.2+543756 & b* & 42 & ${2542} \pm {4}$ & 95 & 6.94 & ${9.45}_{-0.17}^{+0.13}$ & ${8.76}_{-0.27}^{+0.29}$ & ${9.63}_{-0.16}^{+0.12}$ \\
13 & J135938.5+372631 & c* & 42 & ${2710} \pm {4}$ & 51 & 2.95 & ${9.10}_{-0.18}^{+0.13}$ & ${8.43}_{-0.23}^{+0.25}$ & ${9.29}_{-0.16}^{+0.12}$ \\
14 & J220741.6+400853 & - & 56 & ${4702} \pm {4}$ & 129 & 3.00 & ${9.35}_{-0.17}^{+0.13}$ & ${9.46}_{-0.28}^{+0.31}$ & ${9.72}_{-0.22}^{+0.15}$ \\
15 & J220743.8+414343 & d* & 78 & ${5705} \pm {4}$ & 135 & 2.62 & ${9.57}_{-0.18}^{+0.12}$ & ${8.51}_{-0.40}^{+0.45}$ & ${9.72}_{-0.18}^{+0.12}$ \\
16 & J223258.7+393853 & - & 47 & ${3999} \pm {5}$ & 132 & 2.17 & ${9.04}_{-0.18}^{+0.13}$ & ${8.78}_{-0.21}^{+0.21}$ & ${9.32}_{-0.15}^{+0.11}$ \\
17 & J223902.0+383211 & - & 53 & ${4546} \pm {5}$ & 59 & 1.98 & ${9.12}_{-0.17}^{+0.12}$ & ${8.39}_{-0.30}^{+0.35}$ & ${9.30}_{-0.17}^{+0.12}$ \\
18 & J221640.4+402424 & - & 55 & ${4612} \pm {4}$ & 78 & 1.14 & ${8.91}_{-0.18}^{+0.12}$ & ${8.63}_{-0.26}^{+0.27}$ & ${9.18}_{-0.18}^{+0.13}$ \\
19 & J221800.3+405946 & - & 50 & ${4262} \pm {4}$ & 139 & 1.61 & ${8.97}_{-0.18}^{+0.13}$ & ${8.59}_{-0.23}^{+0.23}$ & ${9.22}_{-0.16}^{+0.12}$ \\
20 & J222407.1+411511 & UGC 12027 & 48 & ${4133} \pm {4}$ & 93 & 4.80 & ${9.42}_{-0.18}^{+0.13}$ & ${9.01}_{-0.21}^{+0.21}$ & ${9.66}_{-0.15}^{+0.11}$ \\
21 & J131846.6+274359 & UGC 8363 & 38 & ${2455} \pm {5}$ & 144 & 7.79 & ${9.41}_{-0.18}^{+0.12}$ & ${8.91}_{-0.20}^{+0.20}$ & ${9.64}_{-0.15}^{+0.11}$ \\
22 & J133339.8+602315 & UGCA 363 & 32 & ${2066} \pm {5}$ & 34 & 5.14 & ${9.10}_{-0.18}^{+0.12}$ & ${8.44}_{-0.20}^{+0.19}$ & ${9.30}_{-0.15}^{+0.11}$ \\
23 & J222230.6+360028 & UGC 12005 & 73 & ${5476} \pm {4}$ & 54 & 3.93 & ${9.69}_{-0.17}^{+0.12}$ & ${9.69}_{-0.23}^{+0.23}$ & ${10.05}_{-0.19}^{+0.13}$ \\
24 & J224941.0+394852 & - & 69 & ${5349} \pm {4}$ & 35 & 4.66 & ${9.71}_{-0.17}^{+0.12}$ & ${8.54}_{-0.24}^{+0.26}$ & ${9.86}_{-0.17}^{+0.12}$ \\
\hline
\end{tabular}
\tablefoot{ 
            The first six galaxies are galaxies found in pairs.
            $D$ is distance with errors of 15\%. 
            \vsys\ is systemic velocity given in the optical convention and the barycentric rest frame.
            \w50\ is the width of the global profile at 50\% peak emission, with errors of the same order as \vsys. $F_{\text{\hi}}$ is the total \hi\ flux of the source with errors of 15\%.
            \mhi, \mstar\ and \mbar\ are \hi, stellar and baryonic masses, respectively.\\
            * a) WISEA J020343.19+371442.6, b) SDSS J130830.62+543757.4, c) WISEA J135937.97+372636.1, d) UGC 11919:[SJZ2013] 22.
            }
            \label{tab:general_properties}
\end{table*}

\section{Analysis}
\label{sec:analysis}

\subsection{Summary of \hi\ masks used in this work}
\label{sec:masks_HI}

    In this work, we will use several masking techniques for the \hi\ data, depending on the purpose of the task. In this section, we shortly describe each mask for easier reference.

    As mentioned in Sect. \ref{sec:Apertif_data}, SoFiA source finder produces a mask of the source after each detection. This mask is produced using multiple spatial and spectral resolution kernels and extends far enough out to enclose the galaxy emission down the level of noise. Because of this, it allows us to robustly measure the total flux of the galaxy which we will use for the calculation of the \hi\ mass (see Sect. \ref{sec:dist_and_masses}).

    In contrast, when performing kinematic modeling using the whole 3D cube (see Sect. \ref{sec:analysis-kinematic_model}), we prefer to limit ourselves to regions of high $S/N$ at the full spatial resolution of the data cube, in order to minimize the effect of noise on the final kinematic model. In this case, we will make use of the default masking scheme from \barolo\ (SEARCH option, see \barolo\ documentation\footnote{\url{https://bbarolo.readthedocs.io/en/latest/}}) to produce a mask that includes most of the galaxy's emission, while minimising the influence of external noise peaks.
    
    And lastly, for the estimation of the geometry of the \hi\ disk using the total intensity map (see Sect. \ref{sec:analysis-kinematic_model}), we wish to include low level emission to capture the outskirt of the galaxy, but still stay in the mid-high $S/N$ regime in order to limit the influence of the noise on the derived geometry. To produce this mask, we again use \barolo, but this time we add an option to smooth the cube 1.2 times before selecting regions of mid-high $S/N$.

\subsection{Three-dimensional kinematic modeling}
\label{sec:analysis-kinematic_model}
    
    To obtain the rotational velocities of our sample, we adopt the tilted ring model \citepads{1974ApJ...193..309R,1987PhDT.......209B}. In this model, the gas is assumed to be in pure circular motion. The galaxy is broken down into concentric rings of different radii, each with its own set of parameters. For simplicity, we group parameters into two categories: geometric (coordinates of the center ($x_0$, $y_0$), thickness of the ring ($z_0$), inclination ($i$) and position angle ($PA$)); and kinematic (systemic velocity (\vsys), rotational velocity (\vrot) and velocity dispersion (\disp)). We make distinction between two position angles (both measured anticlockwise from the north direction): the morphological (\PAhi) describing the geometry of the \hi\ disk as projected on the sky, and kinematic (\PAkin) describing the angle with the steepest gradient in velocity. The tilted ring model was historically widely used for studying galaxy kinematics from 2D velocity field maps \citepads[e.g.][]{1987PhDT.......199B}. More recent studies applied the model directly to 3D emission line data cubes \citepads[e.g.][]{2012ascl.soft08008J,2015MNRAS.451.3021D} which proved essential for accurate determination of underlying kinematics by enabling the correction of the beam smearing effect \citepads[e.g.][]{2009A&A...493..871S}.

    \subsubsection{\barolo\ set-up and assumptions}
    \label{sec:barolo_setup_assumptions}
    
    To derive kinematic properties of our sample, we use the \barolo\ \citepads{2015MNRAS.451.3021D} software. \barolo\ was designed for (and well tested on) poorly resolved data \citepads{2015MNRAS.451.3021D,2020MNRAS.495.3636M,2023MNRAS.521.1045R}, making it a well-suited tool for this study.
    The software builds 3D model cubes based on the tilted ring model and compares them with the data cube to find the best fitted model. 
    The main strength of this software is the convolution step where the model is convolved with a Gaussian beam before the calculation of residuals between the model and the data cube, thereby minimizing the effect of the beam smearing on the model.
    
    Here we describe a few key parameters when running \barolo, while the full parameter file is given in Appendix \ref{App:par_file}. As mentioned in \ref{sec:masks_HI}, we produce the mask using the default masking option by \barolo. Furthermore, we choose the azimuthal normalisation for the flux of the model cube, as our measurements do not have enough spatial resolution to trace gas distribution in much detail. We choose $\cos^2(\theta)$ for the weighing function in the computation of residuals, thereby giving priority to the major axis where rotation is better traced. Finally, we put a limit for the minimum velocity dispersion to half of the spectral resolution to avoid unrealistically low dispersions that can arise in low $S/N$ conditions. We note that in our final kinematic models, the obtained velocity dispersions never reach this floor value.
    
    When running \barolo, we make a few assumptions about the geometric parameters. We do not allow $x_0$, $y_0$, $i$ and \PAkin\ to change between different radii, as we cannot track changes in geometric parameters across different rings due to the relatively poor spatial resolution of the galaxies in our sample. The only exception is \mess\ which is sampled by 8 compound beams along the major axis (for the 1.5 times smoothed cube used for the modeling, see Sect. \ref{sec:obtained_kin_models}) and shows signs of a warp. For this reason, we let \PAkin\ and $i$ change between radii in this case. Additionally, any effect of the disk thickness in the data is dominated by the beam. Physical beam diameters in the sample are all $\geq 2.3$ kpc, while dwarf galaxy disk thickness ranges from a few hundred pc to $\sim 1$ kpc \citepads[e.g.][]{2011MNRAS.415..687B,2020ApJ...889...10D,2020A&A...644A.125B}. Even in our physically best resolved case, kinematic models produced with $z_0=0$ pc and $z_0=500$ pc are fully consistent. Therefore, we assume a razor thin rotating \hi\ disk ($z_0=0$ pc).

    \subsubsection{Kinematic modeling procedure}
    
    We constrain the geometric parameters of the \hi\ disk using a publicly available python script CANNUBI\footnote{\url{https://www.filippofraternali.com/cannubi}} (briefly described in \citetads{2022MNRAS.512.3230M,2023MNRAS.521.1045R}). CANNUBI uses a Markov Chain Monte Carlo (MCMC) approach where values of parameters are evaluated based on the residuals between the data and the corresponding model produced with \barolo, made using either total flux maps, or the 3D cubes themselves. We used the total flux maps as the computational time is notably shorter. The mask used in this step is defined in Sect. \ref{sec:masks_HI}. With CANNUBI, we fit $x_0$, $y_0$, $i$, \PAhi\ and the extent of the \hi\ disk. 
    We note that, given the angular resolution of our data, our assumption of the razor thin \hi\ disk does not compromise the derivation of the \hi\ inclination as long as true inclination is below $\sim$80\dg.
    In this regime, the residuals of moment 0 maps between models with $z_0=0$ pc and $z_0=500$ pc are lower than the typical rms noise in the Apertif maps, thereby making the two models indistinguishable.

    For the final kinematic fit with \barolo, we use the median values of posteriors from CANNUBI to constrain the geometric parameters. 
    In one case (\svel), we used the optical inclination, as the optical geometry was significantly better aligned with the galaxy kinematics.
    The morphological position angle is given as an initial guess for the kinematic position angle (\PAkin) (except for 3 galaxies, \pairtworight, \svel\ and \paver, for which we provide no initial guess on \PAkin), but we let \barolo\ fit it as part of a two-stage run. In the two-stage run, \barolo\ first fits unconstrained parameters (\vsys\ and \PAkin\ in our case) together with \vrot\ and \disp\ in each ring, after which it fits a functional form (a constant value in our case, see Sect. \ref{sec:barolo_setup_assumptions}) to the radial distribution of these parameters.
    In the second run, \barolo\ again fits \vrot\ and \disp, now keeping all other parameters fixed.

    We propagate errors on the inclination by running \barolo\ with the same parameter files as in the original run (with inclination $i$), but now changing the inclination to $i\pm \sigma_i$. The error on the final rotational velocity is then taken to be the difference between the original case and the case $i+ \sigma_i$ ($i- \sigma_i$) for lower (upper) error estimate when the difference is higher than the statistical error obtained from \barolo. Otherwise, we adopt the \barolo\ error on rotational velocities.

    \subsubsection{Obtained kinematic models}
    \label{sec:obtained_kin_models}

    Out of 24 initial galaxies, 13 showed a velocity gradient that can be interpreted as regular rotation, and had sufficient $S/N$ per channel that enabled us to perform the modeling. In a few cases with low $S/N$ (\bSNRtwo, \tstd, \mess\ and \dwtm, Figs. \ref{fig:HI-op-bSNR_2}, \ref{fig:HI-op-tstd}, \ref{fig:HI-op-mess} and \ref{fig:HI-op-dwtm}), we smoothed the cubes 1.5 times the synthesized beam before running CANNUBI to increase the $S/N$ and consequently pick up the faint emission otherwise hidden in the noise. We also used these smoothed cubes for the final kinematic modeling in these cases. 
    
    The results of kinematic modeling are listed in Table \ref{tab:HI_par}. We report the maximum value of rotational velocity, and the mean value of velocity dispersion across all rings. \vsys\ is reported in Table \ref{tab:general_properties}. For kinematically modeled galaxies, we report \vsys\ of the best fit model, while for the rest of the sample it is measured as a weighted mean from the global spectral profile. The error corresponds to half of the spectral resolution.
    
\begin{figure*}
        \centering
        \includegraphics[width = 0.95 \textwidth]{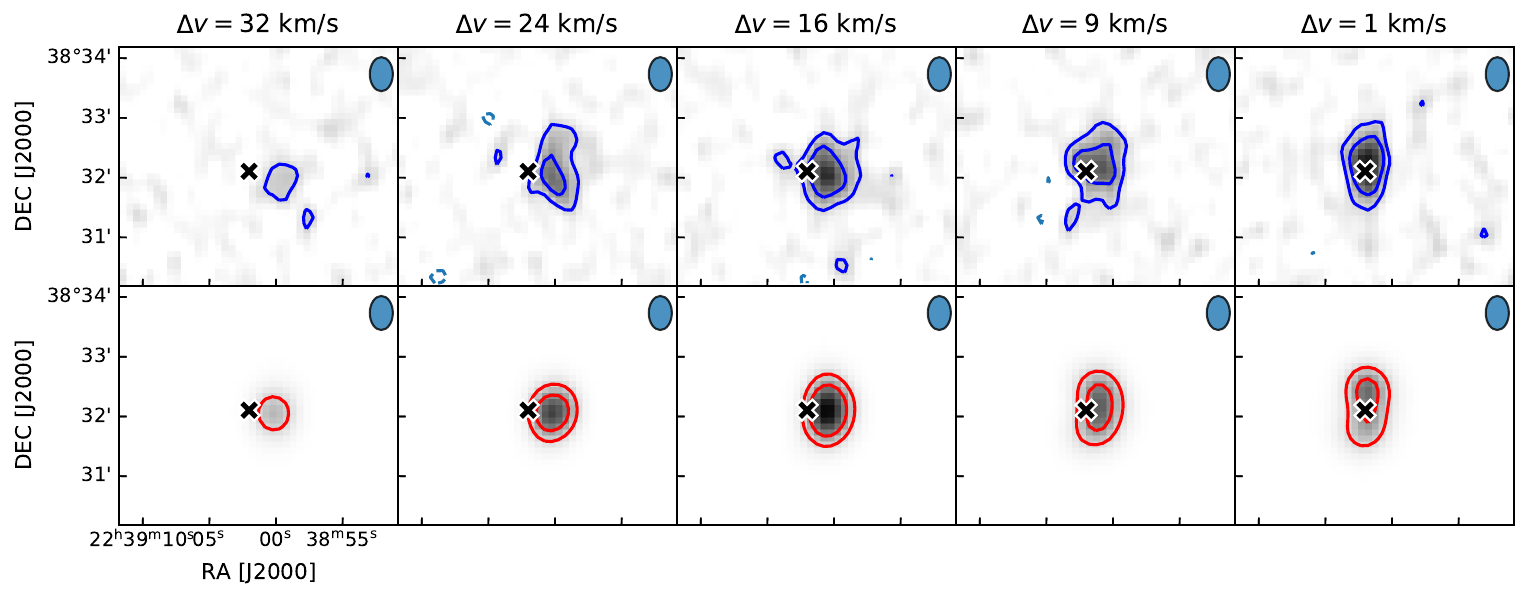}\\
        \centering
        \includegraphics[width = 0.95 \textwidth]{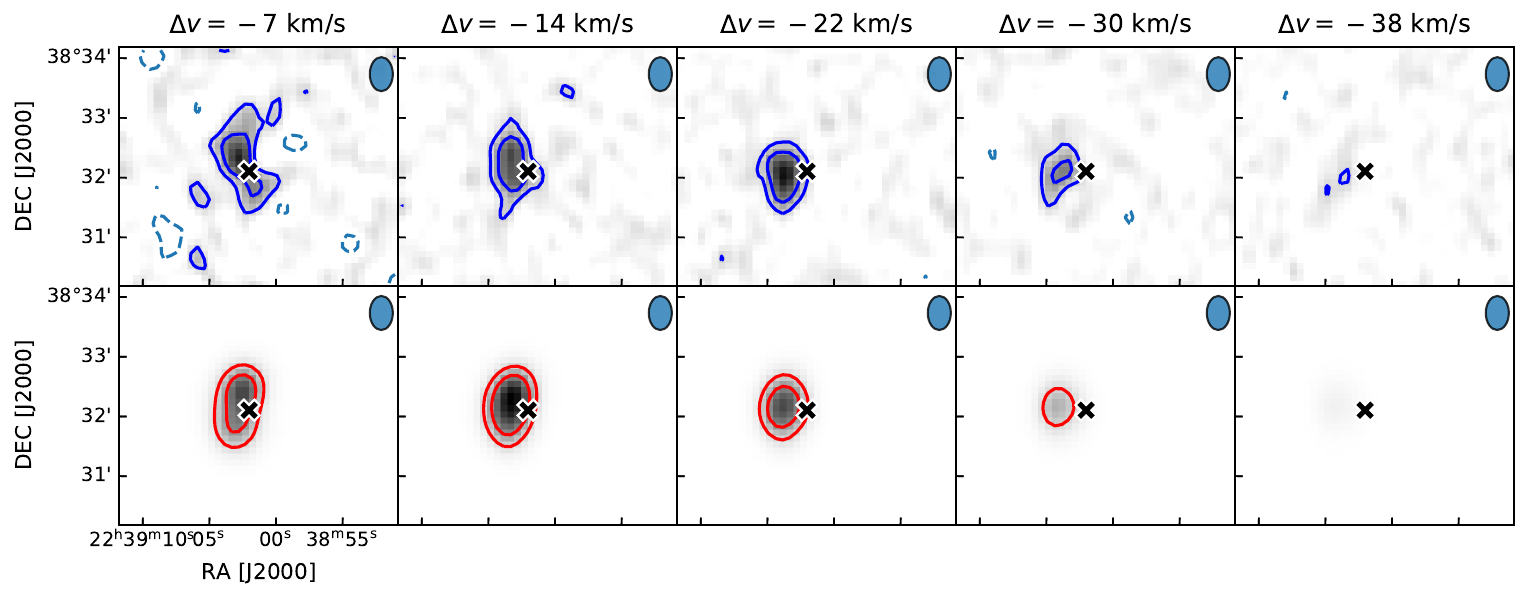}\\
        \centering
        \caption{Comparison between the channel map of the galaxy \pavel\ and its best-fit model. Top panels and blue contours represent the data, while the lower panels and red contours represent the best fit model. Contours are plotted starting from 3 times the noise per channel in the cube reported in Table \ref{tab:technical_params}, and are spaced by a factor of 2 in intensity. The black $X$ indicated the center of the galaxy, as determined by CANNUBI (see Sect. \ref{sec:analysis-kinematic_model}).}
        \label{fig:ch_maps}
    \end{figure*}

\begin{figure*}
        \centering
        \includegraphics[width = 1. \textwidth]{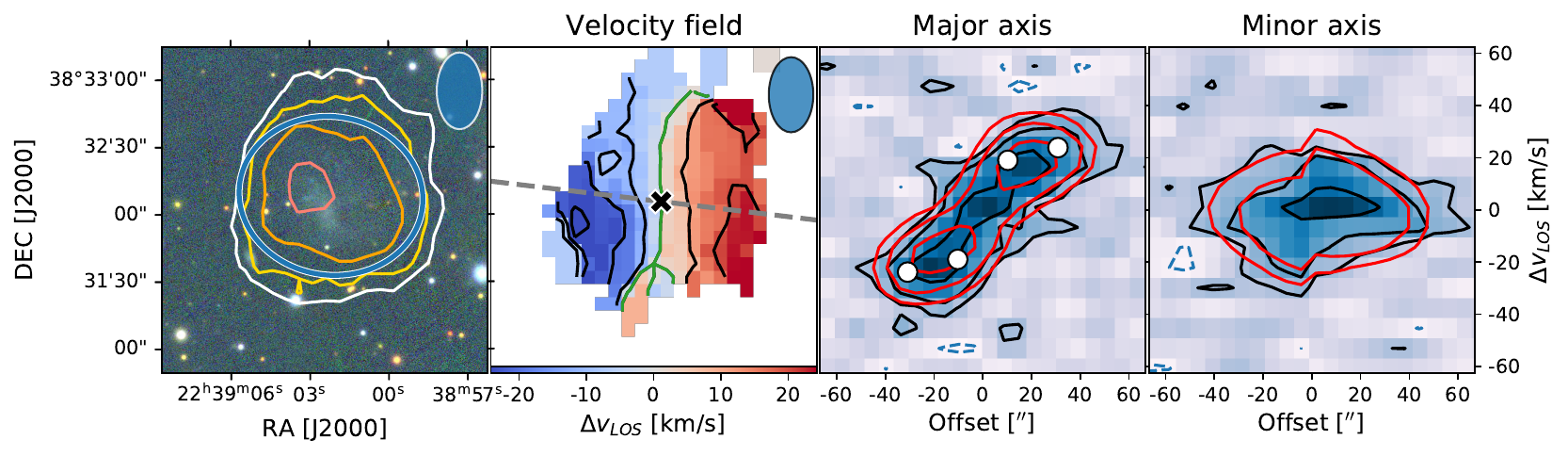}
        \caption{\hi\ kinematics of \pavel. \textit{Leftmost:} Color-composite image from PS1 $g$-, $r$- and $i$-bands overlaid with \hi\ contours. Contours are set to start at the level of intensity corresponding to a pseudo 4$\sigma$ contour (see Sect. \ref{sec:analysis-kinematic_model} for explanation) in the total intensity map (white), and grow by a factor of 2 in intensity towards the redder colors. Lowest contour for \pavel\ corresponds to the column density of $7.2 \times 10^{19}$ cm$^{-2}$. The overplotted ellipse in blue represents the median geometric parameters obtained from CANNUBI (including the size). \textit{Middle left:} Velocity field obtained as weighted-mean value from the \hi\ data from \barolo. The black cross represents the best-fit center position and the gray dashed line the kinematic position angle. Contours are given in spacing of 10 \kms\ with the green contour indicating \vsys. \textit{Middle right:} Position-velocity slice along the major axis with data (blue) and model (red) contours. White points represent the obtained projected rotational velocity of the best fit model. Contours are plotted starting from 2 times the noise per channel in the cube as reported in Table \ref{tab:technical_params}, and grow linearly. \textit{Rightmost:} Position-velocity slice along the minor axis (perpendicular to the dashed line in the middle left panel) with the same color scheme and contour levels as the middle right panel.}
        \label{fig:pavel-kinematics}
    \end{figure*}

    In the description that follows we show the results of kinematic modeling for one galaxy (\pavel), while the rest can be found in Appendix \ref{app:results_sample}. Results for \pavel\ are shown in the form of channel maps in Fig. \ref{fig:ch_maps}, where the top panels represent the data and the bottom panels the best fit model. The model is present in all channels where the galaxy emission is detected and represents the data well. Moment maps and position-velocity (PV) slices for \pavel\ are shown in Fig. \ref{fig:pavel-kinematics}. On the leftmost panel of the figure, the \hi\ contours are overlaid on top of the optical image. We note that the noise in the total intensity map is not uniform because the cube is masked with a 3D mask before producing the map, resulting in a different number of channels contributing per pixel. 
    Therefore, the \hi\ contours in the image are the so-called pseudo $X \sigma$ contours, obtained by selecting pixels with X-0.25 and X+0.25 values in the $S/N$ map, and taking the average of the flux of the corresponding pixels in the total \hi\ map \citepads[see e.g.][]{2001A&A...370..765V}.
    The blue ellipse in the image describes the obtained disk geometry from CANNUBI. Contours are spread outside of the ellipse due to the prolongation of the beam in the north-south direction.
    The velocity field in the middle left panel shows a velocity gradient that suggests the presence of a regularly rotating disk. This is also evident in the PV slices, where data (blue with black contours) shows no major deviation from regular rotation, and is well described by the model (red contours).

    Given the low spatial resolution of most of our sample, it is not straightforward to say whether we are able to trace the flat part of the rotation curve for our galaxies. However, in 5 cases (\inter, \bSNRtwo, \nce, \beaut\ and \mess; Figs. \ref{fig:HI-op-inter}, \ref{fig:HI-op-bSNR_2}, \ref{fig:HI-op-nce}, \ref{fig:HI-op-beaut} and \ref{fig:HI-op-mess}), the PV slices along the major axes show a turnover that suggests that the flat part has been reached. While \mess\ shows this turnover, its \hi\ inclination shows consistencies with inclinations down to 0\dg, making it impossible to robustly constrain the rotational velocity. Therefore, for \mess\ and the other galaxies in the sample that do not show evidence of a turnover in the PV slices, we consider our derived rotational velocities to be lower limits to the rotational velocity in the flat part of the rotation curve.

    From 13 kinematically modeled galaxies, one (\pairtworight; Fig. \ref{fig:HI-op-pair_2-right}) seems to be kinematically lopsided, i.e. its rotational velocity is higher on one side than the other; or one side of the galaxy is not detected far enough out. In either case, this complicates the interpretation of the derived rotational velocity as a tracer of the underlying potential.
    Another two kinematically modeled galaxies show signatures of warps, with \paver\ (Fig. \ref{fig:HI-op-PA_180}) showing a warp in position angle, and \mess\ (Fig. \ref{fig:HI-op-mess}) showing a warp in both inclination and position angle. Unfortunately, we cannot robustly model the warps due to the low spatial resolution of our sample. For these reasons, we flag these 3 galaxies in Table \ref{tab:HI_par}, and do not include them in the BTFR in Sect. \ref{sec:btfr}.

\subsubsection{Asymmetric drift correction}
\label{sec:asym_drift}

    As mentioned before, the major contribution to gas kinematics in disk galaxies comes from rotation. However, gas pressure can still have a significant influence in counteracting the pull of gravity. In order to correct for pressure-support and determine the circular speed (which allows us to characterize the gravitational potential), we follow the procedure from \citetads{2017MNRAS.466.4159I}. In particular, we make use of their equation 9:
    \begin{equation}
        V_{\text{A}}^2 = -R\, \sigma_{\text{V}}^2\, \frac{\partial\, \sigma_{\text{V}} \,\Sigma_{\text{\hi}}\, \cos i}{\partial \, R} 
    \label{eq:ADC}
    \end{equation}
    where $V_{\text{A}}$ is the asymmetric drift correction (ADC), $\sigma_{\text{V}}$ the velocity dispersion, $\Sigma_{\text\hi}$ the projected surface density of \hi, and $R$ the distance from the galaxy center. Given that the rotation curves are sampled by 2 points for most galaxies in our sample, we fit the term within the derivative with a linear function, and multiply its slope by $-R\, \sigma_{\text{V}}^2$ in order to obtain the asymmetric drift. Then, the circular velocity ($V_{\text{circ}}$) is obtained with \citepads[e.g.][]{2017MNRAS.466.4159I}:
    \begin{equation}
        V_{\text{circ}}^2 = \text{\vrot}^2 + V_{\text{A}}^2.
    \end{equation}
    The errors on inclination were propagated to the circular velocity the same way as was done for the rotational velocity (see Sect. \ref{sec:analysis-kinematic_model}).
    
    The maximum circular velocities across the rings and the corresponding ADC terms are given in Table \ref{tab:HI_par}. All kinematically modeled galaxies in our sample seem to be highly rotationally supported with little contribution from pressure support. However, we note that given our resolution, we are not able to perform a robust ADC, e.g. by fitting a more physically motivated function to the term under the derivative in Eq. \ref{eq:ADC} \citepads[e.g.][]{2002AJ....123.1316B,2017MNRAS.466.4159I}.

    \begin{table*}[]
        \caption{The \hi\ parameters and kinematics.}
        \centering

        \resizebox{\textwidth}{!}{%
        \begin{tabular}{llllllllllllll}
        \hline \hline
         & Name & $i$ & \PAhi & Geom. & \PAkin & $\Delta R$ & $N_{\text{rings}}$ & $R_{\text{out}}$ & beam/$\Delta R$ & $\sigma_{V}$ & $V_{\text{rot}}$ & $V_{\text{A}}$ & $V_{\text{circ}}$ \\
         
            &  & [\dg] & [\dg] & & [\dg] & [\arcsec] & & [kpc] & & [\kms] & [\kms] & [\kms] & [\kms]   \\
            \hline
1 & \paironeleft & - & - & - & - & - & - & - & - & - & - & - & - \\
2 & \paironeright & - & - & - & - & - & - & - & - & - & - & - & - \\
3 & \pairtwoleft & ${33}_{-33}^{+12}$ & ${120}_{-5}^{+5}$ & HI & 118 & 20.9 & 2 & 9.7 & - & ${8.8}_{-2.1}^{+1.8}$ & ${42.6}_{-8.6}$ & ${9.0}_{-2.9}^{+3.3}$ & ${43.5}_{-8.8}$ \\
4 & \pairtworight$^{\dagger}$ & ${32}_{-32}^{+12}$ & ${62}_{-5}^{+5}$ & HI & 36 & 20.7 & 2 & 9.6 & 1.2 & ${5.5}_{-3.9}^{+2.9}$ & ${69.1}_{-17.2}$ & ${6.7}_{-3.0}^{+3.7}$ & ${69.4}_{-17.3}$ \\
5 & \pairthrleft & - & - & - & - & - & - & - & - & - & - & - & - \\
6 & \pairthrright & - & - & - & - & - & - & - & - & - & - & - & - \\
7 & \bSNRone & - & - & - & - & - & - & - & - & - & - & - & - \\
8 & \svel & ${31}_{-31}^{+17}$ & ${91}_{-38}^{+29}$ & op & 49 & 12.2 & 3 & 8.2 & 1.9 & ${6.0}_{-3.1}^{+2.9}$ & ${51.3}_{-4.2}^{+10.5}$ & ${5.7}_{-2.0}^{+2.0}$ & ${51.6}_{-4.2}^{+10.4}$ \\
9 & \lsmerge & ${53}_{-13}^{+9}$ & ${159}_{-10}^{+9}$ & - & 171 & - & - & - & - & - & - & - & - \\
10 & \fon & - & - & - & 159 & - & - & - & - & - & - & - & - \\
11 & \inter* & ${64}_{-1}$ & ${23.0}_{-0.2}^{+0.2}$ & HI & 18 & 20.0 & 3 & 10.6 & - & ${16.4}_{-4.5}^{+4.8}$ & ${81.4}_{-4.4}^{+4.6}$ & ${20.3}_{-4.1}^{+4.9}$ & ${83.9}_{-6.0}^{+6.8}$ \\
12 & \bSNRtwo* & ${41}_{-6}^{+4}$ & ${136}_{-7}^{+7}$ & HI & 148 & 28.7 & 3 & 17.3 & - & ${6.9}_{-2.8}^{+2.6}$ & ${70.0}_{-5.0}^{+9.3}$ & ${7.5}_{-1.9}^{+2.0}$ & ${70.4}_{-5.4}^{+9.3}$ \\
13 & \tstd & ${43}_{-7}^{+5}$ & ${136}_{-11}^{+9}$ & - & 173 & - & - & - & - & - & - & - & - \\
14 & \nce* & ${43}_{-11}^{+7}$ & ${75}_{-12}^{+11}$ & HI & 70 & 31.3 & 2 & 17.1 & - & ${8.2}_{-2.7}^{+2.4}$ & ${87.6}_{-9.5}^{+24.1}$ & ${7.3}_{-3.1}^{+3.3}$ & ${87.9}_{-9.5}^{+24.1}$ \\
15 & \beaut* & ${53}_{-15}^{+9}$ & ${313}_{-14}^{+12}$ & HI & 314 & 22.1 & 2 & 16.6 & 1.4 & ${11.9}_{-4.7}^{+5.0}$ & ${75.1}_{-6.0}^{+20.5}$ & ${13.9}_{-4.6}^{+4.4}$ & ${76.4}_{-6.2}^{+20.4}$ \\
16 & \stick & ${53}_{-19}^{+14}$ & ${156}_{-23}^{+22}$ & HI & 153 & 17.9 & 2 & 8.1 & 1.8 & ${12.1}_{-4.3}^{+5.2}$ & ${73.3}_{-13.1}^{+21.0}$ & ${14.5}_{-4.2}^{+4.4}$ & ${74.7}_{-13.3}^{+21.3}$ \\
17 & \pavel & ${31}_{-31}^{+11}$ & ${262}_{-5}^{+5}$ & HI & 263 & 20.6 & 2 & 10.7 & - & ${7.6}_{-2.5}^{+2.7}$ & ${46.6}_{-4.4}$ & ${8.9}_{-2.5}^{+2.6}$ & ${47.4}_{-5.0}$ \\
18 & \reg & ${36}_{-13}^{+12}$ & ${234}_{-5}^{+5}$ & HI & 244 & 16.0 & 2 & 8.5 & 1.2 & ${6.7}_{-2.9}^{+2.3}$ & ${56.5}_{-11.9}^{+22.1}$ & ${7.6}_{-2.8}^{+3.1}$ & ${57.0}_{-11.6}^{+22.0}$ \\
19 & \paver$^{\dagger}$ & ${55}_{-20}^{+10}$ & ${186}_{-13}^{+18}$ & HI & 171 & 18.4 & 2 & 8.9 & 1.4 & ${12.6}_{-4.8}^{+5.2}$ & ${75.2}_{-7.7}^{+23.5}$ & ${15.9}_{-5.5}^{+5.6}$ & ${76.8}_{-9.4}^{+24.3}$ \\
20 & \mess$^{\dagger}$ & ${27}_{-27}^{+8}$ & ${284}_{-5}^{+5}$ & HI & 286 & 25.1 & 4 & 23.4 & - & ${8.0}_{-3.4}^{+3.6}$ & ${80.7}_{-5.5}$ & ${12.2}_{-4.1}^{+4.9}$ & ${81.6}_{-6.8}$ \\
21 & \wtm & ${49}_{-15}^{+11}$ & ${279}_{-14}^{+16}$ & HI & 277 & 25.2 & 3 & 13.8 & - & ${9.2}_{-3.5}^{+4.0}$ & ${93.0}_{-10.3}^{+19.2}$ & ${10.3}_{-3.3}^{+3.5}$ & ${93.6}_{-10.3}^{+19.3}$ \\
22 & \dwtm & ${27}_{-27}^{+9}$ & ${245}_{-25}^{+22}$ & - & 347 & - & - & - & - & - & - & - & - \\
23 & \misal & ${29}_{-8}^{+8}$ & ${197}_{-12}^{+17}$ & - & 269 & - & - & - & - & - & - & - & - \\
24 & \weir & - & - & - & 124 & - & - & - & - & - & - & - & - \\
             \hline
        \end{tabular}}
        \tablefoot{
            Name is given by the name from the literature (when available) or the shortened name from Apertif.
            Inclination is denoted as $i$ and the \hi\ morphological position angle as \PAhi. Geom. gives the set of geometric parameters used for kinematic modeling (either \hi\ or optical). \PAkin\ is the kinematic position angle, $\Delta R$ the separation of the rings used for kinematic modeling, $N_{\rm rings}$ is the number of the rings in the kinematic model, $R_{\rm out}$ the outermost radius of the model (number of rings times $\Delta R$), and beam/$\Delta R$ the oversampling rate of the rotation curve along the kinematic major axis. Finally, \disp\ is the mean velocity dispersion in all rings, and \vrot, $V_{\text{A}}$ and $V_{\text{circ}}$ are the maximum rotational velocity, the asymmetric drift and the circular velocity, respectively. Upper errors on rotational and circular velocities are unconstrained, and therefore not reported, for galaxies whose inclinations are compatible down to inclinations of 0\dg.\\
            (*) Rotation velocity is likely tracing the flat part of the rotation curve, as seen from the shape of the PV slice along the major axis.\\
            ($^{\dagger}$) Galaxy was flagged as having unreliable rotational velocity (see Sect. \ref{sec:obtained_kin_models}).
            }
        \label{tab:HI_par}
    \end{table*}
    
\subsection{Obtaining optical properties}
\label{sec:obtaining_optical_properties}

    Our optical analysis consists of two steps: we first conduct isophotal fitting on the $i-$band image in order to constrain the geometry of the stellar disk, after which we use the obtained stellar geometry to extract surface brightness profiles (in all three bands) following the procedure described in appendix A of \citetads{2023A&A...670A..92M}.
    
    During the isophotal fitting, we mask and slightly smooth the $i-$band images to lower the influence of small scale overdensities on the obtained geometry of the stellar disk. The size of the smoothing kernel depends on the galaxy and can be found in Table \ref{tab:optical_par}. For the fitting we use an Astropy affiliated package \texttt{photutils} \citep{larry_bradley_2022_7419741}. The algorithm fits a set of ellipses of increasing semi-major axes using the position of the center, ellipticity and position angle of the ellipses as free parameters. We run the algorithm two times. The first time we leave all the parameters free to vary, and use the result to constrain the central position by taking the median from all fitted ellipses with semi-major axes larger than half of the full width at half maximum (FWHM) of the PSF. The second time, we fix the central position and fit only for ellipticities and position angles. The final (global) values of ellipticity and position angle are obtained by taking the median value from ellipses which range in estimated surface brightness between 24 and 27 mag arcsec$^{-2}$ (motivated by the classical choice of surface brightness of 25 mag arcsec$^{-2}$ as representative of the outer stellar disks); and which have semi-major axes larger than the total FWHM of the PSF, in order to exclude the effect of the PSF on the derived geometry. To (at least) partially remove the bias towards the inner regions that were fitted with more ellipses, we calculate the median from semi-equidistant ellipses (along the semi-major axis) satisfying these conditions.
    The result of isophotal fitting can be found in Fig. \ref{fig:isophot_fit-Pavel} for \pavel, and in the Appendix \ref{app:results_sample} for the rest of the sample.
    
    To transform from ellipticity ($\epsilon$) to inclination, we use:
    \begin{equation}
        \cos^2(i) = \frac{(1 - \epsilon)^2 - q_0^2}{1 - q_0^2}
    \end{equation}
    where $q_0$ is the intrinsic thickness. We adopt $q_0 = 0.3$ which is a common value used for dwarf irregular galaxies \citepads[e.g.][]{2010MNRAS.406L..65S}.

    A possible caveat to the accurate determination of the geometry of the stellar disk with isophotal fitting in this mass regime is the influence of clumpy star forming regions on the obtained geometries of the isophotes. Bright clumpy regions can dominate over the fainter disk component of the galaxy, which we are trying to constrain. Indeed, some galaxies in our sample might be subject to this effect, as they clearly show many bright clumpy regions in their outskirts (e.g. \inter, \bSNRtwo, \misal; Figs. \ref{fig:HI-op-inter}, \ref{fig:HI-op-bSNR_2} and \ref{fig:HI-op-misal}). On the other hand, in few galaxies with less clumpy morphology (\pavel, \pairtwoleft, \svel, \nce; Figs. \ref{fig:isophot_fit-Pavel}, \ref{fig:HI-op-pair_2-left}, \ref{fig:HI-op-svel} and \ref{fig:HI-op-nce}), we see a clear trend in isophotes becoming more aligned with the \hi\ kinematics as we go further out in radius. This points towards another caveat in the determination of optical morphology: the depth of optical data. 
    \citetads{2024A&A...689A.344M} has shown that the apparent misalignment between stellar and gas geometry in the UDG AGC 114905 seen in data with surface brightness depths of $\mu (3\sigma_{10\text{\arcsec}\times 10\text{\arcsec}}) \sim 28.5$ mag arcsec$^{-2}$ disappears when analyzing ultra-deep imaging reaching $\mu (3\sigma_{10\text{\arcsec}\times 10\text{\arcsec}}) \sim 32$ mag arcsec$^{-2}$. This demonstrates a need for deeper optical observations in order to trace the underlying stellar disk far enough out for a robust comparison of \hi\ and stellar morphologies.
    
    Using the optical geometric parameters, we conduct surface brightness photometry on full resolution images following the procedure described in appendix A of \citetads{2023A&A...670A..92M}. We input the global geometric parameters and create a set of ellipses of the same geometry, but with increasing semi-major axes. The semi-major axis is sampled so that each ellipse has width equal to the FWHM of the PSF. For each ellipse, we calculate the mean value in image units (counts) and correct for inclination by multiplying the obtained values by $\cos i$. For the estimation of the background and noise, we fit the sky pixel intensity distribution using two-components: a Gaussian function and a Schechter function. The latter accounts for the possible residing contamination from (masked) foreground stars. We obtain the image's background value and rms noise from the parameters of the Gaussian component. In some cases the fit does not converge, and we simply use the mean and the standard deviation of the sky pixel intensity distribution of the masked image. In most cases, we have extracted surface brightness profiles radially until $S/N=1$ in a given ring is reached. In some cases\footnote{\paironeleft, \pairthrright, \bSNRtwo, \tstd, \nce, \beaut, \dwtm\ and \misal; Figs. \ref{fig:HI-op-pair_1-left}, \ref{fig:HI-op-pair_3-right}, \ref{fig:HI-op-bSNR_2}, \ref{fig:HI-op-tstd}, \ref{fig:HI-op-nce}, \ref{fig:HI-op-beaut}, \ref{fig:HI-op-dwtm} and \ref{fig:HI-op-misal}.} the obtained profiles extended far outside the galaxy due to image artefacts or nearby foreground stars. In these cases we manually truncated the profiles.
    To convert to magnitudes, we use the calibration from PS1 \citepads{2020ApJS..251....4W}:
    \begin{equation}
        m\, [\rm mag] = -2.5\log_{10}(\Sigma \ counts) + 2.5\log_{10}(T) + 25
    \label{eq:mag_from_count}
    \end{equation}
    where $T$ is the exposure time reported in the header. We fit the obtained surface brightness profile with a Sérsic function from which we derive central surface brightness $\mu_{0,X}$, Sérsic index $n_{X}$, effective radius $R_{e,X}$, mean effective surface brightness \mmuefX\ and the apparent magnitude $m_X$ in each band indicated with $X$ (see Appendix \ref{app:photometry_caveats} for details). We list $m_g$, $R_{e,r}$ and \mmuefr\ in Table \ref{tab:optical_par}. Magnitudes and mean surface brightnesses are corrected for the Galactic extinction using results from \citetads{2011ApJ...737..103S}, which are available through NED. Results of these fits in the $i$-band are presented in Fig. \ref{fig:pavel-optical} for \pavel, and in App.\ \ref{app:results_sample} for the rest of the sample. We note that in some cases the \hi\ center is offset from the optical center, but given the \hi\ beam diameter size of $\gtrsim15$\arcsec, the two centers are compatible. In two cases (\pairtwoleft\ and \bSNRtwo\  shown in Figs. \ref{fig:HI-op-pair_2-left} and \ref{fig:HI-op-bSNR_2} respectively), the extracted surface brightness profiles have irregular shapes due to clumpy star-forming regions, which is why we consider their fitted Sérsic parameters to be less reliable. Additionally, the surface brightness profile of \nce\ (shown in Fig. \ref{fig:HI-op-nce}) clearly shows contribution from two components and cannot be well fitted with a single Sérsic function. These galaxies will be flagged in future plots when these parameters (or properties derived from them) are used. The obtained stellar properties for the sample can be found in Table \ref{tab:optical_par}.

\subsubsection{Reliability checks}
\label{sec:opt_reliability}
    
    We found that the Sérsic function gives a good representation of the galaxy light profile, but it assumes the galaxy to extend towards infinity. As galaxies are finite systems, this could result in the overestimation of the total flux of a galaxy. However, the alternative approach of directly measuring the half-light radius ($R_e^{\text{meas}}$) and total magnitude from the data, suffers from the problem of the depth of the data (the comparison of these two approaches in our sample can be found in Appendix \ref{app:photometry_caveats}). \citetads{2001MNRAS.326..869T} analyzed this problem and showed that the two approaches would converge to the same values if the surface brightness profile is traced far enough out in the galaxy. To check the reliability of our fitted parameters, we used the equation 18 from \citetads{2001MNRAS.326..869T} with our fitted $R_e^{\text{fit}}$ values to find the desired outermost radius (in the $r$-band) for which the two approaches would theoretically give at least 85\% compatibility ($R_e^{\text{meas}}/R_e^{\text{fit}} \geq 85\%$). Galaxies which were traced at least out to this radius are considered to have well constrained Sérsic fit parameters, and consequently, robust photometry. Excluding the galaxies with unreliable Sérsic fit (see Sect. \ref{sec:obtaining_optical_properties}) out of this analysis, we have 17 galaxies with reliable Sérsic fits and well constrained photometry. The other 7 galaxies in the sample are considered to have less reliable derived Sérsic parameters and will be regarded as "optically unreliable". We note that this only refers to parameters corresponding to the shape of the profile, while the obtained stellar masses are considered to be robustly measured (see Appendix \ref{app:photometry_caveats} for the comparison between measured and fitted properties). The optically unreliable galaxies are marked in Table \ref{tab:optical_par} and flagged in plots which use Sérsic parameters or optical properties derived from them.

    In some galaxies, the isophotal fitting showed significant changes of geometric parameters with radius. To test how these variations influence the final results, we repeated the extraction and the fitting of surface brightness profiles using the 16th and the 84th percentile of ellipticity and position angle from the same selected region as before (outside the FWHM of the PSF and between surface brightness values of 24 and 27 mag arcsec$^{-2}$). Each time, we changed one of the parameters (either $i$ or \PAop) to a higher/lower value. The resulting difference between $R_e$ of the initial run and $R_e$ of 4 cases described above (16th or 84th percentile value for either $i$ or \PAop), was shown to be $\lesssim 15$\%, while for the $\langle\mu\rangle_e$ the difference was up to 0.4 mag arcsec$^{-2}$. These variations were taken into account in the errors by taking the mean difference between the initial run and the 4 cases, and summing it in quadrature with the error from the fit for all fitted parameters of the Sérsic profile ($\mu_e$, $n$ and $R_e$). For apparent magnitudes, the errors were calculated only by taking the mean variation for all different cases. We note that even taking the 16th and the 84th percentile from the whole meaningful range of radii (outside the FWHM of the PSF and surface brightness values lower than 27 mag arcsec$^{-2}$), the variations in parameters were $\lesssim 30\%$ in $R_e$ and $\lesssim$1.5 mag arcsec$^{-2}$ in $\langle\mu\rangle_e$.

\subsubsection{The special case of \beaut}

    Optical images of \beaut\ (Fig. \ref{fig:HI-op-beaut}) show a strong non-uniform foreground Galactic emission. This emission manifests as the gradient in the background levels across the optical images of the galaxy. In order to correct for this effect, we have produced a 2D image of the background using the \texttt{Background2D} class from the \texttt{photutils} python package. We then subtracted the background from the original images and proceeded with photometry as described in Sect. \ref{sec:obtaining_optical_properties}. 

    To characterize the errors of the background subtraction at the position of the galaxy, we additionally extracted surface brightness profiles from the original images (without the subtracted background) and used the difference in the obtained parameters as an additional error estimate. In particular, we updated the errors of Sérsic fit parameters of the profile from the background subtracted image by taking half of the difference between the two cases (subtracted and non-subtracted image) and summing it in quadrature with the statistical error from the fit. For the apparent magnitudes, we have calculated the errors by only taking the half of the difference in the obtained magnitudes between the two cases. This step was performed before the propagation of errors of geometric parameters (See Sect. \ref{sec:opt_reliability}). The rest of the analysis stays the same as for the rest of the sample.

        \begin{figure*}
        \centering

            \includegraphics[width=0.8\linewidth]{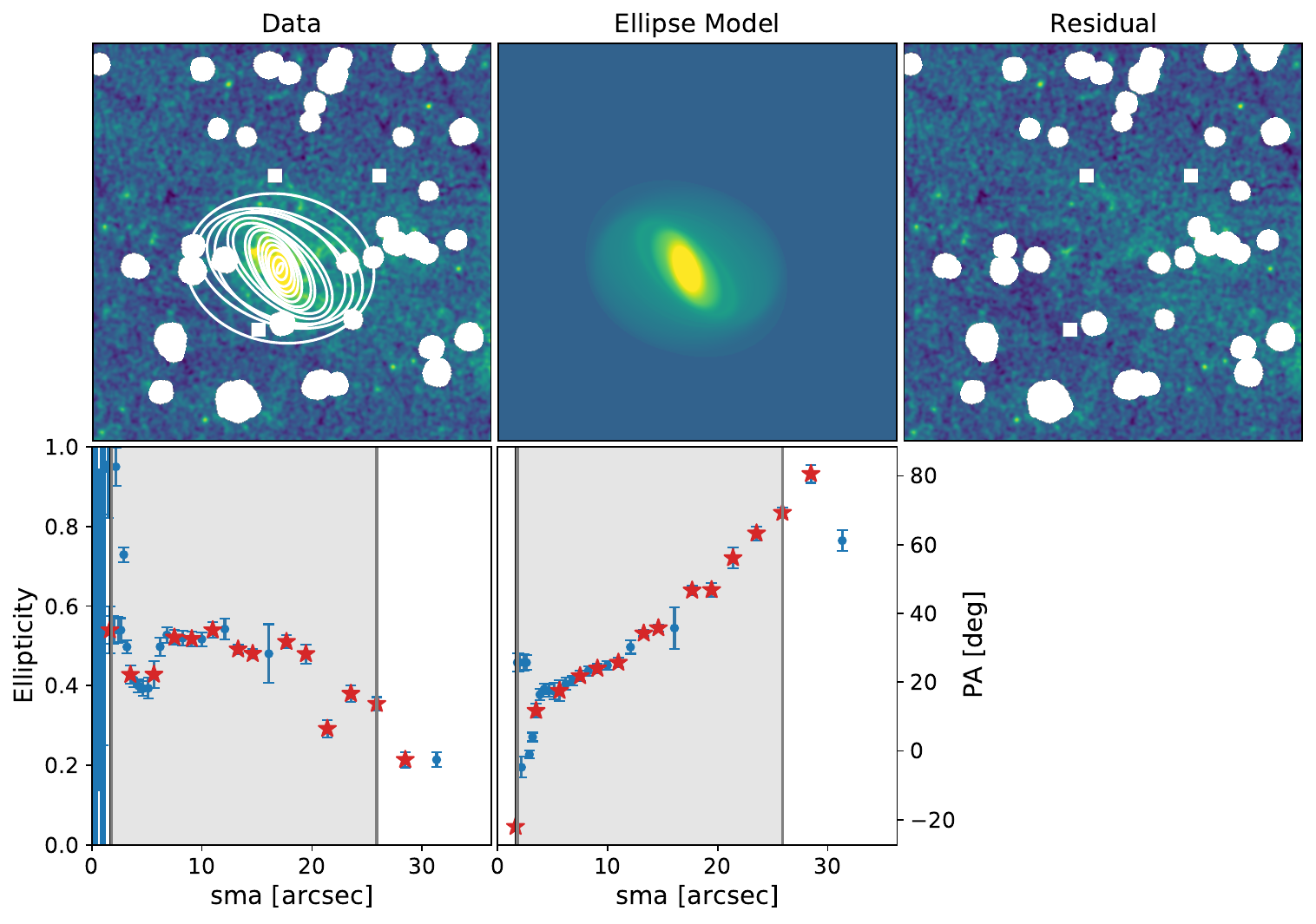}

        \caption{Isophotal fitting result for \pavel. \textit{Upper panels:} The leftmost panel shows the smoothed data overlaid with semi-equidistant ellipses along the semi-major axis length whose parameters are marked with red stars in the lower panels. White regions denote masked pixels in the image. The middle panel shows the model built from all fitted ellipses (blue circles in the lower panels), and the rightmost panel shows the residual of the data image and the model image.
        \textit{Lower panels:} Ellipticity and position angle as a function of the semi-major axis length from the second run of the fitting algorithm (see Sect. \ref{sec:obtaining_optical_properties}) are shown as blue circles, with respect to radius from the fixed center position. Red stars denote parameters of semi-equidistant ellipses plotted in the upper leftmost panel. The black vertical line is located at a distance from the center that corresponds to the FWHM of the PSF after the smoothing, and the gray shaded region corresponds to the region in which the final global geometry was measured (using only the semi-equidistant ellipses, see Sect. \ref{sec:obtaining_optical_properties}).}
        \label{fig:isophot_fit-Pavel}
    \end{figure*}
    
    \begin{table*}[]
        \caption{Optical parameters.}
        \centering
        
        \resizebox{\textwidth}{!}{%
        \begin{tabular}{lllllllllllll}
            \hline
            & Name  & $\sigma_{\rm S}$ & RA, DEC & $i$ & \PAop & $R_{\text{out},r}$ & $g-r$ & $m_g$ & $R_{e,r}$ & \mmuefr \\
            & & [pix] & (hh:mm:ss dd:mm:ss) & [\dg] & [\dg] & [kpc] &  & & [kpc] & [mag arcsec$^{-2}$]  \\
    
            \hline
1 & \paironeleft & 2 & 02:03:50.9 +37:14:56 & ${61} \pm {3}$ & ${173} \pm {3}$ & 3.7 & ${0.42} \pm {0.11}$ & ${17.605} \pm {0.003}$ & ${1.50} \pm {0.02}$ & ${23.03} \pm {0.01}$ \\
2 & \paironeright & 2 & 02:03:43.1 +37:14:43 & ${69} \pm {6}$ & ${18} \pm {3}$ & 11.2 & ${0.49} \pm {0.09}$ & ${16.75} \pm {0.01}$ & ${3.5} \pm {0.3}$ & ${23.90} \pm {0.16}$ \\
3 & \pairtwoleft $^{\dagger}$ * & 6 & 13:36:54.0 +32:05:44 & ${43} \pm {13}$ & ${95} \pm {21}$ & 9.8 & ${0.29} \pm {0.08}$ & ${15.99} \pm {0.11}$ & ${5.7} \pm {0.8}$ & ${24.82} \pm {0.14}$ \\
4 & \pairtworight * & 2 & 13:36:45.5 +32:05:35 & ${57} \pm {19}$ & ${23} \pm {18}$ & 11.6 & ${0.53} \pm {0.12}$ & ${16.86} \pm {0.03}$ & ${4.4} \pm {0.4}$ & ${24.71} \pm {0.24}$ \\
5 & \pairthrleft & 2 & 10:10:24.9 +58:28:30 & ${44} \pm {8}$ & ${21} \pm {8}$ & 5.2 & ${0.31} \pm {0.03}$ & ${15.74} \pm {0.01}$ & ${1.20} \pm {0.04}$ & ${21.59} \pm {0.06}$ \\
6 & \pairthrright & 2 & 10:10:13.6 +58:29:19 & ${57} \pm {8}$ & ${157} \pm {6}$ & 8.4 & ${0.33} \pm {0.04}$ & ${15.22} \pm {0.01}$ & ${2.7} \pm {0.1}$ & ${22.91} \pm {0.12}$ \\
7 & \bSNRone & 6 & 13:30:44.9 +32:45:39 & ${59} \pm {13}$ & ${101} \pm {5}$ & 22.7 & ${0.19} \pm {0.07}$ & ${15.14} \pm {0.02}$ & ${7.6} \pm {0.5}$ & ${23.86} \pm {0.13}$ \\
8 & \svel * & 2 & 13:30:42.1 +29:47:34 & ${46} \pm {9}$ & ${47} \pm {35}$ & 8.7 & ${0.41} \pm {0.14}$ & ${16.79} \pm {0.08}$ & ${3.4} \pm {0.3}$ & ${24.27} \pm {0.12}$ \\
9 & \lsmerge & 2 & 13:35:07.3 +31:31:18 & ${68} \pm {6}$ & ${147} \pm {3}$ & 9.0 & ${0.14} \pm {0.06}$ & ${16.94} \pm {0.01}$ & ${2.9} \pm {0.1}$ & ${23.48} \pm {0.11}$ \\
10 & \fon * & 2 & 13:37:04.6 +31:53:39 & ${51} \pm {8}$ & ${36} \pm {23}$ & 7.8 & ${0.12} \pm {0.14}$ & ${17.29} \pm {0.04}$ & ${2.9} \pm {0.4}$ & ${24.73} \pm {0.16}$ \\
11 & \inter & 2 & 10:16:55.2 +58:23:41 & ${70} \pm {9}$ & ${34} \pm {5}$ & 10.6 & ${0.08} \pm {0.04}$ & ${15.04} \pm {0.03}$ & ${3.6} \pm {0.4}$ & ${23.51} \pm {0.21}$ \\
12 & \bSNRtwo $^{\dagger}$ * & 4 & 13:08:30.4 +54:37:59 & ${51} \pm {13}$ & ${55} \pm {55}$ & 10.1 & ${0.45} \pm {0.09}$ & ${16.17} \pm {0.18}$ & ${4.8} \pm {0.9}$ & ${24.66} \pm {0.26}$ \\
13 & \tstd * & 2 & 13:59:38.0 +37:26:35 & ${56} \pm {9}$ & ${3} \pm {14}$ & 10.3 & ${0.23} \pm {0.12}$ & ${16.35} \pm {0.07}$ & ${4.4} \pm {0.2}$ & ${24.75} \pm {0.09}$ \\
14 & \nce $^{\dagger}$ * & 4 & 22:07:41.4 +40:08:52 & ${56} \pm {7}$ & ${53} \pm {15}$ & 13.7 & ${0.41} \pm {0.13}$ & ${15.50} \pm {0.15}$ & ${9.6} \pm {1.5}$ & ${24.91} \pm {0.16}$ \\
15 & \beaut $^{\dagger}$ & 2 & 22:07:43.7 +41:43:43 & ${58} \pm {4}$ & ${133} \pm {3}$ & 4.3 & ${0.45} \pm {0.24}$ & ${19.04} \pm {0.35}$ & ${1.9} \pm {0.4}$ & ${23.81} \pm {0.05}$ \\
16 & \stick & 2 & 22:32:58.9 +39:38:48 & ${66} \pm {4}$ & ${155} \pm {3}$ & 11.3 & ${0.28} \pm {0.06}$ & ${16.40} \pm {0.01}$ & ${3.5} \pm {0.1}$ & ${23.95} \pm {0.07}$ \\
17 & \pavel * & 2 & 22:39:02.4 +38:32:02 & ${64} \pm {8}$ & ${34} \pm {25}$ & 11.6 & ${0.30} \pm {0.20}$ & ${17.48} \pm {0.06}$ & ${4.6} \pm {0.5}$ & ${25.27} \pm {0.17}$ \\
18 & \reg $^{\dagger}$ * & 4 & 22:16:40.2 +40:24:22 & ${45} \pm {11}$ & ${27} \pm {12}$ & 9.6 & ${0.17} \pm {0.15}$ & ${16.76} \pm {0.06}$ & ${4.7} \pm {0.6}$ & ${24.79} \pm {0.15}$ \\
19 & \paver & 2 & 22:18:00.4 +40:59:47 & ${63} \pm {5}$ & ${15} \pm {3}$ & 7.7 & ${0.29} \pm {0.09}$ & ${17.11} \pm {0.04}$ & ${1.7} \pm {0.1}$ & ${22.92} \pm {0.07}$ \\
20 & \mess $^{\dagger}$ & 2 & 22:24:06.6 +41:15:08 & ${33} \pm {9}$ & ${81} \pm {21}$ & 7.2 & ${0.29} \pm {0.05}$ & ${15.80} \pm {0.01}$ & ${3.4} \pm {0.2}$ & ${23.24} \pm {0.06}$ \\
21 & \wtm & 2 & 13:18:46.3 +27:43:57 & ${53} \pm {2}$ & ${92} \pm {5}$ & 10.2 & ${0.29} \pm {0.04}$ & ${15.107} \pm {0.001}$ & ${3.3} \pm {0.1}$ & ${23.09} \pm {0.01}$ \\
22 & \dwtm & 2 & 13:33:39.1 +60:23:14 & ${54} \pm {7}$ & ${108} \pm {5}$ & 8.8 & ${0.14} \pm {0.07}$ & ${15.49} \pm {0.02}$ & ${2.9} \pm {0.1}$ & ${23.82} \pm {0.10}$ \\
23 & \misal $^{\dagger}$ & 2 & 22:22:30.2 +36:00:35 & ${57} \pm {23}$ & ${14} \pm {15}$ & 17.6 & ${0.29} \pm {0.04}$ & ${15.00} \pm {0.23}$ & ${9.0} \pm {0.8}$ & ${23.71} \pm {0.21}$ \\
24 & \weir & 2 & 22:49:41.6 +39:49:02 & ${44} \pm {6}$ & ${37} \pm {12}$ & 7.1 & ${0.22} \pm {0.13}$ & ${17.48} \pm {0.04}$ & ${3.0} \pm {0.1}$ & ${23.92} \pm {0.02}$ \\
             \hline
        \end{tabular}}
        \tablefoot{
            Name is the same as in Table \ref{tab:HI_par}.
            The $\sigma_{\rm S}$ is the standard deviation of the Gaussian kernel used for smoothing the images for isophotal fitting.
            RA and DEC are obtained by isophotal fitting; 
            $i$ is inclination, \PAop\ is the optical morphological position angle, $R_{\text{out},r}$ the outermost radius included in the Sérsic fit of the surface brightness profile, $g-r$ the color, $m_g$ magnitude in the $g$-band, $R_{e,r}$ effective radius in the $r$-band, and \mmuefr\ the mean surface brightness inside $R_e$ in the $r$-band. Reported values are not corrected for Galactic extinction (except color). \\
            ($^{\dagger}$) Galaxy was flagged as the extent of the surface brightness profile is not being traced far enough out for reliable Sérsic fit (see Sect. \ref{sec:opt_reliability}).\\
            (*) Galaxy is a UDG.
            }
        \label{tab:optical_par}
    \end{table*}
    
    \begin{figure*}
        \centering
        \includegraphics[width = 0.8 \textwidth]{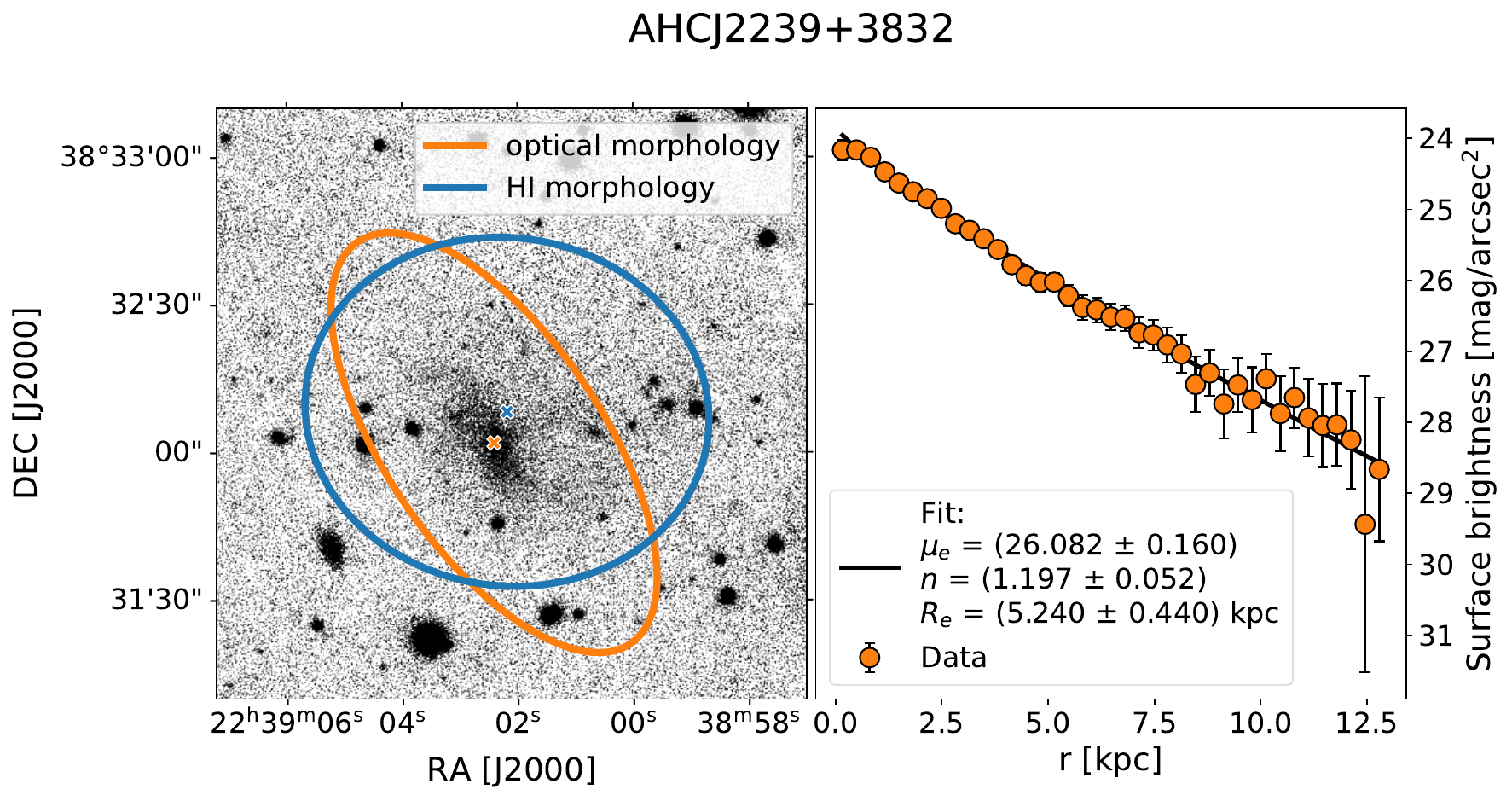}
        \caption{Photometry in the $i$-band of \pavel. \textit{Left:} Optical $i$-band image overlaid with measured geometries from \hi\ morphology (blue) and from $i$-band isophotal fitting (orange). The radius at which the optical geometry is plotted corresponds to the outermost data point in the surface brightness profile. The radius at which the \hi\ geometry is plotted corresponds to the extent of the disk as obtained from CANNUBI. The crosses denote the obtained optical (orange) and \hi\ (blue) centers. \textit{Right:} Surface brightness profile (orange) and the corresponding best fit with a Sérsic profile (black).}
        \label{fig:pavel-optical}
    \end{figure*}

\subsection{Obtaining the distance and masses}
\label{sec:dist_and_masses}
\subsubsection{Distance}
\label{sec:distance}

    We calculate distances using the Extragalactic Distance Database (EDD) \citepads{2009AJ....138..323T}. The EDD provides two calculators \citepads{2020AJ....159...67K} based on two different models of local motions: the Numerical Action Methods (NAM) model of \citetads{2017ApJ...850..207S} and the linear density field model of \citetads{2019MNRAS.488.5438G}. As most of our galaxies are outside of the range of the NAM model ($\lesssim$ 38 Mpc or $\lesssim$ 3000 \kms), for consistency, we will use the linear model for determination of distances of all our sample. 
    As is reported in \citetads{2019MNRAS.488.5438G}, the linear model has accuracy of 15\% or better, depending on the sky region. Unfortunately, the calculators do not provide errors for an individual position which is why we adopt errors of 15\% on all our distances. 
    
    Obtained distances are reported in Table \ref{tab:general_properties}. In most cases, the Hubble flow distance used for the selection is within a few Mpc of the EDD obtained distance, well within the 15\% errors. However, in 6 cases, the difference between the two distances is $\sim 10$ Mpc. This corresponds to $\sim20$\% in absolute error which we find significant enough to adopt the EDD distances in the rest of the paper. Nonetheless, we note that this difference would not significantly change the outcome of our selection procedure, it would only add two more galaxies to our sample. These galaxies were not initially selected using the Hubble flow distance due to our absolute magnitude cut. We do not include these two galaxies to our sample.

\subsubsection{Masses (\hi, stellar and baryonic)}
\label{sec:masses}
    
    We calculate \hi\ masses using a relation from \citetads{2012ARA&A..50..531K}:
    
    \begin{equation}
        \frac{M_{\text{\hi}}}{\rm{M_{\odot}}} = 2.343\cdot 10^5\cdot \left(\frac{D}{\rm Mpc}\right)^2 \cdot \frac{F_{\text{\hi}}}{\rm Jy\, km/s}
        \label{eq:HI_mass}
    \end{equation}
    where $F_{\text{\hi}}$ is the total \hi\ flux, and $D$ the distance to the galaxy. For the calculation of the flux, we use the mask provided by SoFiA which was produced using multiple spatial and spectral resolution kernels in order to capture all the galaxy emission down to the level of noise. We assume a 15\% error on $F_{\text{\hi}}$ coming from the calibration of the flux scale, primary beam correction and mosaicking of Apertif compound beams \citepads{2022A&A...667A..39K}.
    The obtained \hi\ masses are listed in Table \ref{tab:general_properties}.

    We obtain stellar masses by applying the mass-to-light color relation for dwarf irregular galaxies from \citetads{2016AJ....152..177H}:
    \begin{equation}
        \log_{10}(M/L)_{g'} = -0.601(\pm 0.090) + 1.294(\pm 0.401)\,(g'-r')
    \end{equation}
    where apparent magnitudes are given in the standard photometric system. We transform the PS1 extinction corrected (see Sect. \ref{sec:obtaining_optical_properties}) $g-$ and $r-$band magnitudes by applying the conversion from equation 6 in \citetads{2012ApJ...750...99T}. For a robust estimation of galaxy color, we extracted additional surface brightness profiles using the largest FWHM of the PSFs between the three bands for each galaxy, truncated them at the same radius, and directly measured apparent magnitudes by integrating the profiles. We prefer this direct approach exclusively for the estimation of color, where only the relative difference in magnitudes is important. In all other situations when we make use of magnitudes, we adopt the value from the Sérsic fit, including the transfer from the $g-$band magnitude to $g-$band luminosity in the calculation of the mass. Colors and $g-$band magnitudes are given in Table \ref{tab:optical_par}, while stellar masses are listed in Table \ref{tab:general_properties}.

    Finally, for the estimation of baryonic mass, we use:
    \begin{equation}
        M_b = 1.33\,M_{\text{\hi}} + M_*
    \end{equation}
    where factor $1.33$ corrects for the helium contribution to the gas mass of galaxies. Molecular gas mass in low-mass galaxies has been shown to be significantly smaller than \hi\ and stellar masses (around 10\% of either) \citepads[see e.g.][]{2009AJ....137.4670L,2014MNRAS.445.2599B,2017MNRAS.470.4750A,2018MNRAS.474.4366P,2018MNRAS.476..875C}. Therefore, we neglect it in the calculation of our baryonic mass. Obtained masses are listed in Table \ref{tab:general_properties}.

\section{Properties of the sample}
\label{sec:results}

    The final sample consists of 24 galaxies whose global properties can be found in Table \ref{tab:general_properties}. We discuss and provide specific notes on individual galaxies in Appendix \ref{app:results_sample}.
    
    We note that 6 galaxies are in pairs, with each galaxy having a distinct optical counterpart. We present their results individually throughout this section, but note that two pairs, consisting of \paironeleft\ with \paironeright\ and \pairthrleft\ with \pairthrright, are clearly interacting as seen from the \hi\ bridges in Fig. \ref{fig:kinematics_pairs}. Consequently, their properties might be influenced by the interaction. The third pair (\pairtwoleft\ with \pairtworight) does not show a detectable direct interaction, allowing us to obtain reliable kinematic models for each of them.  

    One galaxy in the sample (\bSNRone) is positioned at the edge between two separate Apertif fields, and was detected two times (once in every field). Unfortunately, the \hi\ data cubes have low $S/N$ and the two detections show highly inconsistent morphology, which is why we only use these data in order to estimate the \hi\ mass of the source, but do not proceed with the characterisation of the \hi\ disk nor the kinematic modeling.

\subsection{Global properties of the sample}
\label{sec:global_properties}

    We put our sample in the context of The Arecibo Legacy Fast ALFA (ALFALFA) \hi\  survey \citepads{2005AJ....130.2598G}. We use the completed $\alpha$.100 \citepads{2018ApJ...861...49H} catalog for the comparison of \hi\ properties. For consistency, in this comparison we will regard our 3 pairs of galaxies as single sources because they would not be resolved by the Arecibo beam. For comparison of stellar masses, we use the stellar mass of the brighter galaxy in the pair (but the \hi\ mass of the whole pair) in order to be consistent with the ALFALFA-SDSS catalog \citepads{2020AJ....160..271D}. Additionally, we select a subsample of ALFALFA galaxies inside the same \vsys\ range as our sample, i.e. between $2000<\text{\vsys [\kms]}<5750$, in order to mitigate the bias towards high \hi\ mass galaxies which are detectable to larger distances.
    
\begin{figure*}
        \centering
        \includegraphics[width = 0.95 \textwidth]{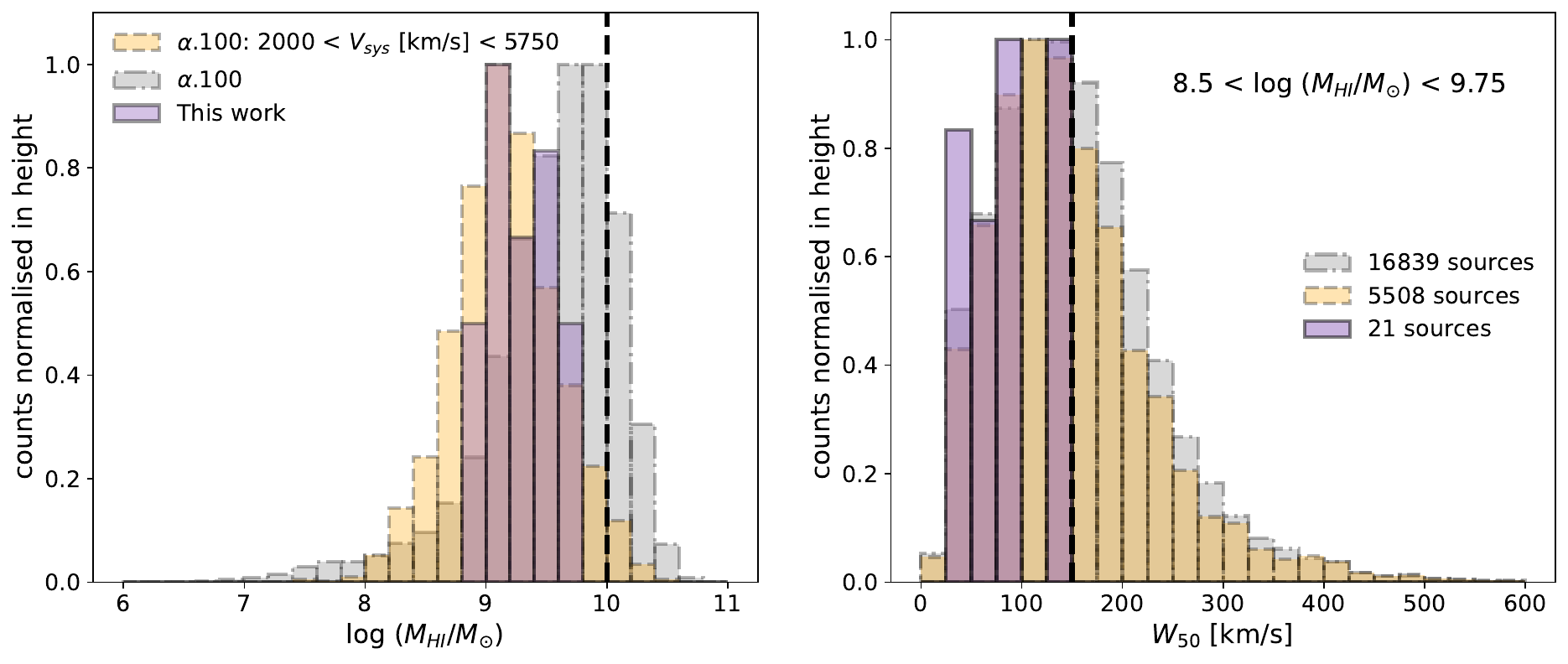}
        \caption{\textit{Left:} Histogram of \mhi\ normalised to the same maximum bin height with $\alpha$.100 sample in grey, the $\alpha$.100 volume limited subsample with $2000<\text{\vsys [\kms]}<5750$ in orange, and our sample in purple. \textit{Right:} Histogram of log \w50\ normalised to the same maximum bin height for the \hi\ mass range between $10^{8.5} - 10^{9.75}$\msun. Colors are the same as in the left panel.
        }
        \label{fig:M_HI_W50_hist}
    \end{figure*}

    The left panel in Fig. \ref{fig:M_HI_W50_hist} shows a histogram of \mhi. Compared to the total $\alpha$.100 catalog, our sample is probing the regime of lower \hi\ masses, by design with our selection of dwarf galaxies. When considering the ALFALFA subsample of limited volume, our sample peaks at the same \hi\ mass range, but does not show tails towards lower and higher \hi\ masses present in the ALFALFA sample. The absence of the tail towards lower masses is expected as our selection procedure requires galaxies to be resolved, thereby excluding lower \hi\ mass galaxies. The absence of the tail towards larger \hi\ masses is due to the exclusion of high stellar mass galaxies in our selection procedure (by imposing the absolute magnitude cut, see Sect. \ref{sec:source_selection}) as well as to our \snrgp\ and \w50\ cuts which exclude high mass edge-on galaxies (see Sect. \ref{sec:discussion_HI_sample}). Furthermore, by considering only the \mhi\ range of our sample (8.50 < log (\mhi/\msun) < 9.75) for ALFALFA, we plot a histogram of \w50\ shown on the right hand side of Fig. \ref{fig:M_HI_W50_hist}. As expected, our sample peaks at lower \w50\ than the corresponding ALFALFA samples due to the additional \w50\ cut in our selection. Assuming all samples are randomly oriented, the large tails of the ALFALFA samples point towards the presence of larger total mass galaxies that we have excluded in our selection.

\begin{figure*}
        \centering
        \includegraphics[width = 0.95 \textwidth]{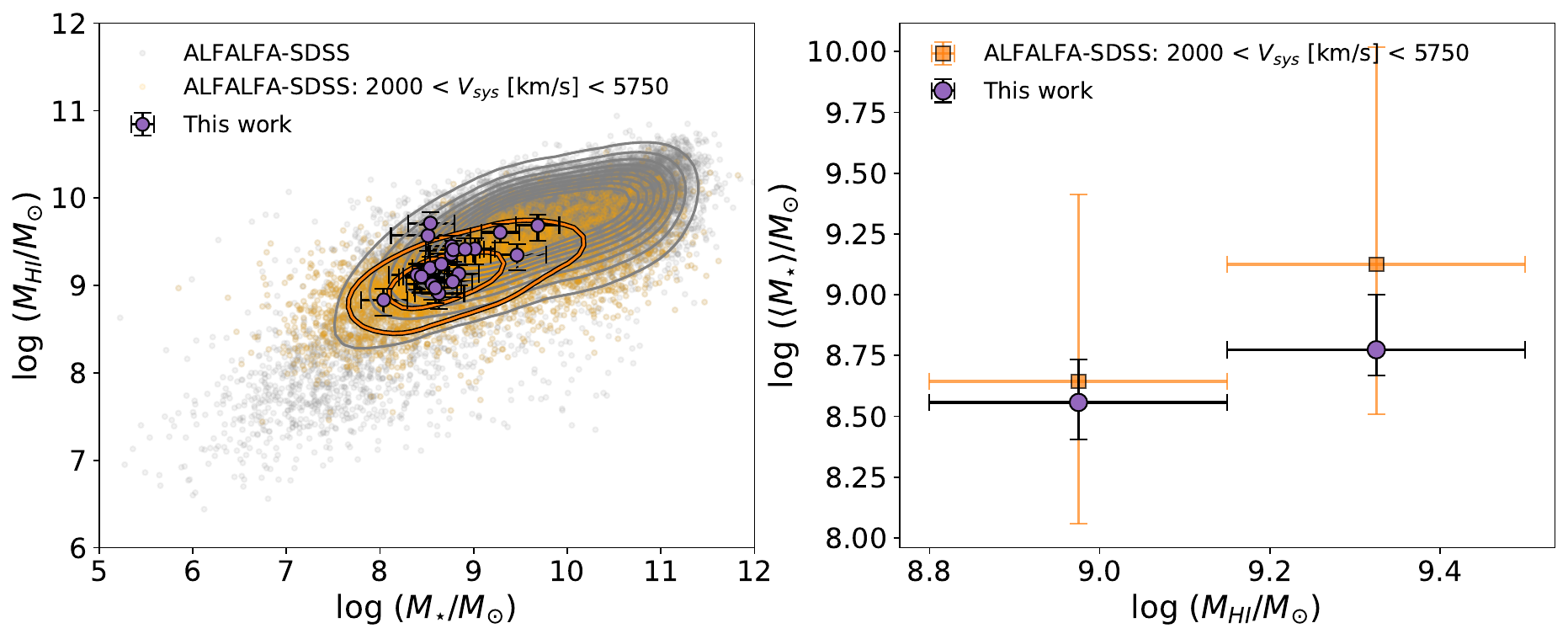}
        \caption{\textit{Left:} \mhi\ - \mstar\ relation.
        Our sample is shown as purple circles, the total ALFALFA-SDSS sample is given by gray points and contours, and the volume limited subsample with $2000<\text{\vsys [\kms]}<5750$ is given by orange points and contours. \textit{Right:} Median values of \mstar\ for \mhi\ bins between $8.8 < \rm log (\text{\mhi/\msun}) < 9.5$. Our sample is represented by purple circles, while the volume limited subsample from ALFALFA-SDSS is represented by orange squares.
        }
        \label{fig:M_star-M_HI}
    \end{figure*}

    The left panel of Fig. \ref{fig:M_star-M_HI} shows the \mstar\ - \mhi\ relation where we compare our sample to the ALFALFA-SDSS catalog \citepads{2020AJ....160..271D}. They report three different measurements of stellar masses for the sample. We choose the optically based method from \citetads{2011MNRAS.418.1587T}, which (with the translation given in equation 3 from \citetads{2020AJ....160..271D}) was shown to be the most consistent with the Spectral Energy Distribution (SED) fitting (see \citetads{2020AJ....160..271D} for more details). Our sample populates the same area in the graph as the volume limited ALFALFA sample, while the total ALFALFA sample tends to higher \hi\ masses for the same stellar mass due to the bias described before. The right panel of Fig. \ref{fig:M_star-M_HI} shows the median \mstar\ for two \mhi\ bins within $8.8 < \rm log (\text{\mhi/\msun}) < 9.5$. We have 9 and 8 galaxies in the lower and upper \mhi\ bin, respectively. As expected from our selection procedure, our sample has slightly smaller median stellar masses than the ALFALFA-SDSS sample for the highest \mhi\ bin due to our \w50\ cut as well as the cut on absolute magnitude (a proxy for stellar mass).

\subsection{Comparison of optical and \hi\ geometries}
\label{sec:results-comparison-geometries}
    
    We derived geometric parameters ($x_0$, $y_0$, $i$ and $PA$) for both the stellar (24 galaxies) and gaseous components (17 galaxies) independently. The ellipses describing disk geometries are visually compared in Fig. \ref{fig:pavel-optical} on the left panel for \pavel, and in Appendix \ref{app:results_sample} for the rest of the sample. 
    
\begin{figure*}
        \centering
        \includegraphics[width = 0.32 \textwidth]{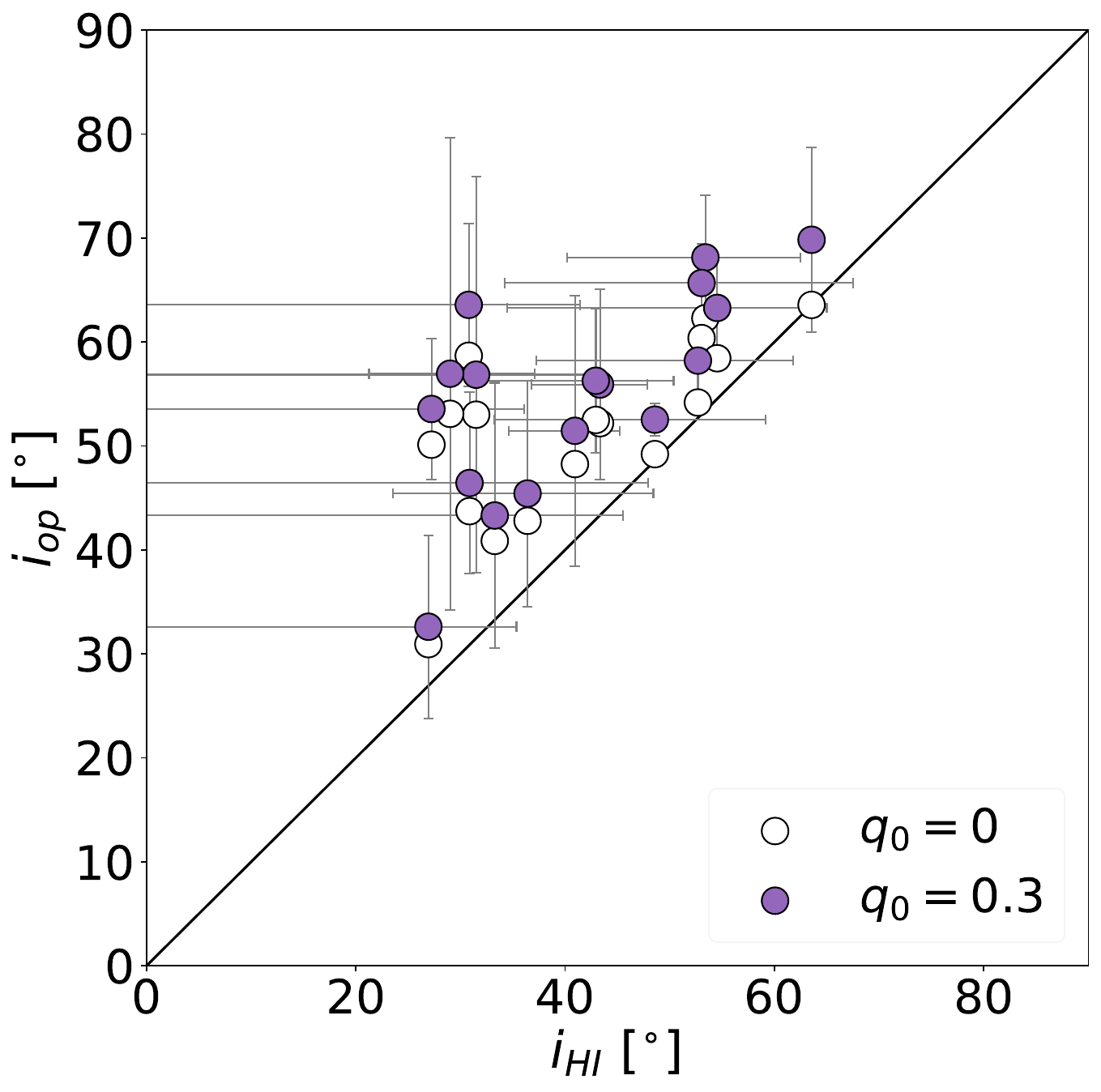}
        \centering
        \includegraphics[width = 0.32 \textwidth]{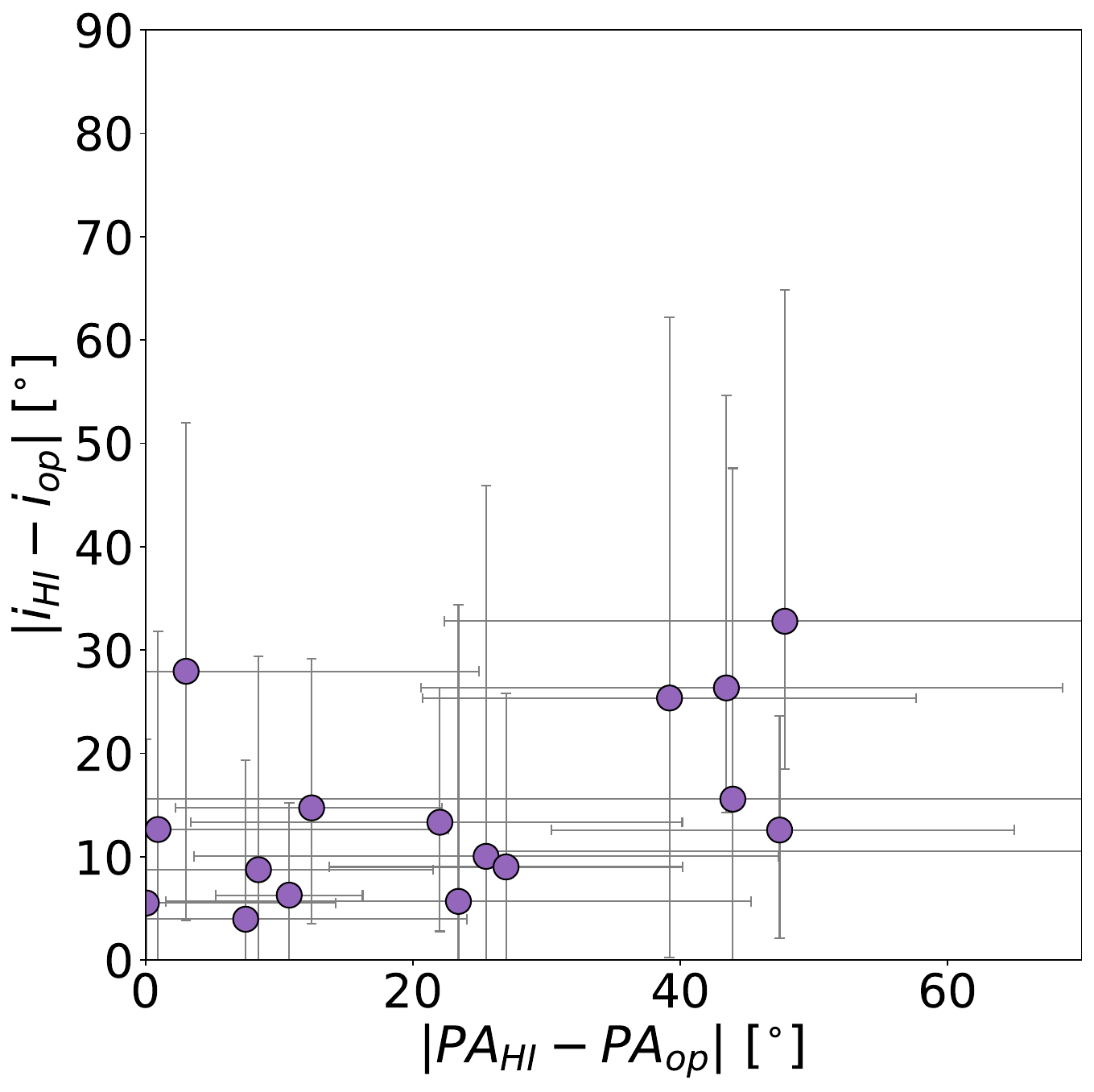}
        \centering
        \includegraphics[width = 0.32 \textwidth]{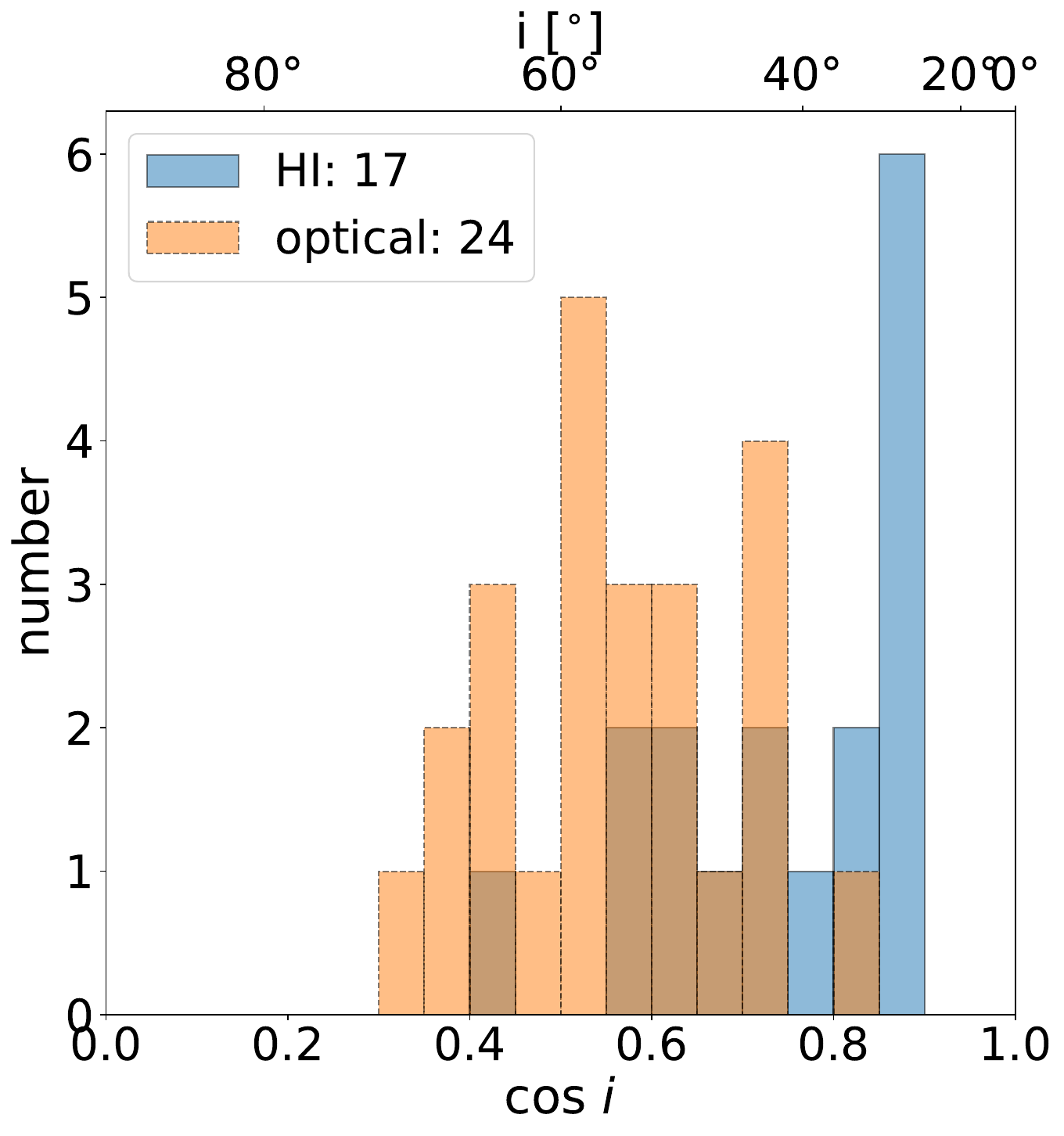}
        \caption{Comparison between \hi\ and optical geometric parameters. 
        \textit{Left}: Comparison of optical and \hi\ inclinations with purple circles corresponding to optical inclinations obtained by assuming intrinsic thickness of $q_0=0.3$, and white circles corresponding to optical inclinations for a razor thin stellar disk.
        \textit{Middle:} Difference in optical and \hi\ inclinations with respect to the difference in the optical and \hi\ morphological position angles.
        \textit{Right:} Histogram of the cosine values of inclinations obtained from \hi\ in blue and optical in orange. The number in the legend corresponds to the number of galaxies with well constrained inclinations. We note that all 6 galaxies in the last \hi\ bin to the right show consistency with inclinations down to 0\dg.}
        \label{fig:inc_pa}
    \end{figure*}

    The left panel of Fig. \ref{fig:inc_pa} shows the comparison between optical and \hi\ inclinations for the galaxies in our sample. For 10 galaxies, the optical and \hi\ inclinations are compatible within the errors, while for the other 7 galaxies, the discrepancy is significantly larger. Generally, the \hi\ determined inclinations are systematically lower than optically determined ones. A possible explanation for this discrepancy could be the assumption of a finite thickness for the stellar disk when deriving the optical inclination, and a razor thin \hi\ disk when deriving the \hi\ inclination. However, even when assuming a razor thin stellar disk (shown as white circles in the plot), the discrepancy is still present, although mitigated. 

    The middle panel of Fig. \ref{fig:inc_pa} shows the difference in inclination versus the difference in morphological position angle. For 7 galaxies we see good consistency between geometric parameters with $\Delta PA \lesssim 25$\dg\ and $\Delta i \leq 10$\dg. For cases with large differences in position angle ($\gtrsim$40\dg), inclinations cannot be directly compared as optical and \hi\ morphologies are misaligned. Therefore, it is not surprising to find large differences in inclinations in these cases. However, even with $\Delta PA \lesssim 25$\dg, we find significant differences in inclinations with 4 cases having $\Delta i > 10$\dg. We conclude that in these cases it is not straightforward to use one set of parameters for both the gas and stars. These comparisons are, however, subject to some caveats. As mentioned in Sect. \ref{sec:obtaining_optical_properties}, possible caveats in the isophotal fitting procedure are the influence of clumpy star forming regions which might dominate over the underlying fainter disk component, as well as the finite depth of optical images which might not allow us to trace the disk component far enough out for a robust comparison with \hi\ morphology. 
    
    On the right panel of Fig. \ref{fig:inc_pa}, we show a histogram for \hi\ and optically determined inclinations. 
    We point out that all 6 galaxies in the rightmost \hi\ bin show consistency with inclinations down to 0\dg\ which would potentially flatten the peak at these values. 
    Both distributions seem to lack high inclinations. While \hi\ source finding favors lower inclinations, the majority of this bias is likely a consequence of our selection procedure where we have a cut of \w50\ < 150 \kms, and a cut of \snrgp\ < 3 (see Sect. \ref{sec:discussion_HI_sample} for more details). However, we note that 3 galaxies (\lsmerge, \paver\ and \misal\ in Figs. \ref{fig:HI-op-lsmerge}, \ref{fig:HI-op-PA_180} and \ref{fig:HI-op-misal}, respectively) might be edge-on. This is seen from the isophotal fitting which showed an inner region (compared to the one used to infer the global geometry of the system) with ellipticities around 0.7, corresponding to inclinations of 90\dg\ for $q_0=0.3$. In comparison, the obtained optical (\hi) inclinations are 68\dg\ (53\dg), 63\dg\ (55\dg) and 57\dg\ (29\dg). These galaxies also show broad \hi\ emission region in the PV slices along the major axis which could be a signature of edge-on galaxies, but we note that these cases are not well spatially resolved, which could also produce broad PV profiles. If these galaxies were edge-on, the distribution of inclinations would be flatter than it appears.

\subsection{Properties of UDGs in our sample}
\label{sec:results_UDGs}

    For the UDG classification, we adopt conditions from Sect. \ref{sec:introduction} (\mmuefX > 24 mag arcsec$^{-2}$, $R_{e,X} > 1.5$ kpc) applied to the $r$-band of PS1 due to better quality of the data compared to the $g$-band ($i$-band is not commonly used in the literature for this classification, see Sect. \ref{sec:introduction}). According to this, our sample contains 9 UDGs in total, but with 4 of them being optically unreliable (see Sect. \ref{sec:obtaining_optical_properties}). The 5 reliable UDGs are \pavel, \pairtworight, \svel, \fon\ and \tstd\ (Figs. \ref{fig:pavel-optical}, \ref{fig:HI-op-pair_2-right}, \ref{fig:HI-op-svel}, \ref{fig:HI-op-fon} and \ref{fig:HI-op-tstd}). They are all marked in Table \ref{tab:optical_par}.

    Throughout the paper, we will compare our UDG sample to the one from \citetads{2019ApJ...883L..33M,2020MNRAS.495.3636M} (hereafter M20). M20 have used similar analysis to ours for both the optical photometry, and the \hi\ kinematic modeling. For one of their galaxies (AGC 114905), we will use the results from \citetads{2022MNRAS.512.3230M} which have been obtained using higher quality \hi\ data. This sample is particularly interesting to use for comparison as it has been shown to systematically deviate from the BTFR.
    
\begin{figure}
        \centering
        \includegraphics[width = 0.48 \textwidth]{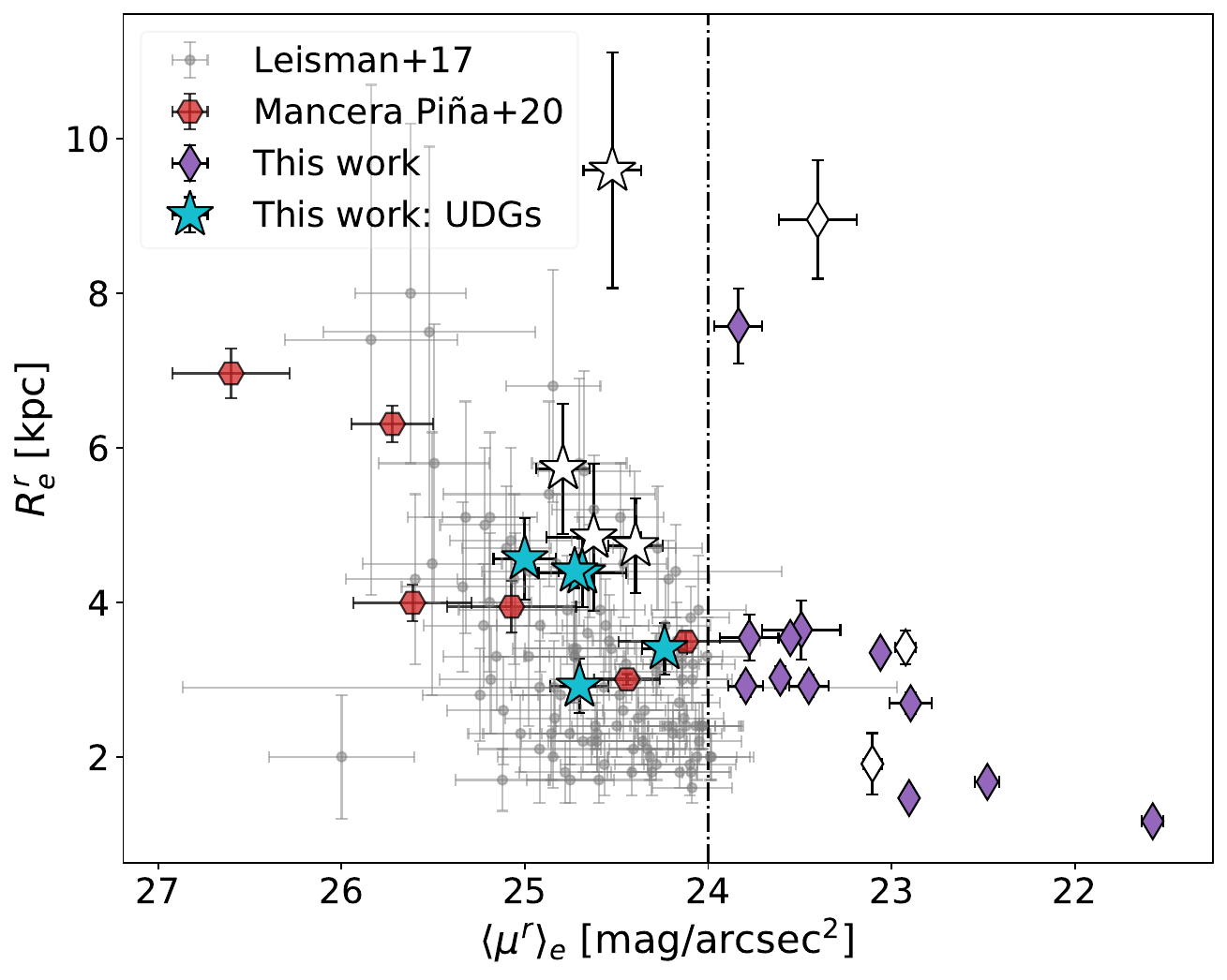}
        \caption{Mean effective surface brightness vs. the effective radius in the $r$-band. Galaxies with well constrained optical parameters (see Sect. \ref{sec:obtaining_optical_properties} for more details) are in color, with UDGs as cyan stars and standard dwarf galaxies as purple diamonds. The rest of the sample with less reliable parameters are shown as white markers, with UDGs as stars and standard dwarfs as diamonds. The \hi\ selected UDG sample from \citetads{2017ApJ...842..133L} is plotted as gray circles, and the sample from M20 as red hexagons. The dash dotted line corresponds to the threshold of surface brightness used for the classification of UDGs.
        }
        \label{fig:mu_re}
    \end{figure}

    In Fig. \ref{fig:mu_re}, we show the \mmuefr\ - $R_{e,r}$ relation. We transferred our \mmuefr\ to SDSS band (using equation 6 in \citetads{2012ApJ...750...99T}) in order to compare with the \hi\ selected UDG sample from \citetads{2017ApJ...842..133L} shown as gray circles, and the sample from M20 shown as red hexagons. \citetads{2017ApJ...842..133L} had an additional constraint on the absolute magnitude $M_r > -17.6$, but all our UDGs (except one optically unreliable) also satisfy that criteria. We note that they do not report the band in which the $R_e$ is measured.  Almost all our UDGs populate the same area as UDGs from \citetads{2017ApJ...842..133L}, with only one optically unreliable UDG having a significantly higher $R_{e,r}$. This outlier is \nce\ whose best fit Sérsic function had a Sérsic index of 2.5, unusually high for previously known UDGs. Compared to the M20 sample, our UDG sample has higher \mmuefr\ on average, with optically reliable subsample populating similar area as the three brightest UDGs from M20.

    As rotational velocity of a galaxy strongly depends on its inclination, with lower inclinations introducing larger uncertainties, we compare inclinations of our UDG subsample to inclinations of our standard dwarfs in order to better understand the precision of our derived rotational velocities between the two samples. For \hi-derived inclinations, 5 out of 8 UDGs have inclinations lower than 40\dg, with 4 of these having inclinations consistent with zero. In comparison, 3 out of 9 standard dwarfs have \hi\ inclination less than 40\dg, with only two consistent with zero.
    Taking the optical inclinations, 3 out of 4 galaxies with inclinations lower than 50\dg\ are labeled as UDGs. Therefore, UDGs in our sample seem to have lower inclinations than standard dwarfs, possibly leading to higher uncertainties in derived rotational velocities.

\section{Scaling relations}
\label{sec:results_scaling_relations}
\subsection{Stellar mass-size relation}
\label{sec:results_mass_size}

    We look at the stellar mass-size relation to study the position of our \hi-selected galaxies with respect to the optically selected ones. 
    We make our comparison with the sample of late-type galaxies from \citetads{2013MNRAS.434..325F}. Their sample is part of the Analysis of the interstellar Medium of Isolated GAlaxies (AMIGA) project, specifically selected for high isolation. This ensures low-to-none contribution of environment on the intrinsic galaxy properties, making it a favorable sample for comparison with the \hi\ selected sample which naturally contains more isolated systems.

    In \citetads{2013MNRAS.434..325F}, they report effective radii defined in a circular aperture\footnote{Sometimes used for early-type galaxies for consistency between non-spherical systems. Calculated by multiplying the effective radii along the semi-major axis of an early-type galaxy ($a_e$) with the square root of the axial ratio ($r_e = a_e\times \sqrt{b/a}$).} ($r_e$), which we convert to elliptical apertures ($R_e$, defined along the semi-major axis, i.e. along the disk) using $R_e = r_e\times\sqrt{a/b}$, where $a$ and $b$ are semi-major and semi-minor axes. 
    
\begin{figure}
        \includegraphics[width = 0.48 \textwidth]{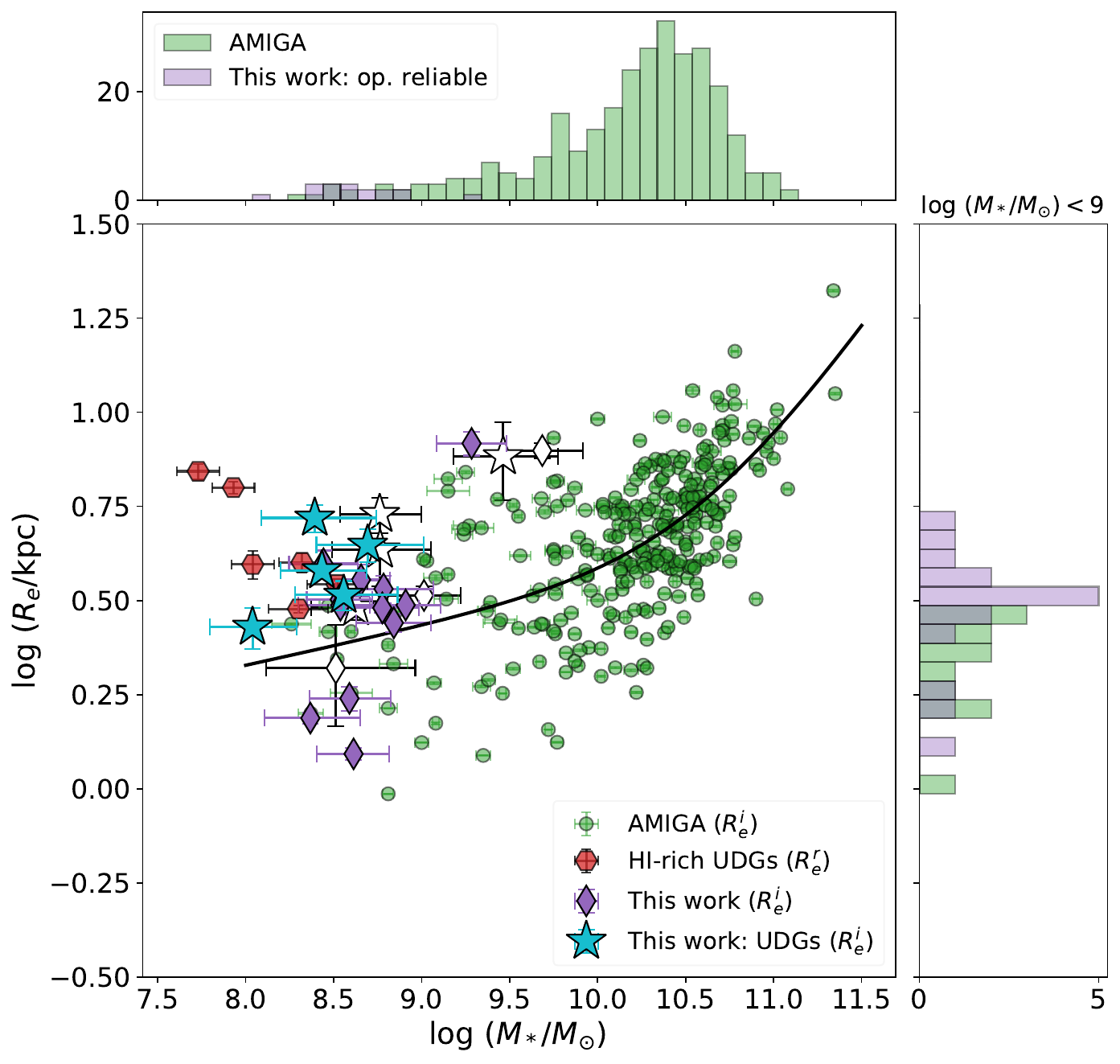}
        \caption{Stellar mass-size relation. Our sample is given with the same markers as in Fig. \ref{fig:mu_re} and is plotted with $R_e$ values from the $i-$band. Galaxies from the AMIGA sample with $i-$band $R_e$ values are given in green, and the corresponding fit (see Sect. \ref{sec:results_mass_size}) is given in black. The \hi-rich UDG sample from M20 is plotted with $R_e$ values from the $r-$band and is denoted as red hexagons. Histogram on the top shows the distribution of stellar masses, with our sample plotted in purple and the AMIGA in green. The histogram on the right shows distributions of effective radii for \mstar< $10^9$\msun\ in our and the AMIGA samples with the same color scheme.}
        \label{fig:mass-size}
    \end{figure}

    The stellar mass-size plot is shown in Fig. \ref{fig:mass-size}. To compare where our sample lies with respect to the AMIGA sample, we fit the AMIGA sample with the functional form from eq. 2 in \citetads{2013MNRAS.434..325F}. In contrast to their work, we leave all parameters free to fit in order to better follow the trend in their data for our comparison. We see that our sample is generally more diffuse (tends to larger $R_e$ at the same stellar mass) than the AMIGA sample. This is likely a consequence of basing our selection on \hi\ detections and thereby allowing lower surface brightness (but same stellar mass) galaxies into the selection, as well as of having a cut in absolute magnitude as part of our selection procedure, which could potentially exclude higher surface brightness galaxies of the same stellar mass. 
    This is particularly interesting as the isolated AMIGA sample used for this comparison has been shown to be systematically more diffuse than an optically selected sample without a strict isolation criterion \citepads{2003MNRAS.343..978S}. Quantitatively, AMIGA galaxies are $\sim$1.2 times larger, for the same stellar mass.

    We also plot UDGs from M20 in our stellar mass-size plot in Fig. \ref{fig:mass-size}, in order to compare with our UDG subsample. The M20 sample does not have $R_e$ measurements in the $i-$band, so we plot their $r-$band $R_e$ values. On average, our sample has higher stellar masses than M20, but similar $R_e$ as their most massive UDGs. Generally, the M20 sample seems to populate more extreme regime of UDG population having lower surface brightness (see Sect. \ref{sec:results_UDGs}), lower stellar masses, and extending to larger effective radii.
    
\subsection{Baryonic Tully-Fisher relation (BTFR)}
\label{sec:btfr}

    Here we look at the BTFR connecting baryonic mass and rotational velocity. We want to explore if the UDGs in our sample show offsets from the relation when compared to the rest of the sample, as was indicated by previous works \citepads[e.g.][]{2017ApJ...842..133L,2020MNRAS.495.3636M}. We compare our sample to various samples from the literature. From the SPARC sample \citepads{2016AJ....152..157L}, we took 126 galaxies with reliable rotation curves (quality flag $Q=1$ or 2) and inclinations $i>25$\dg. From these, we excluded 3 galaxies that are part of the LITTLE THINGS subsample of 17 galaxies from \citetads{2017MNRAS.466.4159I} that have more detailed analysis and use a similar approach to ours. Additionally, we add the SHIELD sample \citepads{2016ApJ...832...89M} of 12 low-mass galaxies. We also compare to two resolved samples of \hi-rich UDGs: 6 UDGs from M20, and 11 edge-on UDGs from \citetads{2019ApJ...880...30H}.

\begin{figure}
        \centering
        \includegraphics[width = 0.49 \textwidth]{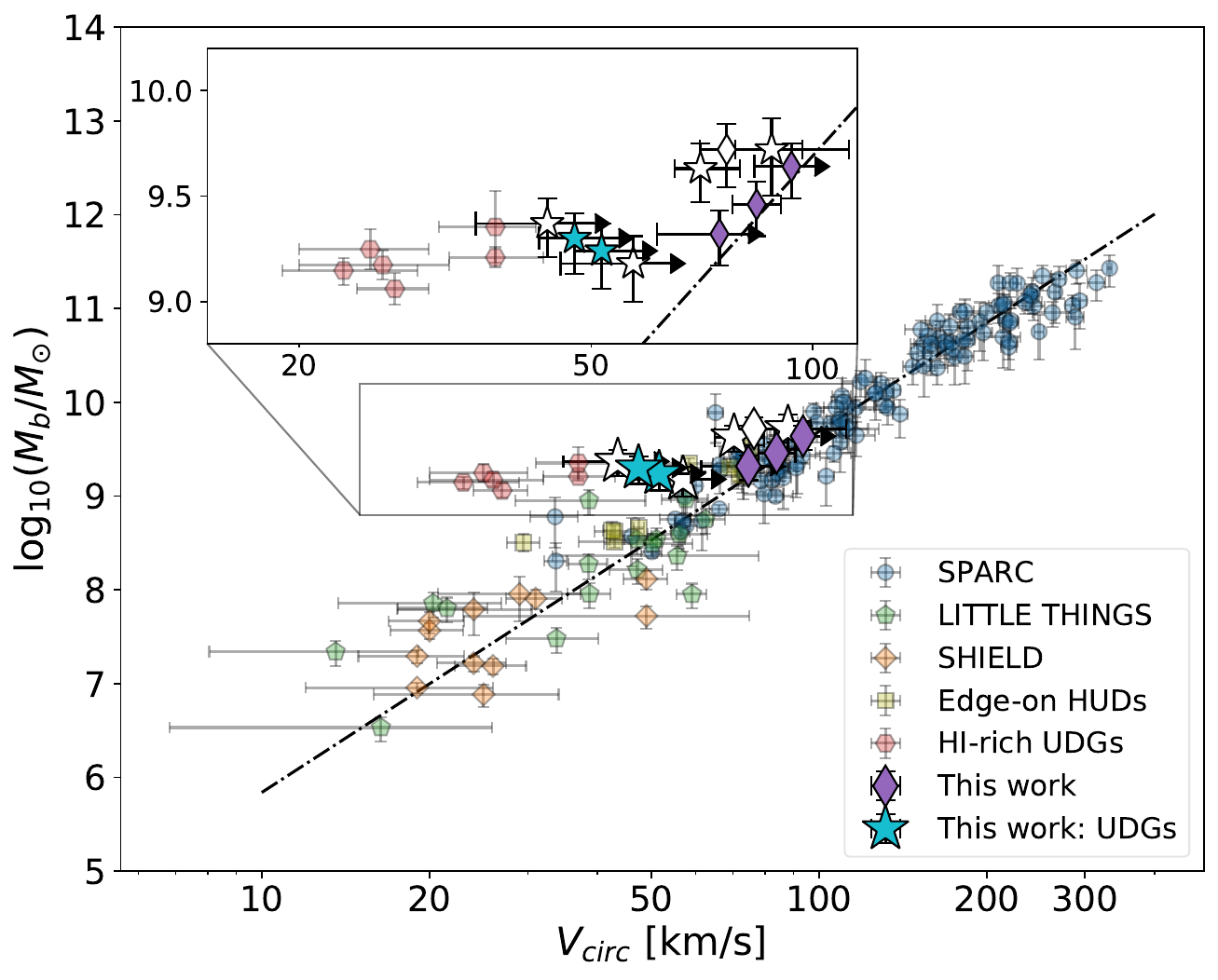}
        \caption{The BTFR comparing the placement of our sample with respect to various samples from the literature. 
        Galaxies in our sample with well constrained rotational velocity (see Appendix \ref{app:results_sample} for more details) are denoted as in Fig \ref{fig:mu_re}. We note that the placement on the BTFR is robust for all galaxies (both optically reliable and unreliable) as derived stellar mass is not significantly impacted by the shape of the Sérsic fit (see Sect. \ref{sec:opt_reliability} for more details); only the UDG classification of the optically unreliable ones is less certain.
        For galaxies which show a flattening in the PV slice along the major axis, we provide errors on circular velocities, while others we plot as lower limits. Black dash dotted line is the best fit model to the SPARC sample from \citetads{2016AJ....152..157L}.}
        \label{fig:BTFR}
    \end{figure}

    The BTFR is shown in Fig. \ref{fig:BTFR}. As mentioned in Sect. \ref{sec:analysis-kinematic_model}, we have used \hi\ geometric parameters (except for the position angle, which is derived from the kinematics) for the final kinematic model for all galaxies except \svel. In this case, we used optical parameters due to the high misalignment of \hi\ morphological and kinematic position angles, making the obtained \hi\ (morphological) inclination non-applicable. A more detailed discussion on the inclination impact for our sample in the BTFR can be found at the end of this section.

    All the standard dwarf galaxies in our sample lie on the relation defined by the SPARC sample, while the UDGs seem to be slightly offset. However, due to the low resolution of our sample, it is not straightforward to say if we are able to trace the flat part of the rotation curve for our galaxies. \citetads{2019MNRAS.484.3267L} explored the connection of the physical extent at which the rotation is measured and the reached rotational velocity, pointing out that velocities measured at $2R_e$ (and lower) might not always be a good tracer of the flat part of the rotation curve for low-mass galaxies, systematically underestimating the $V_{\rm flat}$. When we apply this threshold to our sample, we find that 7 out of 10 galaxies in the BTFR are traced outside of $2R_e$, 3 of which are UDGs (two of them are optically reliable). As mentioned in Sect. \ref{sec:analysis-kinematic_model}, we also independently selected 4 galaxies with potentially flat rotation curve based on the shape of the PVs, all of which also have rotation curves extending beyond 2$R_e$, and two of which are UDGs.
    We also note that 3 UDGs offset from the relation have inclinations consistent with 0\dg, which could move them to higher rotational velocities in the plot.

    When compared to UDGs from M20, our UDGs have higher circular velocities for the same baryonic mass. As seen in Sect. \ref{sec:results_mass_size}, the M20 sample has on average lower stellar masses than our sample, leading to higher gas fractions for the same baryonic mass.
    We explored possible correlations between the increase in $R_e$ and \mhi/\mstar\ fraction with the offset from the relation, but no evident trend is seen. However, we note that the M20 sample represents a more extreme UDG population on average, with the main difference between the two samples being the gas fraction. As high gas fractions of UDGs are likely driven by systematic differences in specific angular momentum of these systems \citepads{2021A&A...651L..15M}, they might explain the relative difference in systematic offsets between the two samples, when taken on average. However, more statistically significant samples are needed to confirm this trend.

\begin{figure}
    \centering
    \includegraphics[width = 0.49 \textwidth]{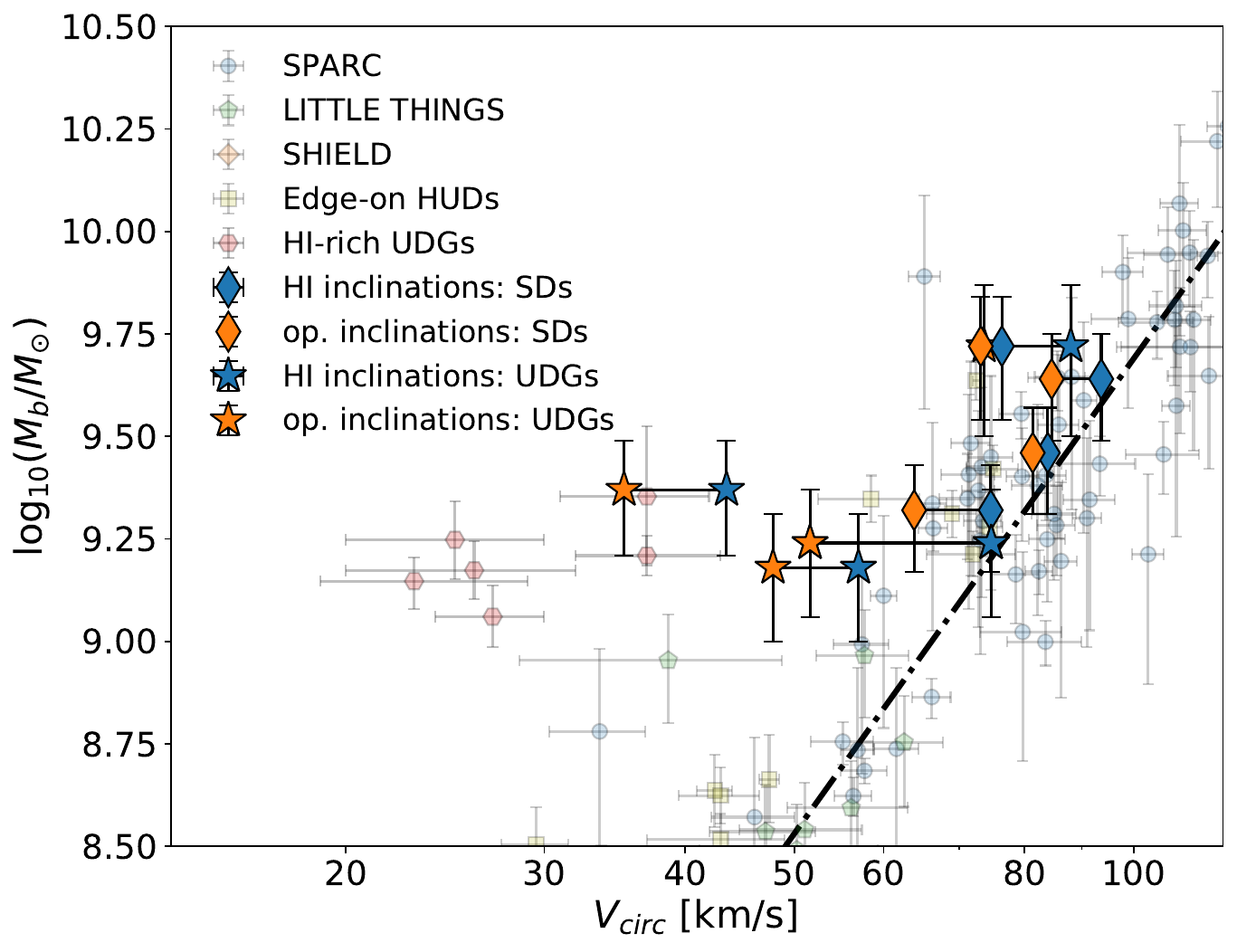}
    \caption{BTFR where galaxies from our sample are plotted based on the inclination used for kinematic modeling. For both UDGs (stars) and standard dwarfs (diamonds), rotational velocities obtained using \hi\ inclinations are given in blue, while the ones obtained using optical inclinations are given in orange. Note that here we do not make a distinction between the optically reliable vs. unreliable and/or rotationally flagged vs. unflagged galaxies, as long as their \PAhi\ - \PAop\ < 45\dg.}
    \label{fig:inc-BTFR}
\end{figure}

    As shown in Sect. \ref{sec:results-comparison-geometries}, our sample shows significant differences in \hi\ and optical inclinations. To explore the impact of this difference on the galaxies' position in the BTFR, we ran \barolo\ again, this time using optical inclinations for galaxies whose optical and \hi\ morphological position angles do not differ by more than 45\dg. Fig. \ref{fig:inc-BTFR} shows the difference of the two cases in the resulting BTFR. While standard dwarf galaxies in our sample are very close to the BTFR from literature in both cases, the UDGs tend to lie off the relation, but their precise locations vary significantly depending on the inclination used. Generally, optical inclinations give systematically lower rotational velocities, as expected from the left panel of Fig. \ref{fig:inc_pa}.

    Taking only the optical or only the \hi\ inclination at the low-mass scales can give a significantly different result in terms of the BTFR. This is especially important as most works up until now have used either the \hi\ \citepads[e.g.][]{2017MNRAS.466.4159I,2020MNRAS.495.3636M} or optical \citepads[e.g.][]{2020ApJ...902...39K,2022ApJ...940....8M,2023ApJ...947L...9H} geometry. Our sample demonstrates that at these low masses, the measurements of \hi\ and optical morphologies can give inconsistent results (possibly due to the insufficient depth of optical images, see Sect. \ref{sec:obtaining_optical_properties}). Therefore, care should be taken when interpreting \hi\ kinematics based on optical morphologies in the low surface brightness regime, and placing galaxies on the BTFR.

\section{Discussion}
\label{sec:discussion}

\subsection{Incidence of pairs}
\label{sec:incidence_pairs}
    We found 3 pairs in our sample of 24 galaxies. This makes the percentage of galaxies in pairs to be 25\% (6/24) in our sample. All of our pairs have stellar mass ratios larger than 0.3. In comparison, \citetads{2013MNRAS.428..573S} studied the frequency of dwarf satellite galaxies as a function of the primary galaxy stellar mass in the Sloan Digital Sky Survey (SDSS).
    The fraction of galaxies in pairs for primary masses and absolute magnitude differences of pairs in our sample (up to one magnitude difference) would be $\sim 5\%$ (see figure 3 from \citetads{2013MNRAS.428..573S}).
    Another work by \citetads{2018MNRAS.480.3376B} compared the SDSS catalog with the \textit{Illustris-1} cosmological simulation \citepads{2014Natur.509..177V,2015A&C....13...12N} and found a similar result ($\sim 4\%$) for the primary stellar mass range of $(2\times 10^8 - 5\times 10^9)$ \msun, but could potentially go up to $\sim 6$\% for future surveys with better completeness at these masses. 
    As we are finding a significantly larger fraction of dwarf galaxy pairs compared to previous studies on optical spectroscopic data, we explore the potential biases of our selection procedure and the \hi\ source finding technique. If galaxies in pairs were detected as individual objects in the initial source finding catalog from Apertif, there would be no bias in our procedure. However, two pairs in the sample were detected as a single source, possibly introducing the biases. We refer to these cases as close pairs, in contrast to pairs which were not recognised as a single source.

    To test if our selection criteria have biased the selection of close pairs in our sample, we apply each criteria again on each galaxy in the two pairs. For consistency, we used the Hubble flow distance and a beam size taken as the geometric mean of the major and minor beam axes (as was used in the selection, see Sect. \ref{sec:source_selection}) in order to test the \mhi\ condition, and the condition that the galaxy is resolved by 3 beams. 
    The results are summarized in Table \ref{tab:pairs_test}. In both pairs, there is one galaxy that passes all criteria. 
    Given that one of the galaxies of the pair always passes the selection criteria, we would have seen the companion galaxy in the data products either way, and thus also include it in the sample. We conclude that our selection did not bias us towards finding more pairs than we would obtain if each galaxy was selected as a single source.

    On the other hand, the source finding itself could introduce additional biases. 
    We note that the Apertif survey is surface brightness limited for resolved sources, but flux limited for unresolved ones (hence also \hi\ mass limited, see Eq. \ref{eq:HI_mass}). 
    As galaxies in our sample are by construction resolved by 3 or more Apertif beams, they would all be resolved with the largest beam used in the source finding (39\arcsec\ beam, see Sect. \ref{sec:Apertif_data}), thereby putting us in the surface brightness limited regime.
    Thus there should be no bias in the detection of a galaxy separately or as part of a close pair.
    
    We conclude that our procedure cannot explain the high frequency of dwarf pairs in the sample. However, our sample is small and a more indepth study using a larger sky area from Apertif or other wide-field resolved \hi\ surveys such as WALLABY \citepads{2020Ap&SS.365..118K}, is advisable for the robust quantification of the incidence of gas-rich dwarf pairs.

    \begin{table*}[]
        \caption{Test of selection criteria for galaxies in close pairs.}
        \centering
        \begin{tabular}{lllll}
        \hline \hline
         Name & log (\mhi/\msun) < 10 & \w50\ < 150 \kms & \snrgp\ > 3 & resolved with 3 beams \\
            \hline
        \paironeleft & 9.00 (\cmark) & $\sim$ 36 \kms (\cmark) & $\sim$ 4.5 (\cmark) & \xmark \\
        \paironeright & 9.54 (\cmark) & $\sim$ 36 \kms (\cmark) & $\sim$ 6.8 (\cmark) & \cmark \\
        \pairthrleft & 8.44 (\cmark) & $\sim$ 70 \kms (\cmark) & $\sim$ 2.8 (\xmark) & \xmark  \\
        \pairthrright & 8.89 (\cmark) & $\sim 110$ \kms (\cmark) & $\sim 4.7$ (\cmark) & \cmark \\
             \hline
        \end{tabular}
        \tablefoot{
        Tick marks mean the galaxy passes the given criterion, while the crosses mean it fails the criterion.
            }
        \label{tab:pairs_test}
    \end{table*}

\subsection{Impact of our \hi-based selection}
\label{sec:discussion_HI_sample}

    Basing our selection on the \hi\ properties of galaxies, we are subject to some biases. As seen in Sect. \ref{sec:results-comparison-geometries}, we are missing high inclination galaxies in our sample. This is partly a consequence of the \hi\ survey which is biased towards lower inclinations. The bias comes from the galaxy emission being distributed over a smaller number of spectral channels (increasing the $S/N$ in each individual channel) than for higher inclination galaxies. Additionally, \snrgp\ threshold in the selection criteria enhances this bias towards lower inclinations for the same reason, as well as the imposed threshold on the \w50. We tested the effects of these selection criteria on our sample by exploring how the \snrgp\ and \w50\ values obtained from the preliminary Apertif source list would change for a range of inclinations (taking the \hi\ measured inclination as the true one). Only $\sim 30$\% of the tested galaxies (17 galaxies, those that have \hi\ inclinations measured) would have been selected for the whole range of inclinations. From the remaining galaxies, $\sim$30\% would be selected only for inclinations $\lesssim$40\dg. We note that in 16/17 cases, the \snrgp\ criterion was the more stringent one, thereby far more dominant for introducing biases than the imposed \w50\ cut. This analysis demonstrates that we have a significant bias towards low inclinations in our sample, but a proper quantification is beyond the scope of this work.
    
    Furthermore, as shown in Sect. \ref{sec:results_mass_size}, our sample is shifted towards higher effective radii in the stellar mass-size relation compared to the isolated AMIGA sample. This could also be a consequence of the \hi-based selection as it is by construction optically independent and allows us to pick galaxies of lower surface brightness for the same stellar mass. This effect is especially important for galaxies with lower inclinations present in our sample as they appear less bright when projected on the sky and hence have lower probability of being detected in optical surveys. 

    We compare the number of UDGs in our sample to expectations based on the UDG \hi\ mass function (HIMF) of \citetads{2018A&A...614A..21J}. They constructed the HIMF and the velocity width function (made using \w50) of UDGs (selected by \citetads{2017ApJ...842..133L}) from the $\alpha.70$ ALFALFA \hi\ untargeted survey catalog \citepads{2011AJ....142..170H}. They report that the fraction of UDGs to total number of galaxies peaks at log \mhi/\msun $\sim 8.8$ with a contribution of 6\%, and lowers towards both higher and lower \hi\ masses. In comparison to our optically reliable subsample, we have 5 UDGs, making $\sim30$\%.  However, \citetads{2018A&A...614A..21J} do not make a distinction between large stellar mass galaxies with lower gas fractions and dwarf galaxies with lower masses in both stars and gas. 
    Therefore, their comparison for UDG frequency is against a large range of stellar masses and hence total galaxy masses, bringing the final UDG fractions lower.
    In addition, the contribution of UDGs is seen to steeply rise (more steeply than the total velocity width function) towards the narrower \w50\ profiles. This goes in accordance with possibly lower rotational velocities as well as low \hi\ inclinations of UDGs compared to standard dwarfs. Given that we have put an upper limit on \w50\ in our selection, we may have introduced some preference for selecting UDGs in our sample.

\section{Conclusions}
\label{sec:conclusion}

In this work, we have presented a sample of 24 dwarf galaxies with spatially resolved \hi\ data from the Apertif imaging survey \citepads{2022A&A...667A..38A}. Our selection procedure ensured that galaxies are resolved by at least 3 beams in \hi, that the average signal-to-noise ratio per channel in the global profile (\snrgp) is larger than 3, and that their \hi\ and stellar masses are \mhi<$10^{10}$\msun\ and \mstar$\lesssim 10^{9}$\msun, respectively. We measured the geometry of the \hi\ disk for 17 galaxies (out of 24) in the sample and have successfully produced kinematic models of 13 of them using \barolo. We also studied properties of their stellar disks (24 galaxies) by conducting isophotal fitting and extracting surface brightness profiles from \textit{g-, r-} and \textit{i-}bands of the Pan-STARRS 1 photometric survey. We identified 5 UDGs, and 4 candidate UDGs in the sample. Using the above results, we position our sample with respect to other samples from the literature on the stellar mass-size relation and the BTFR. 

In the following, we summarize our main conclusions:
\begin{itemize}

    \item Our \hi\ selected sample seems to be more diffuse (has larger $R_{e,X}$ at given stellar mass) than the optically selected sample of isolated galaxies from the AMIGA project (Fig. \ref{fig:mass-size}). This shows that an \hi-based selection returns a different population from an optical selection, and hence should be taken into account in statistical studies of dwarf galaxies.
    
    \item We find apparent misalignments between the derived optical and \hi\ morphologies for 9 (out of the 17 measured) galaxies in our sample.
    However, this comparison is subject to caveats originating from the influence of clumpy star forming regions on the derived stellar morphology, as well as the insufficient depth of PS1 images which might not trace the underlying stellar disk far enough out for a robust comparison with the \hi\ morphology.

    \item Standard dwarf galaxies in our sample lie on the BTFR determined for higher mass galaxies, but the UDGs in our sample seem to be slightly shifted towards lower rotational velocities (Fig. \ref{fig:BTFR}). However, we are limited by a small number of galaxies in our sample, and in most cases we cannot guarantee that the measured rotational velocities are fully representative of the flat part of the rotation curves.

    \item Inclination has a large impact on the measured rotational velocity, and consequently on the position of a galaxy in the BTFR (Fig. \ref{fig:inc-BTFR}). Given the significant misalignments between the measured \hi\ and stellar disk morphologies, care should be taken when deciding which geometric parameters to use for the derivation of the kinematics.

    \item We find a larger fraction (25\%) of dwarf galaxies in pairs than studies based on optical spectroscopic data. We consider possible biases in our selection procedure and the \hi\ source-finding, and find it unlikely to account for this. 
    Our findings suggest that \hi\ surveys might be detecting more dwarf galaxy pairs than are found in optical spectroscopic surveys. This can have relevant implications for the formation and evolution of these low-mass galaxies and will be further explored in an upcoming publication.
    
\end{itemize}

The above conclusions are based on an early \hi\ source list from Apertif data, which is limited in both sensitivity and coverage. Future work is undertaking source finding on more sensitive data and over a larger survey area. This will enable detection of many more sources passing our selection criteria and could put stronger constraints on our final conclusions. A more statistical study of the fraction of dwarf pairs and/or UDGs would also benefit from dropping some of our current selection criteria, such as spatially-resolving the \hi\ disk or the \snrgp\ > 3. Furthermore, new upcoming facilities such as the Square Kilometre Array (SKA) will enable the detection of even higher number of dwarf galaxies with superior spatial and spectral resolution. The present study and future follow-ups will clarify the complexity of the present-day dwarf galaxy population, permitting to obtain key constraints of theoretical models of galaxy formation and evolution at the low-mass end.

\begin{acknowledgements} \\
We thank the anonymous referee for a valuable report that improved this manuscript. We thank Sabrina Stierwalt for pointing us towards previous studies on dwarf galaxy pairs. We also greatly thank Jackson Fuson and John M. Cannon for sharing their asymmetric drift correction code with us. KMH acknowledges financial support from the grant CEX2021-001131-S funded by MCIN/AEI/ 10.13039/501100011033 from the coordination of the participation in SKA-SPAIN funded by the Ministry of Science and Innovation (MCIN); and from grant PID2021-123930OB-C21 funded by MCIN/AEI/ 10.13039/501100011033 by “ERDF A way of making Europe” and by the "European Union". Authors KMH and BŠ acknowledge the Spanish Prototype of an SRC (SPSRC) service and support funded by the Ministerio de Ciencia, Innovación y Universidades (MICIU), by the Junta de Andalucía, by the European Regional Development Funds (ERDF) and by the European Union NextGenerationEU/PRTR. The SPSRC acknowledges financial support from the Agencia Estatal de Investigación (AEI) through the "Center of Excellence Severo Ochoa" award to the Instituto de Astrofísica de Andalucía (IAA-CSIC) (SEV-2017-0709) and from the grant CEX2021-001131-S funded by MICIU/AEI/ 10.13039/501100011033. KMH and JMvdH acknowledge funding from the ERC under the European Union’s Seventh Framework Programme (FP/2007–2013)/ERC Grant Agreement No. 291531 (‘HIStoryNU’). AAP acknowledges support of the STFC consolidated grant [ST/S000488/1]. PEMP acknowledges the support from the Dutch Research Council (NWO) through the Veni grant VI.Veni.222.364.

This work makes use of data from the Apertif system installed at the Westerbork Synthesis Radio Telescope owned by ASTRON. ASTRON, the Netherlands Institute for Radio Astronomy, is an institute of the Dutch Research Council (“De Nederlandse Organisatie voor Wetenschappelijk Onderzoek”, NWO). The Pan-STARRS1 Surveys (PS1) have been made possible through contributions of the Institute for Astronomy, the University of Hawaii, the Pan-STARRS Project Office, the Max-Planck Society and its participating institutes, the Max Planck Institute for Astronomy, Heidelberg and the Max Planck Institute for Extraterrestrial Physics, Garching, The Johns Hopkins University, Durham University, the University of Edinburgh, Queen's University Belfast, the Harvard-Smithsonian Center for Astrophysics, the Las Cumbres Observatory Global Telescope Network Incorporated, the National Central University of Taiwan, the Space Telescope Science Institute, the National Aeronautics and Space Administration under Grant No. NNX08AR22G issued through the Planetary Science Division of the NASA Science Mission Directorate, the National Science Foundation under Grant No. AST-1238877, the University of Maryland, and Eotvos Lorand University (ELTE). This work made use of Astropy:\footnote{http://www.astropy.org} a community-developed core Python package and an ecosystem of tools and resources for astronomy \citepads{astropy:2013, astropy:2018, astropy:2022}. This research made use of Photutils, an Astropy package for
detection and photometry of astronomical sources \citep{larry_bradley_2023_1035865}. This research has made use of the NASA/IPAC Extragalactic Database (NED),
which is operated by the Jet Propulsion Laboratory, California Institute of Technology,
under contract with the National Aeronautics and Space Administration.

\end{acknowledgements}

\bibliographystyle{aa} 
\bibliography{refs.bib} 
\newpage

\begin{appendix}

\section{$^{3D}$Barolo Parameter file example}
\label{App:par_file}

Here we give an example of the parameter file for kinematic modeling used in this work. For additional information, please visit the \barolo\ documentation\footnote{\url{https://bbarolo.readthedocs.io/en/latest/index.html}}.
\newline
\newline
\noindent FITSFILE...; data cube name\\
3DFIT True; perform 3D fitting \\
NRADII...; number of rings in the model \\
RADSEP...; separation between rings in arcsec\\
VSYS...; systemic velocity in \kms \\
XPOS...; X axis position of the center of the galaxy in pixels \\
YPOS...; Y axis position of the center of the galaxy in pixels \\
INC...; inclination \\
PA...; kinematic position angle \\
Z0 0; thickness of the disk in arcsec \\
DISTANCE...; distance to the galaxy in Mpc \\
FREE VROT VDISP VSYS PA; parameters for fitting \\
NORM AZIM; azimuthal normalisation of the model cube \\
MASK SEARCH; method for producing a 3D mask \\
THRESHVELOCITY 1; keyword for masking which ensures to not take into account emissions with 2 or more channels apart \\
FLAGROBUSTSTATS True; use median and MADFM for calculating cube statistics \\
SNRCUT 3.3; initial signal-to-noise ratio cut for producing a mask \\
FLAGGROWTH True; enable enlarging the initial mask \\
GROWTHCUT 2.5; enlarge the mask until the S/N reaches 2.5 \\
BWEIGHT 0; not penalising models extending outside the mask\\
WFUNC 2; choosing $\cos^2 \theta$ as a weighing function \\
LINEAR ...; 0.5 times spectral resolution, in units of channels \\
NOISE... ; rms noise per channel in the cube \\
TWOSTAGE True; enable two-stage fitting \\
POLYN 0; fit functional form to selected parameters across the rings, in this case VSYS and PA are fitted with a constant value \\
FLAGERRORS True; calculate and report errors on rotational velocity and velocity dispersion \\
MINVDISP...; minimal value of velocity dispersion, in our case half of the spectral resolution of the cube \\
SIDE B; fit both receding and approaching sides of the galaxy\\

\section{Truncated vs. interpolated photometry}
\label{app:photometry_caveats}

\begin{figure}
    \centering
    \includegraphics[width = 0.40 \textwidth]{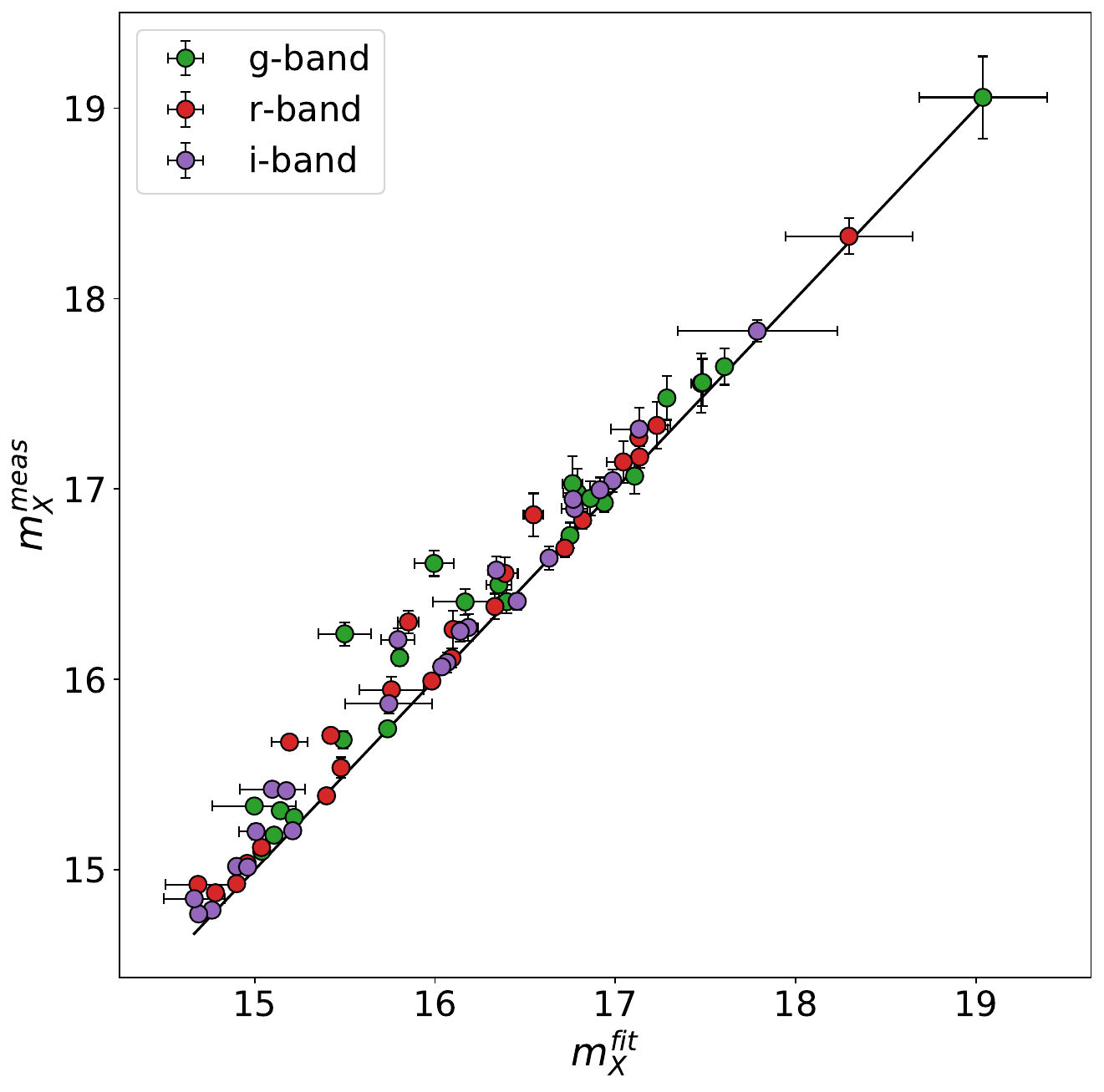}
    \caption{Comparison of apparent magnitudes in different bands for values obtained from the fit of the Sérsic profiles and the directly measured ones.}
    \label{fig:mags_compare}
\end{figure}

\begin{figure*}
   
    \begin{minipage}{.48\textwidth}
    \centering
    \includegraphics[width = 0.9 \textwidth]{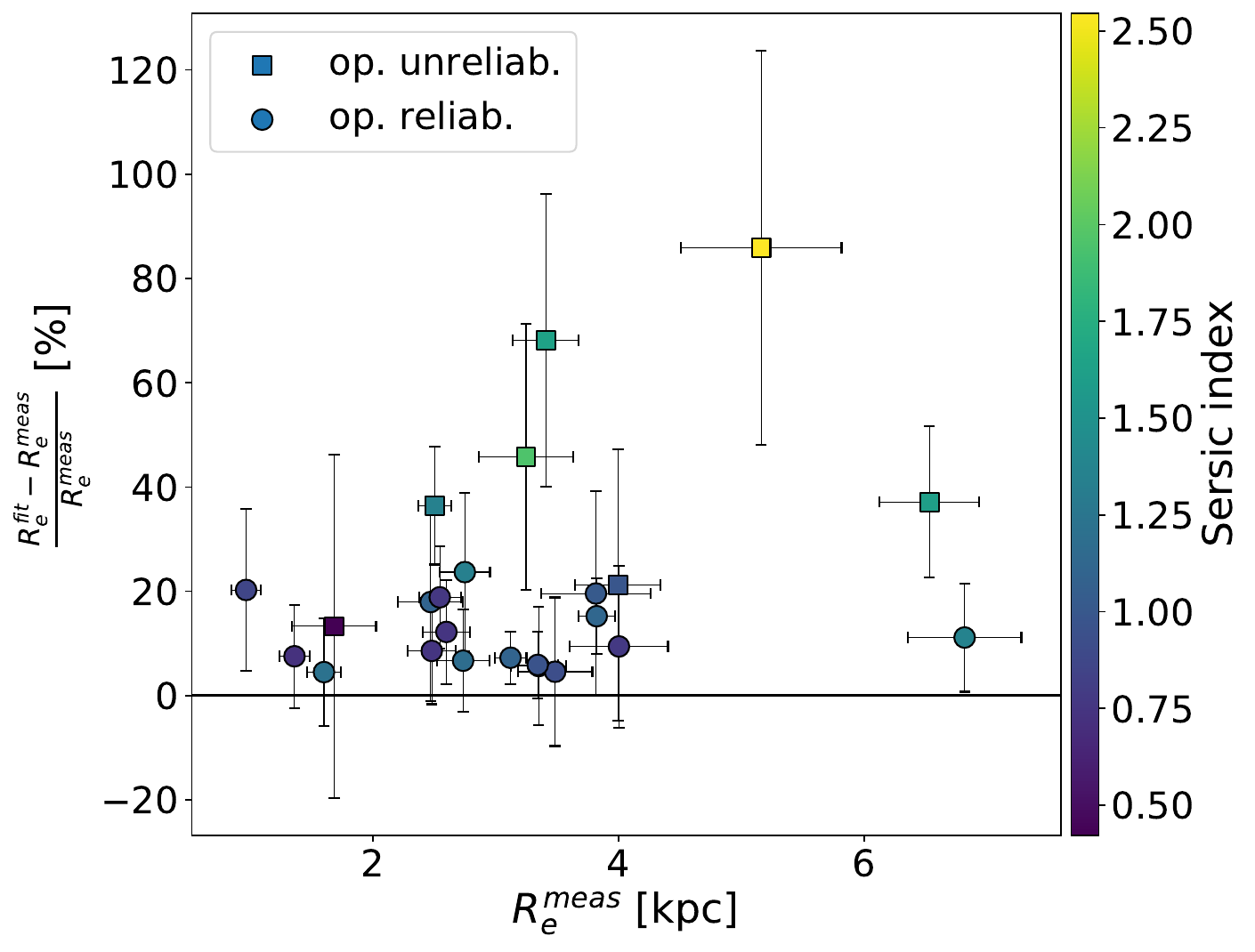}
    \captionof{figure}{Difference of effective radii obtained from the fit of Sérsic profiles and the directly measured ones for the $r-$band, given in percentages. The points are colorcoded using the Sérsic index, with optically reliable galaxies represented as circles and optically unreliable as squares.}
    \label{fig:re_compare}
    \end{minipage}%
    \hfill
    \begin{minipage}{.48\textwidth}
    \centering
    \includegraphics[width = 0.75 \textwidth]{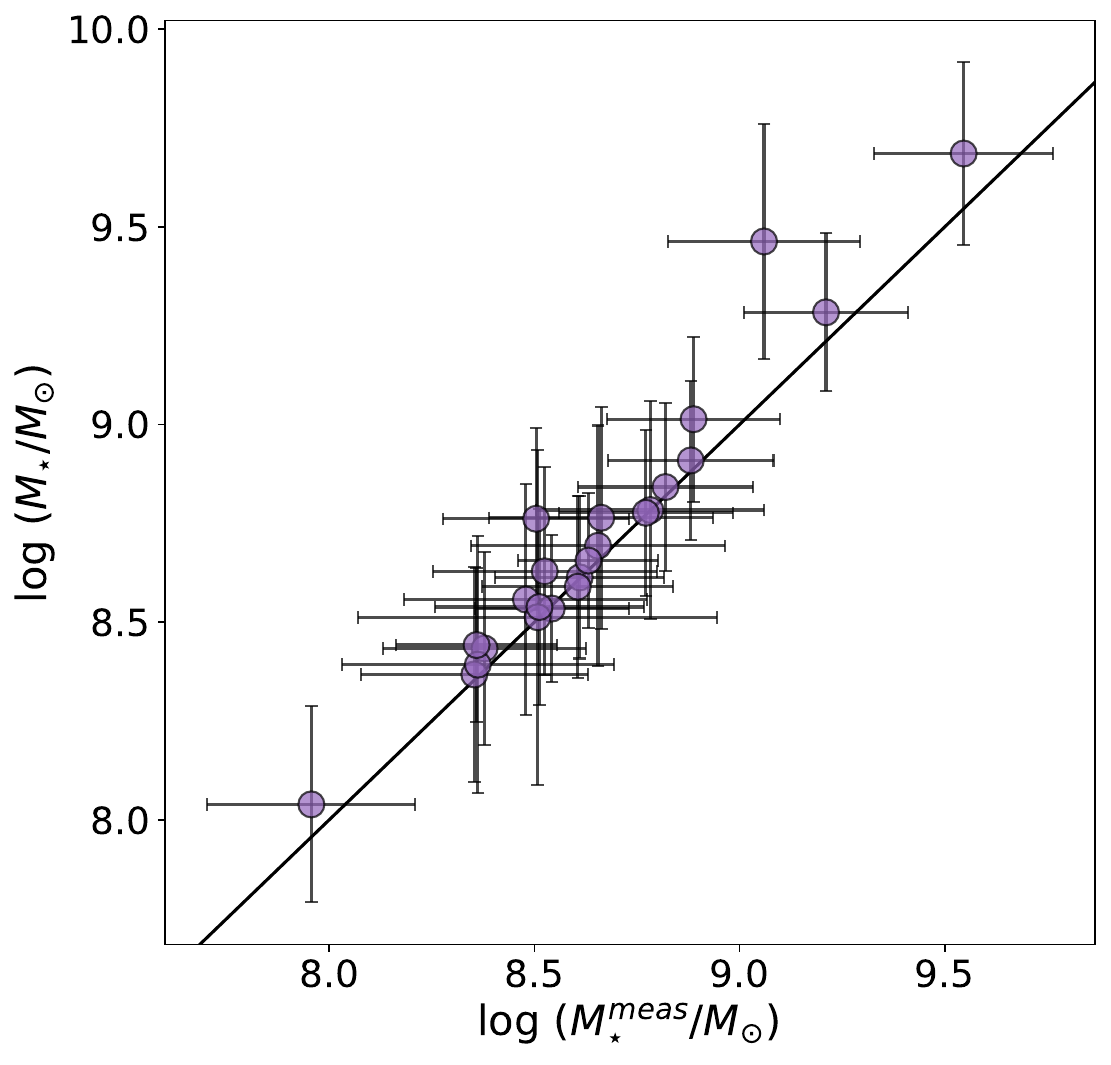}
    \captionof{figure}{Comparison of stellar masses calculated using either the magnitude obtained from the fitted Sérsic function, or using the directly measured magnitude from the extracted surface brightness profile (corresponding to $M^{\rm meas}_{\star}$).}
    \label{fig:stellar_mass_comparison}
    \end{minipage}
\end{figure*}

    In this work, we have obtained optical parameters ($R_{e,X}$, $m_X$ and \mmuefX) from the fits of the extracted surface brightness profiles for which we used the Sérsic function defined by:
    \begin{equation}
        \mu_{X}(r) = \mu_{e,X} + \frac{2.5b}{\ln(10)}\left[\left(\frac{r}{R_{e,X}}\right)^{1/n_{X}}-1\right],\ \ \ \ b = 2n_{X} - \frac{1}{3} + \frac{4}{405n_{X}}
    \end{equation}
    where $\mu_{e,X}$ is surface brightness at $R_{e,X}$, and $n_X$ the Sérsic index. The $m_X$ and \mmuefX\ are then obtained using equations \citepads[see e.g.][]{2005PASA...22..118G}:
    \begin{equation}
        \text{\mmuefX} = \mu_{e,X} - 2.5\log \left( \frac{n_{X}e^b}{b^{2n_{X}}}\Gamma(2n_{X})\right)
    \end{equation}
    \begin{equation}
        m_{X} = \text{\mmuefX} - 2.5\log (2\pi (R_{e,X})^2)
        \label{eq:magnitude}
    \end{equation}
    where $\Gamma(x)$ is the gamma function. $R_e$, as defined by the Sérsic function, is theoretically the same as half-light radius. However, the Sérsic function extends to infinity in $r$, while galaxies do not. Because of this, Eq. \ref{eq:magnitude} could be overestimating $m_X$ and thus also overestimating $R_{e,X}$, or rather $R^{\rm fit}_e> R^{\rm meas}_e$, where $R^{\rm meas}_e$ is the half-light radius. 

    On the other hand, we can directly measure $m_X$ and $R_{e,X}$ from the profile. 
    To do this, we sum the intensity in all rings that were used in the fitting of the profiles to get the total flux and convert it to $m_X$ using Eq. \ref{eq:mag_from_count}. We provide these measured values in Table \ref{tab:phot_meas}. The comparison of values obtained by the fit and those from direct measurement can be found in Fig. \ref{fig:mags_compare}. Similarly, we find the half-light radii by taking the radius containing half of the total flux. The comparison of obtained measured and fitted $R_{e,X}$ can be found in Fig. \ref{fig:re_compare}. We computed \mmuefX\ by summing the flux inside the measured $R_e$, converting it to a magnitude and using Eq. \ref{eq:magnitude} in reverse. Finally, we compare the obtained stellar masses in Fig. \ref{fig:stellar_mass_comparison}, computed using either the fitted magnitude, or the measured magnitude in the $g-$ band (the color is always computed in the same way). The obtained masses are consistent within the errors for all galaxies, except \pairtwoleft\ and \nce\ for which they are compatible within 2$\sigma$. 

    Directly measured magnitudes are systematically fainter than ones obtained by the fit. This is not surprising as $m^{fit}_X$ are obtained by integration of the Sérsic function to infinity, as discussed before. Similarly, measured $R_{e,X}$ are systematically smaller than fitted values with differences being more pronounced for Sersic indices above 1.

\begin{table*}
\tiny
\centering
\caption{Directly measured photometric properties.}
\begin{tabular}{llllll}
\hline \hline
& Name  & $m_g$ & $R_{e,r}$ & \mmuefr & log\\
& & & [kpc] & [mag arcsec$^{-2}$]  & (\mstar/\msun)\\
\hline
1 & \paironeleft & ${17.64} \pm {0.10}$ & ${1.4} \pm {0.1}$ & ${22.84} \pm {0.20}$ & ${8.35}_{-0.27}^{+0.29}$ \\
2 & \paironeright & ${16.76} \pm {0.07}$ & ${3.4} \pm {0.2}$ & ${23.79} \pm {0.15}$ & ${8.78}_{-0.27}^{+0.29}$ \\
3 & \pairtwoleft $^{\dagger}$ * & ${16.61} \pm {0.07}$ & ${3.4} \pm {0.3}$ & ${24.17} \pm {0.18}$ & ${8.50}_{-0.22}^{+0.23}$ \\
4 & \pairtworight * & ${16.95} \pm {0.10}$ & ${4.0} \pm {0.4}$ & ${24.59} \pm {0.22}$ & ${8.65}_{-0.30}^{+0.32}$ \\
5 & \pairthrleft & ${15.74} \pm {0.02}$ & ${1.0} \pm {0.1}$ & ${21.27} \pm {0.28}$ & ${8.61}_{-0.21}^{+0.20}$ \\
6 & \pairthrright & ${15.28} \pm {0.03}$ & ${2.5} \pm {0.2}$ & ${22.78} \pm {0.17}$ & ${8.82}_{-0.21}^{+0.21}$ \\
7 & \bSNRone & ${15.31} \pm {0.04}$ & ${6.8} \pm {0.5}$ & ${23.70} \pm {0.15}$ & ${9.21}_{-0.20}^{+0.20}$ \\
8 & \svel * & ${16.98} \pm {0.14}$ & ${2.7} \pm {0.2}$ & ${23.99} \pm {0.18}$ & ${8.48}_{-0.28}^{+0.31}$ \\
9 & \lsmerge & ${16.93} \pm {0.05}$ & ${2.6} \pm {0.2}$ & ${23.31} \pm {0.17}$ & ${8.54}_{-0.19}^{+0.19}$ \\
10 & \fon * & ${17.48} \pm {0.12}$ & ${2.5} \pm {0.3}$ & ${24.51} \pm {0.26}$ & ${7.96}_{-0.25}^{+0.26}$ \\
11 & \inter & ${15.10} \pm {0.03}$ & ${3.5} \pm {0.3}$ & ${23.46} \pm {0.19}$ & ${8.63}_{-0.17}^{+0.17}$ \\
12 & \bSNRtwo $^{\dagger}$ * & ${16.41} \pm {0.11}$ & ${4.0} \pm {0.3}$ & ${24.43} \pm {0.22}$ & ${8.66}_{-0.26}^{+0.28}$ \\
13 & \tstd * & ${16.50} \pm {0.10}$ & ${3.8} \pm {0.1}$ & ${24.59} \pm {0.12}$ & ${8.38}_{-0.24}^{+0.26}$ \\
14 & \nce $^{\dagger}$ * & ${16.24} \pm {0.26}$ & ${5.2} \pm {0.7}$ & ${24.06} \pm {0.30}$ & ${9.06}_{-0.23}^{+0.24}$ \\
15 & \beaut $^{\dagger}$ & ${19.06} \pm {0.43}$ & ${1.7} \pm {0.3}$ & ${23.54} \pm {0.59}$ & ${8.51}_{-0.41}^{+0.46}$ \\
16 & \stick & ${16.41} \pm {0.06}$ & ${3.3} \pm {0.2}$ & ${23.81} \pm {0.11}$ & ${8.77}_{-0.21}^{+0.21}$ \\
17 & \pavel * & ${17.55} \pm {0.16}$ & ${3.8} \pm {0.4}$ & ${25.03} \pm {0.28}$ & ${8.36}_{-0.31}^{+0.35}$ \\
18 & \reg $^{\dagger}$ * & ${17.03} \pm {0.16}$ & ${3.2} \pm {0.4}$ & ${24.29} \pm {0.27}$ & ${8.53}_{-0.27}^{+0.28}$ \\
19 & \paver & ${17.07} \pm {0.10}$ & ${1.6} \pm {0.1}$ & ${22.74} \pm {0.19}$ & ${8.61}_{-0.23}^{+0.24}$ \\
20 & \mess $^{\dagger}$ & ${16.11} \pm {0.04}$ & ${2.5} \pm {0.1}$ & ${22.88} \pm {0.12}$ & ${8.89}_{-0.21}^{+0.21}$ \\
21 & \wtm & ${15.18} \pm {0.03}$ & ${3.1} \pm {0.1}$ & ${23.00} \pm {0.09}$ & ${8.88}_{-0.20}^{+0.20}$ \\
22 & \dwtm & ${15.68} \pm {0.06}$ & ${2.7} \pm {0.2}$ & ${23.69} \pm {0.19}$ & ${8.36}_{-0.20}^{+0.20}$ \\
23 & \misal $^{\dagger}$ & ${15.34} \pm {0.17}$ & ${6.5} \pm {0.4}$ & ${23.30} \pm {0.20}$ & ${9.55}_{-0.22}^{+0.21}$ \\
24 & \weir & ${17.56} \pm {0.13}$ & ${2.5} \pm {0.2}$ & ${23.71} \pm {0.16}$ & ${8.51}_{-0.25}^{+0.26}$ \\

\hline
\label{tab:phot_meas}
\end{tabular}
\tablefoot{Column names are the same in Table \ref{tab:optical_par}, with the additional column of the stellar mass. Reported values are not corrected for Galactic extinction (except color). \\
($^{\dagger}$) Galaxy was flagged as the extent of the surface brightness profile is not being traced far enough out for reliable Sérsic fit (see Sect. \ref{sec:opt_reliability}).\\
(*) Galaxy is a UDG.     }
\end{table*}

\FloatBarrier
\section{Notes on individual galaxies}
\label{app:results_sample}
    
    In this section we provide detailed comments for each galaxy considering both gas and stellar properties.

    \textbf{\pavel.} The galaxy is shown in Figs. \ref{fig:pavel-kinematics}, \ref{fig:isophot_fit-Pavel} and \ref{fig:pavel-optical}. The \hi\ disk shows an apparent decrease in inclination with radius, as seen from the total intensity map in Fig. \ref{fig:pavel-kinematics}. Therefore, the output from CANNUBI showed the galaxy being consistent with a wide range of inclinations down to the inclination of 0\dg. The kinematics are regular with no apparent signs of non-circular motions. The stellar counterpart is irregular with a slight excess of emission towards the west compared to the east side of the galaxy in the outer parts. Interestingly, fitted isophotes seem to align more with the \hi\ geometry as they go further out of the galaxy, both by the shifting of the \PAop\ and by slightly decreasing ellipticity.

\begin{figure*}
        \centering
        \includegraphics[width = 0.7 \textwidth]{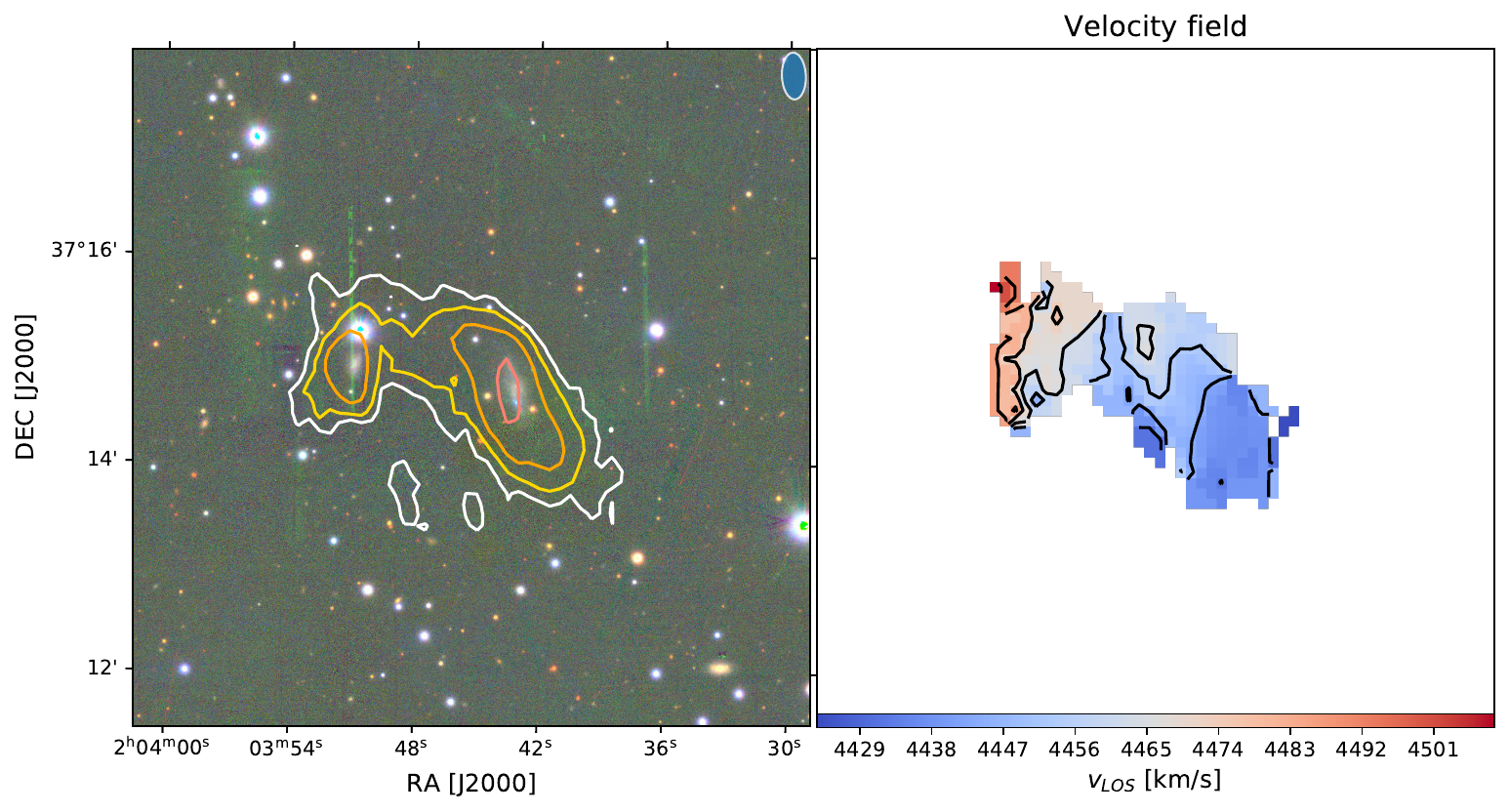}\\
        \centering
        \includegraphics[width = 0.7 \textwidth]{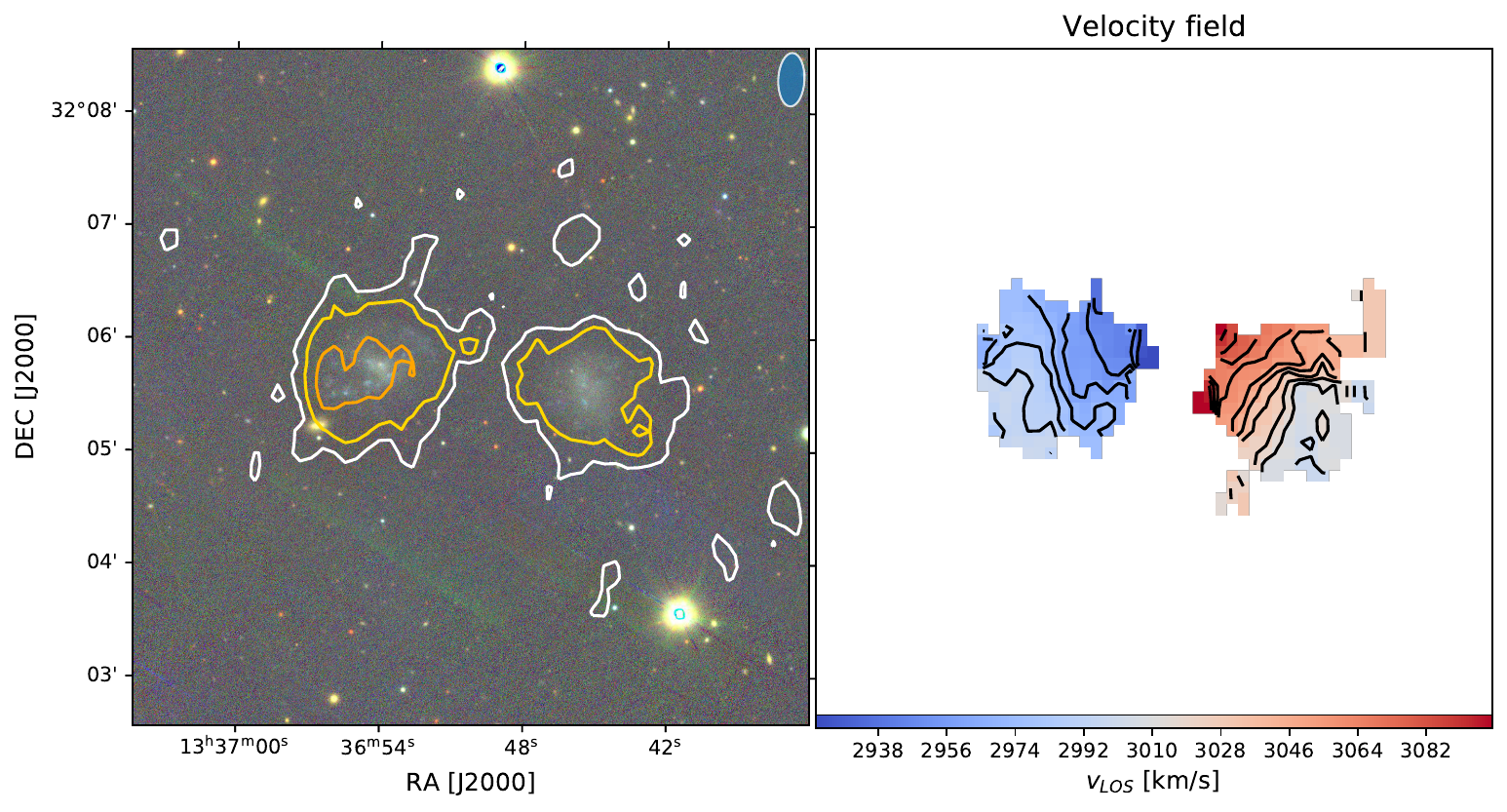}\\
        \centering
        \includegraphics[width = 0.7 \textwidth]{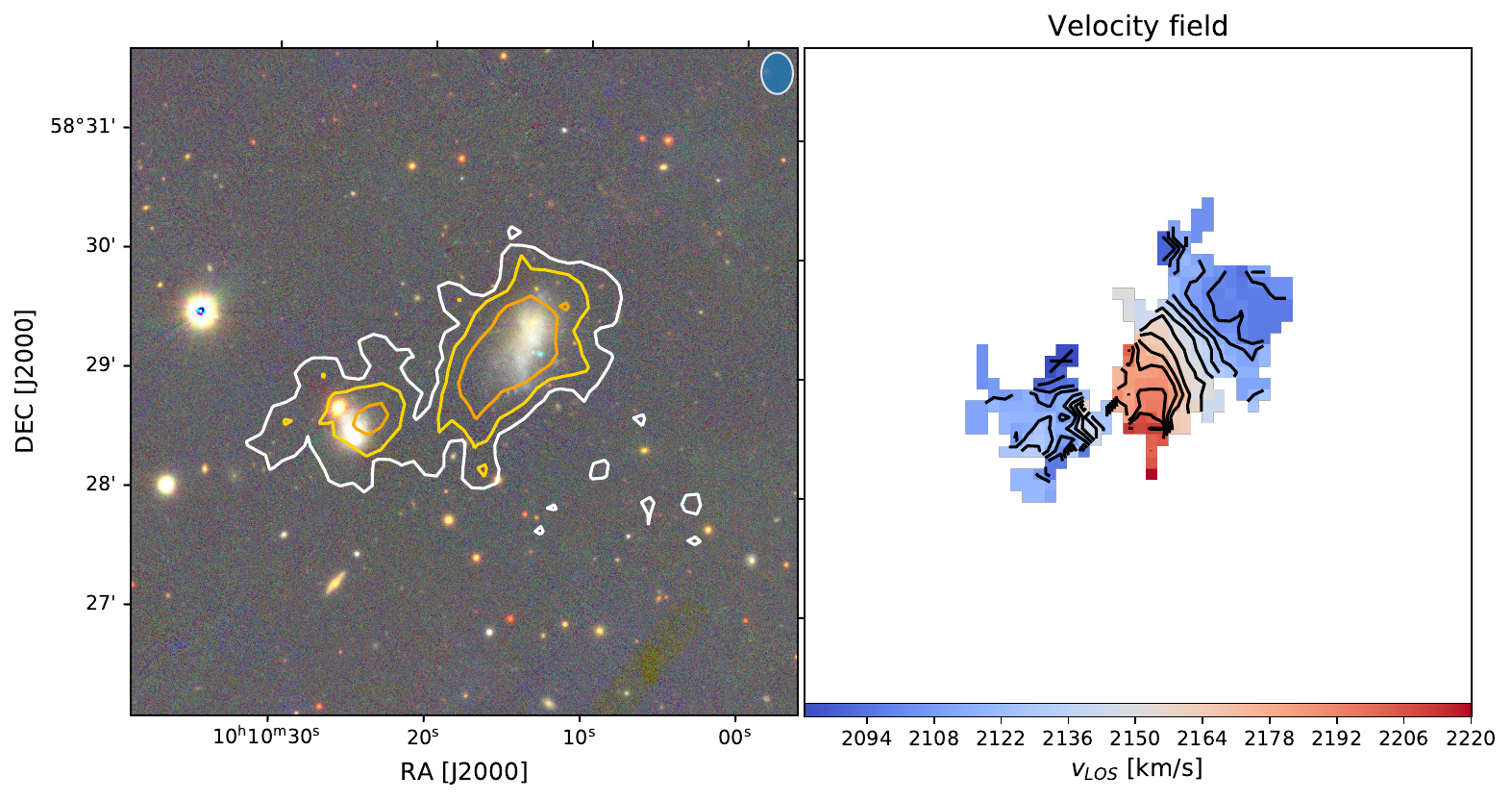}\\
        \caption{Kinematics of pairs. First row shows \paironeleft\ and \paironeright, the second shows \pairtwoleft\ and \pairtworight, and lastly, the third row shows \pairthrleft\ and \pairthrright. \textit{Left panels:} Color-composite image from PS1 $g-$, $r-$ and $i-$bands overlaid with \hi\ contours. The \hi\ contours correspond to column densities starting from $1.2 \times 10^{20}$ cm$^{-2}$ (for \paironeleft\ and \paironeright), $1.8 \times 10^{20}$ cm$^{-2}$ (for \pairtwoleft\ and \pairtworight) and $2.2 \times 10^{20}$ cm$^{-2}$ (for \pairthrleft\ and \pairthrright) shown in white, and growing by a factor of 2 in intensity towards contours in redder colors. \textit{Right panels:} Velocity fields of pairs.}
        \label{fig:kinematics_pairs}
    \end{figure*}

\begin{figure*}
        \centering
        \includegraphics[width = 0.7 \textwidth]{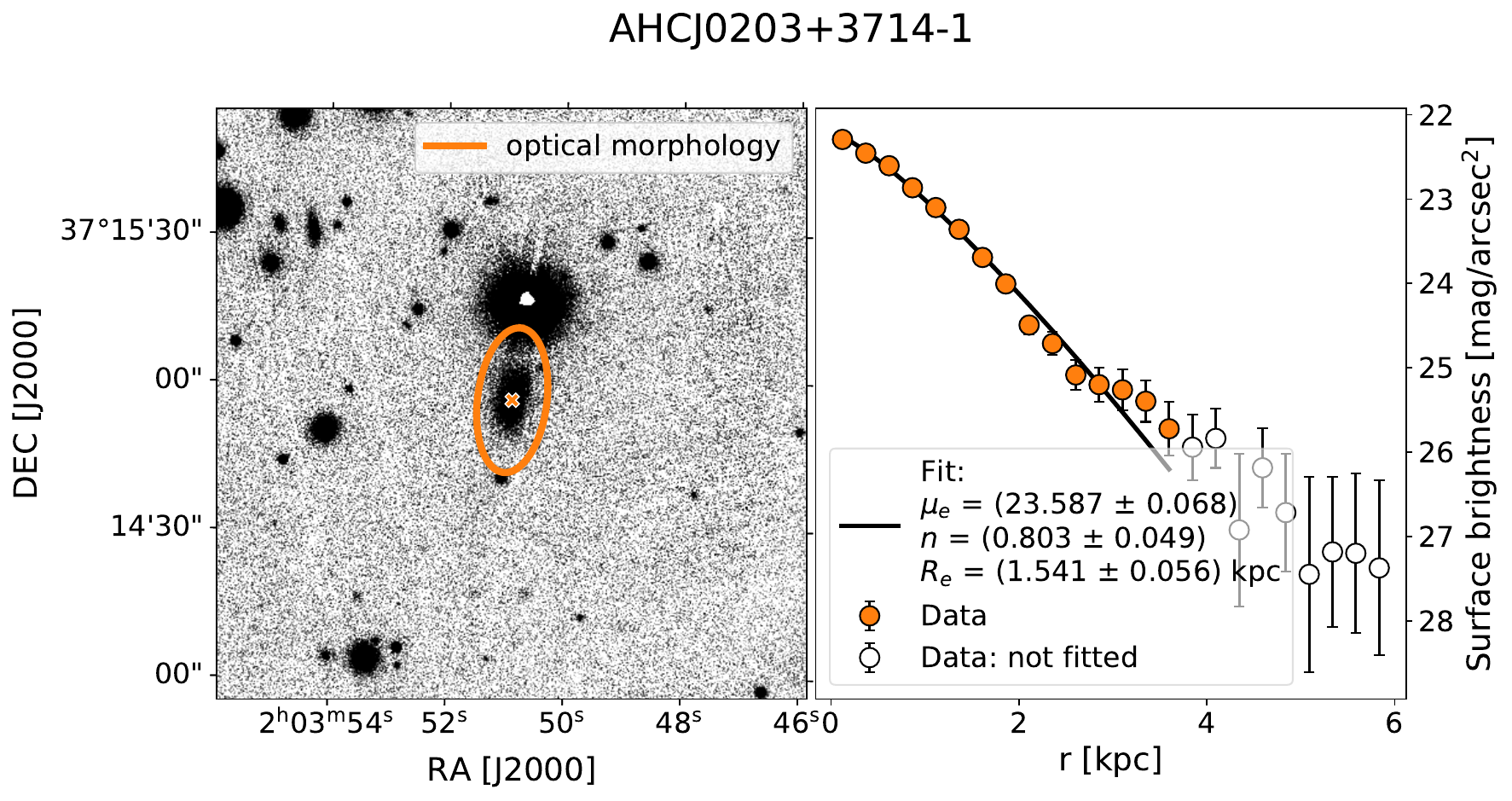}\\
        \includegraphics[width=0.89\linewidth]{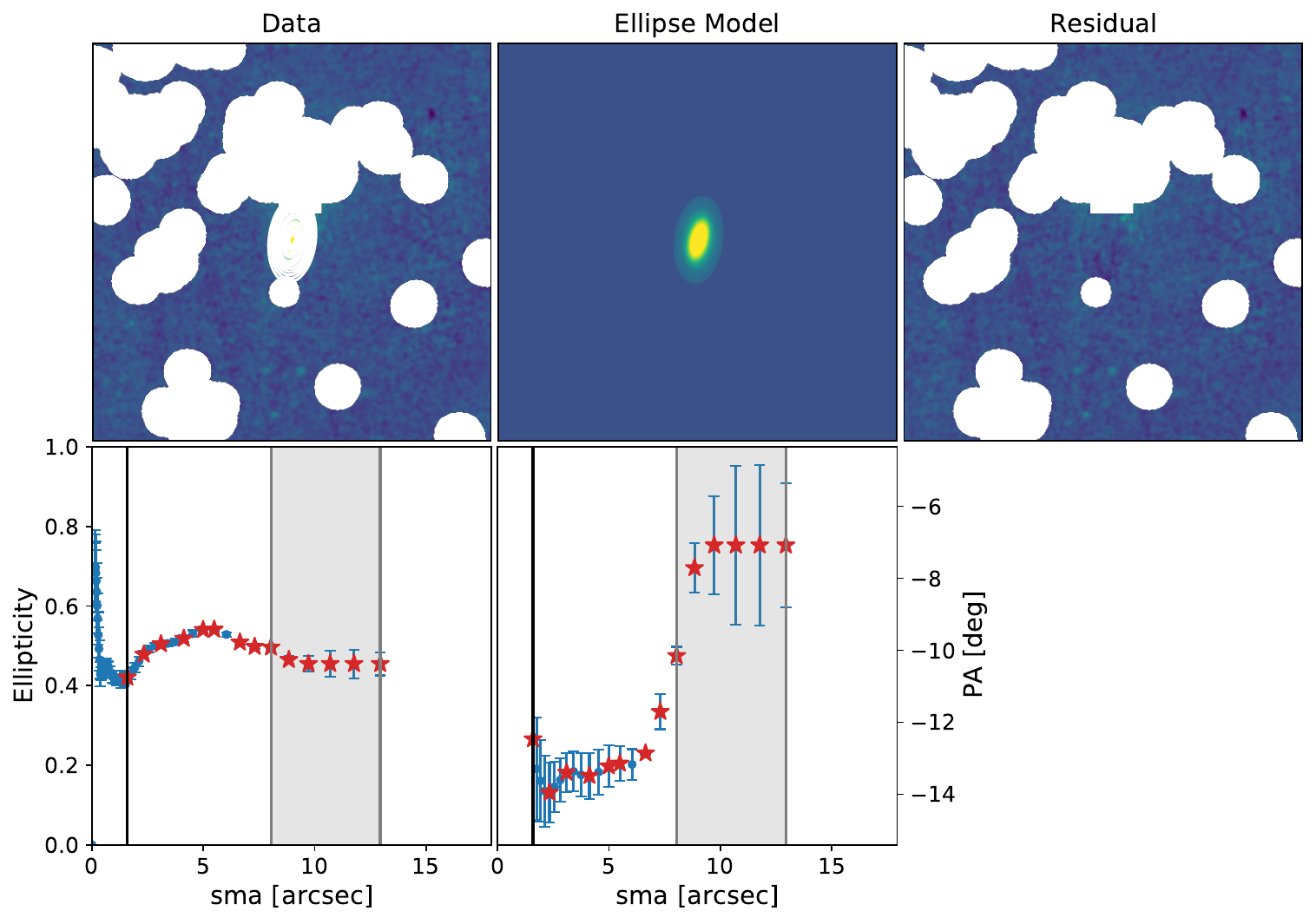}\\
        \caption{\paironeleft. First row is the same as in Fig. \ref{fig:pavel-optical}. Bottom two rows are the same as in Fig. \ref{fig:isophot_fit-Pavel}.}
        \label{fig:HI-op-pair_1-left}
    \end{figure*}

\begin{figure*}
    \centering
        \includegraphics[width = 0.7 \textwidth]{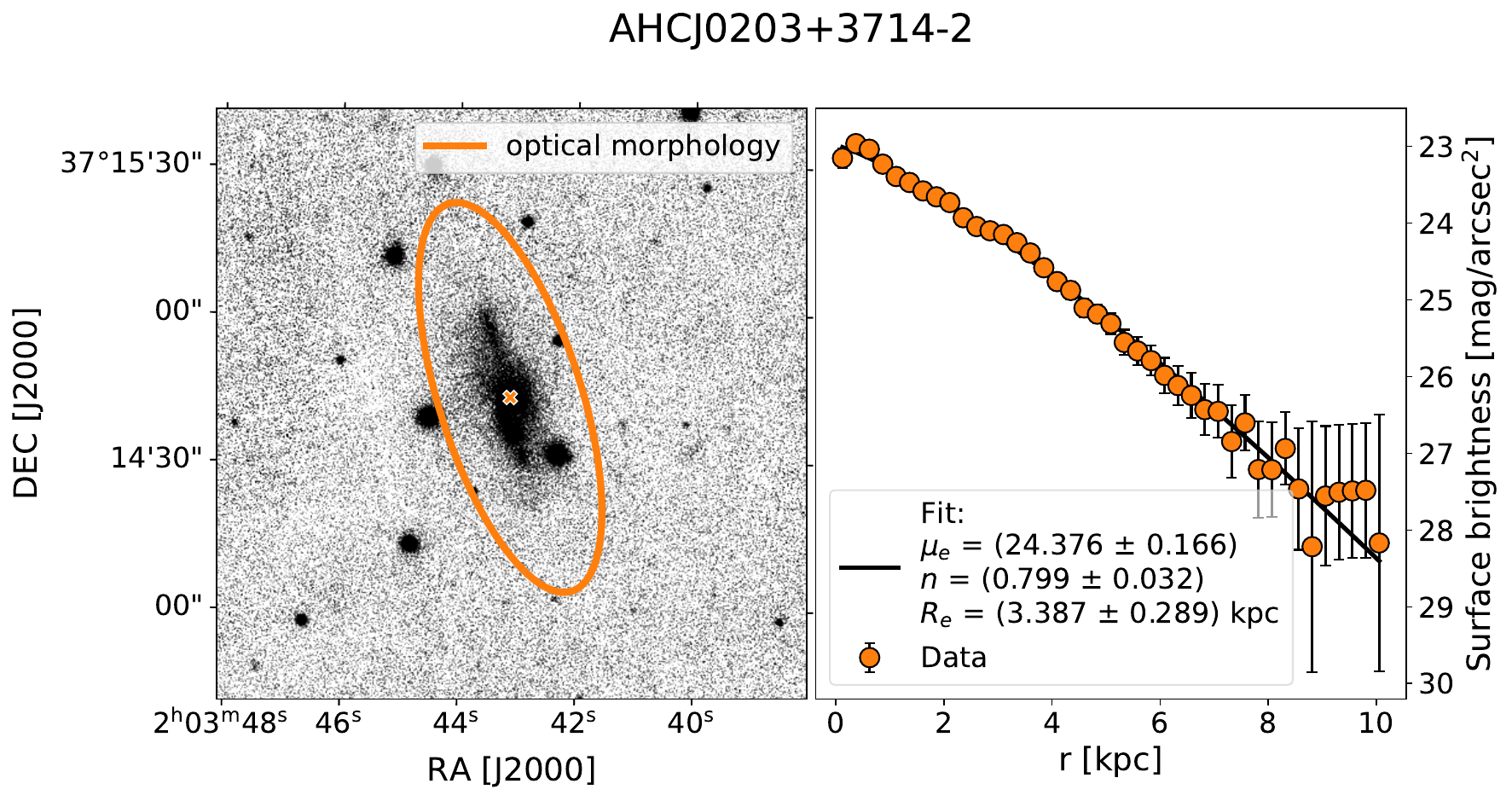}\\
        \includegraphics[width=0.89\linewidth]{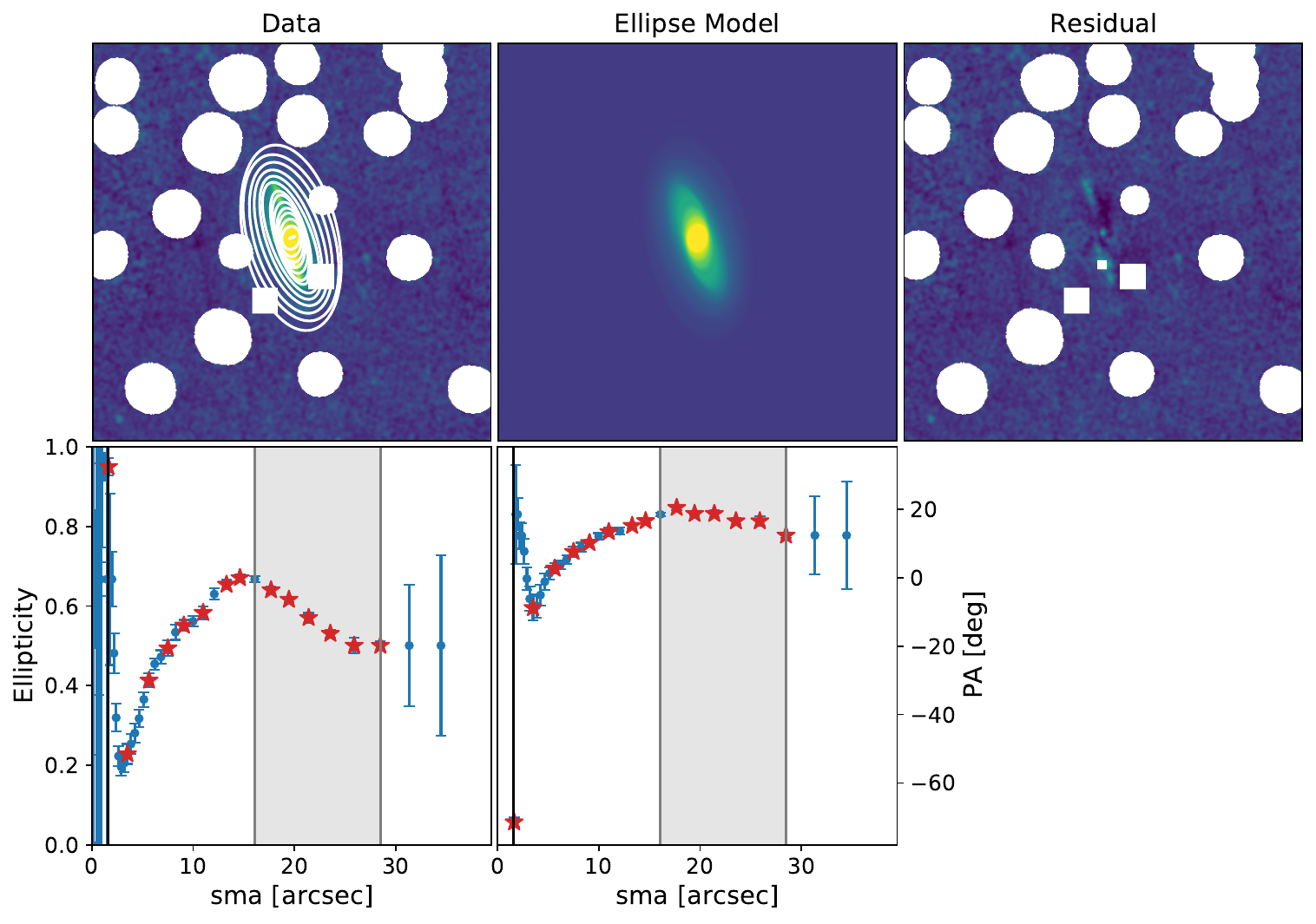}
        \caption{\paironeright. First row is the same as in Fig. \ref{fig:pavel-optical}. Bottom two rows are the same as in Fig. \ref{fig:isophot_fit-Pavel}.}
        \label{fig:HI-op-pair_1-right}
    \end{figure*}

    \textbf{\paironeleft\ and \paironeright}. These two galaxies are part of a pair shown on the uppermost panel of Fig. \ref{fig:kinematics_pairs} with \paironeleft\ being eastward of the two and \paironeright\ westward. The two galaxies are connected by a gas bridge as seen from the 3D cube, and consequently on the total intensity and velocity maps. The internal kinematics of both galaxies seem to be synchronised with their common motion as seen by a continuous velocity gradient in the velocity map. The stellar counterpart of \paironeleft, shown in better detail in Fig. \ref{fig:HI-op-pair_1-left}, is partly obscured by a nearby bright star which might have slightly influenced the isophotal fitting of the galaxy and the extracted surface brightness profile. 
    On the other hand, the stellar counterpart of \paironeright, shown in Fig. \ref{fig:HI-op-pair_1-right}, shows variation in geometric parameters with radius. However, the change is small in our region of interest and the resulting geometry seems to well represent the orientation of the disk when compared to the orientation of \hi\ (Fig. \ref{fig:kinematics_pairs}). 

\begin{figure*}
        \centering
        \includegraphics[width = 0.98 \textwidth]{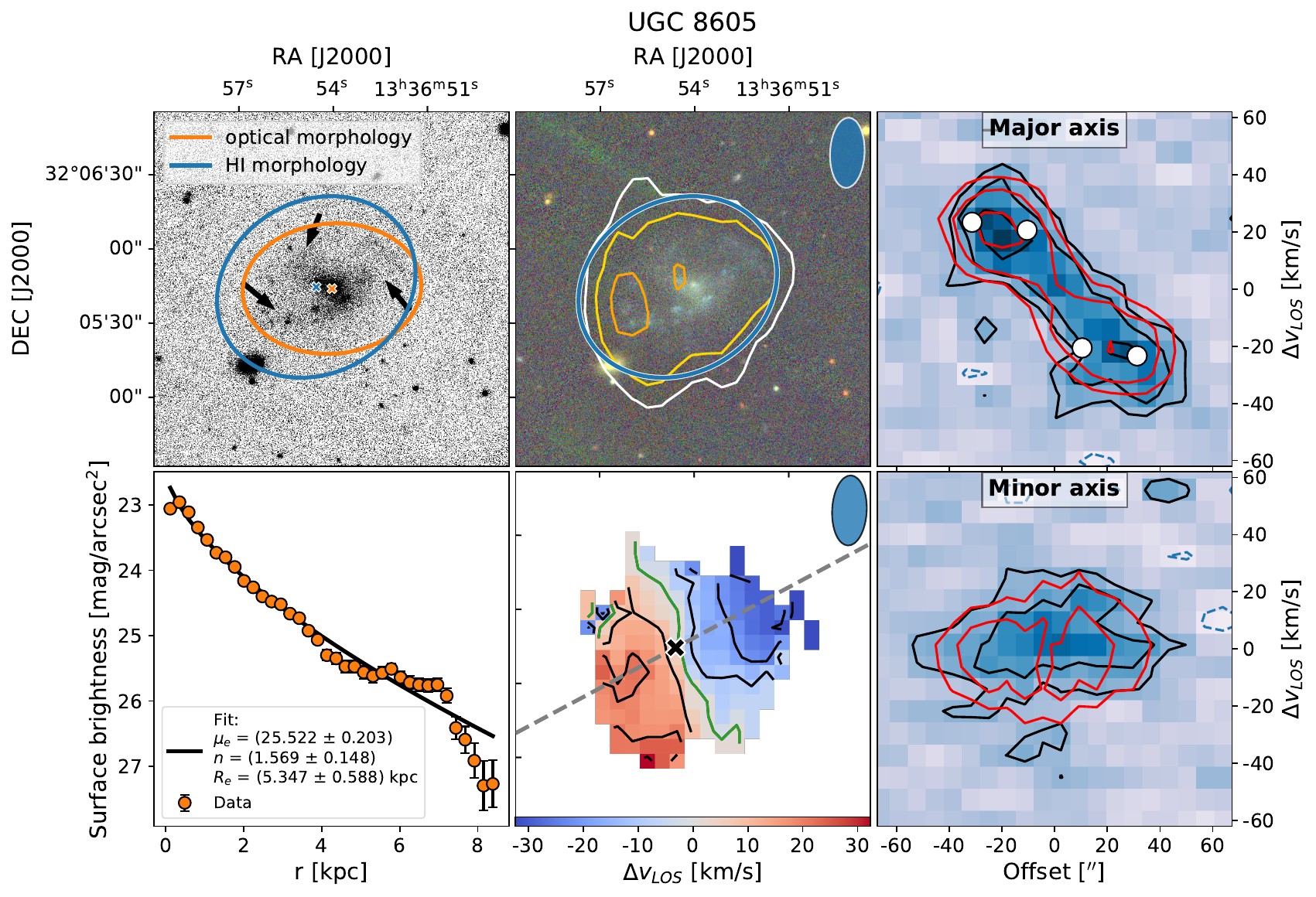}
        \includegraphics[width=0.87\linewidth]{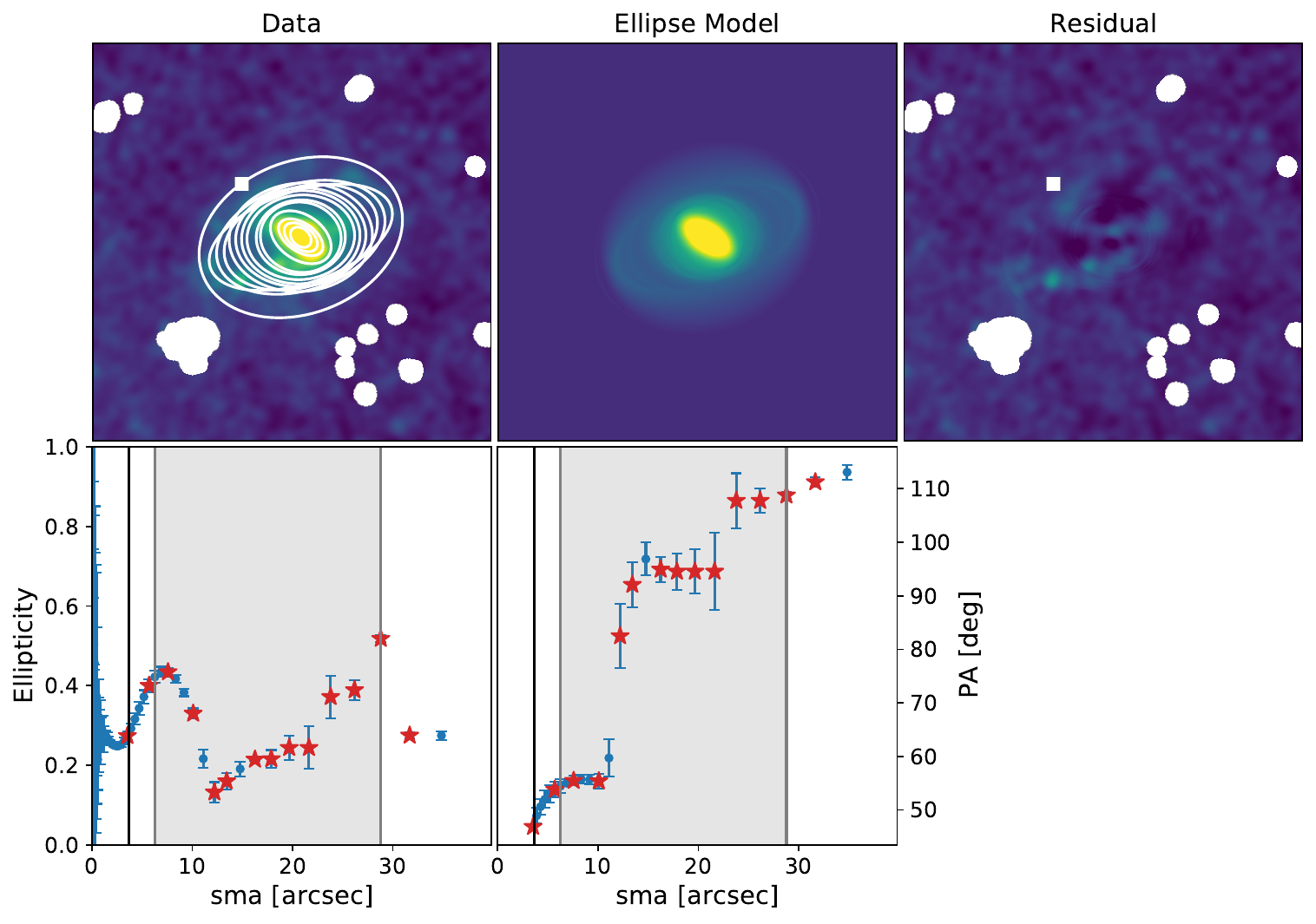}
        \caption{\pairtwoleft. First two rows are the same as Figs. \ref{fig:pavel-kinematics} (middle and right column) and \ref{fig:pavel-optical} (left column) combined. Black arrows in the upper left panel denote extended stellar structures in the outskirt of the galaxy (see the text). The \hi\ contours in the middle upper panel correspond to column densities starting from $1.9 \times 10^{20}$ cm$^{-2}$ (in white) and growing by a factor of 2 in intensity towards contours in redder colors. Bottom two rows are the same as in Fig. \ref{fig:isophot_fit-Pavel}.}
        \label{fig:HI-op-pair_2-left}
    \end{figure*}

\begin{figure*}
    \centering
        \includegraphics[width = 0.98 \textwidth]{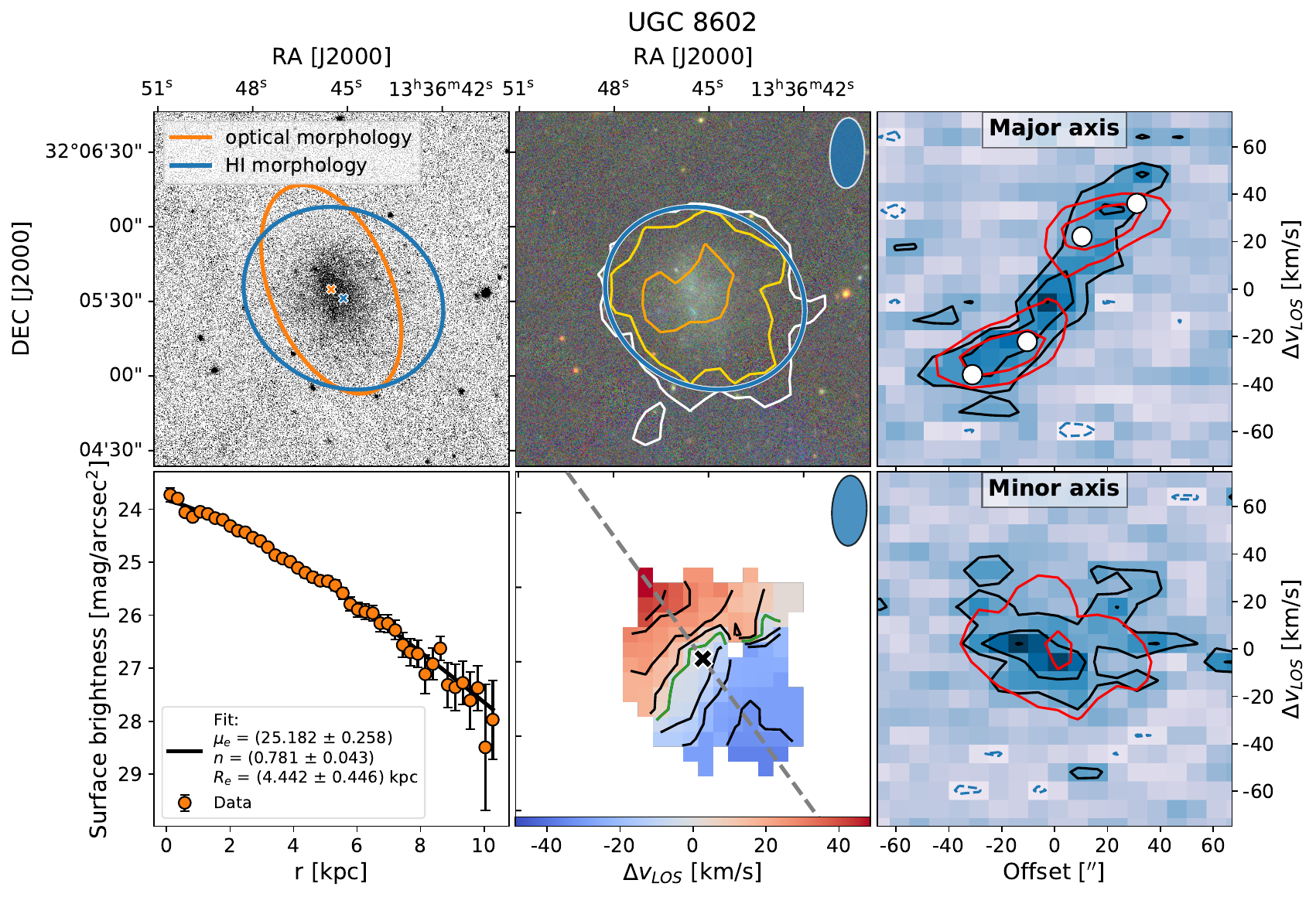}
        \includegraphics[width=0.87\linewidth]{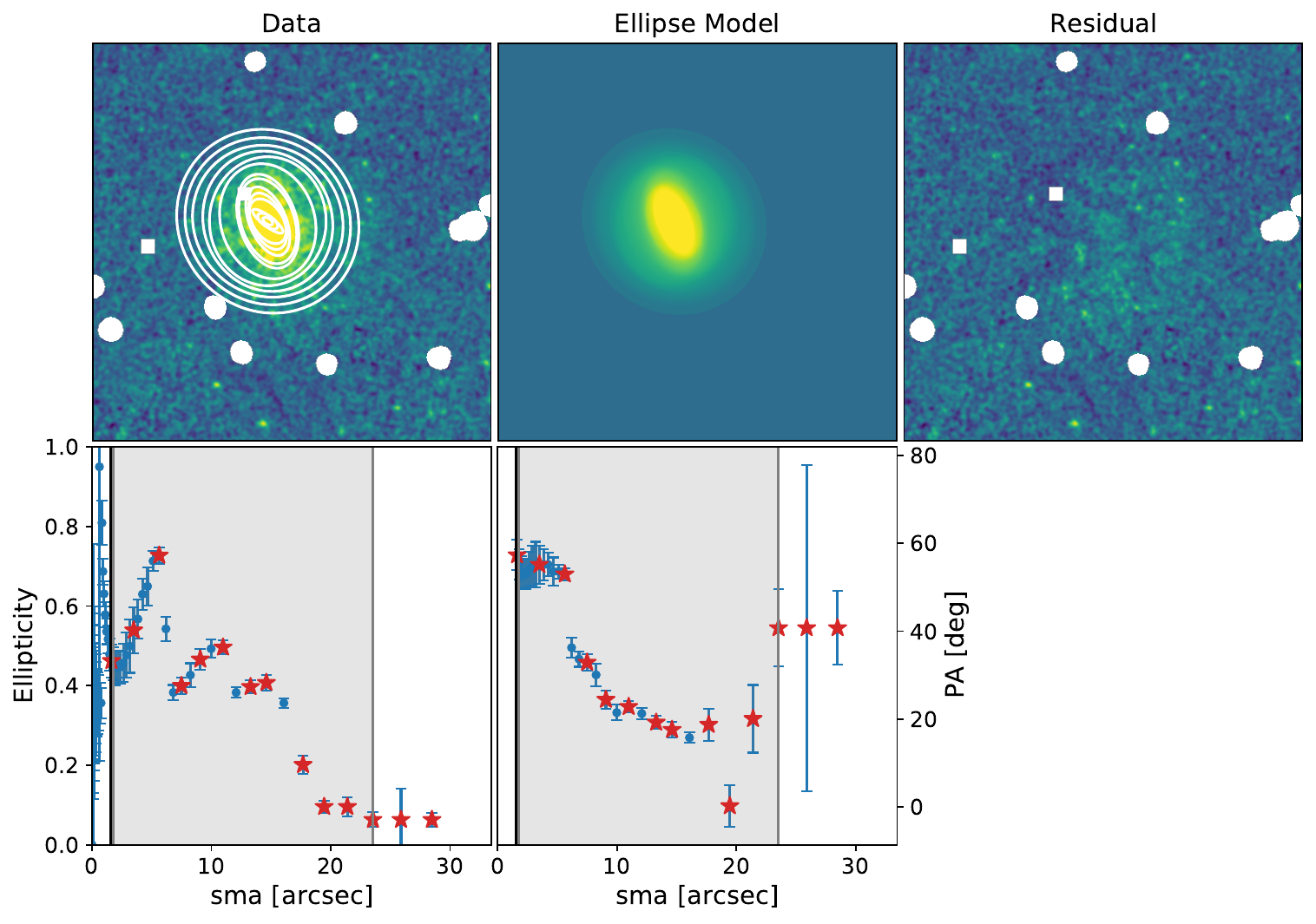}
        \caption{\pairtworight. First two rows are the same as Figs. \ref{fig:pavel-kinematics} (middle and right column) and \ref{fig:pavel-optical} (left column) combined. The \hi\ contours in the middle upper panel correspond to column densities starting from $1.2 \times 10^{20}$ cm$^{-2}$ (in white) and growing by a factor of 2 in intensity towards contours in redder colors. Bottom two rows are the same as in Fig. \ref{fig:isophot_fit-Pavel}.}
        \label{fig:HI-op-pair_2-right}
    \end{figure*}

    \textbf{\pairtwoleft\ and \pairtworight.} These two galaxies also form a pair as is shown in the middle panel of Fig. \ref{fig:kinematics_pairs} with \pairtwoleft\ being the eastward and \pairtworight\ the westward galaxy. These galaxies do not show a detectable gas bridge and are rotating in opposite directions as seen from the velocity map. For this pair, we were able to produce independent kinematic models. The obtained \hi\ geometric parameters of \pairtwoleft, seen in the first two panels of Fig. \ref{fig:HI-op-pair_2-left}, seems to well describe the \hi\ morphology. However, CANNUBI was not able to fully constrain the inclination of the disk showing consistencies down to the inclination of 0\dg. The kinematics of the galaxy seems regular and the kinematic model well reproduces the data. The stellar counterpart shows complex morphology with the three apparent stellar arms going out of the inner disk towards the southeast, northwest and northeast. This features influenced the isophotal fitting as seen from the large change in the position angle and the increase of ellipticity with radius. As in the case of \pavel, the fitted isophotes seem to align more with the \hi\ geometry as they extend further to the ourskirts of the galaxy. The extracted surface brightness profile could not be robustly fitted with a Sérsic function due its irregular shape.
    In the case of \pairtworight, shown in Fig. \ref{fig:HI-op-pair_2-right}, the \hi\ geometric parameters could also not be well constrained, showing consistency with inclinations down to 0\dg and thereby making the morphological position angle less constrained. The galaxy also seems to be kinematically lopsided as seen from its asymmetric PV diagram along the major axis. This means that the derived rotational velocity might not be a reliable tracer of the total underlying potential. 
    Its stellar counterpart has irregular morphology seen by the large change in ellipticity and position angle with radius.

\begin{figure*}
        \centering
        \includegraphics[width = 0.7 \textwidth]{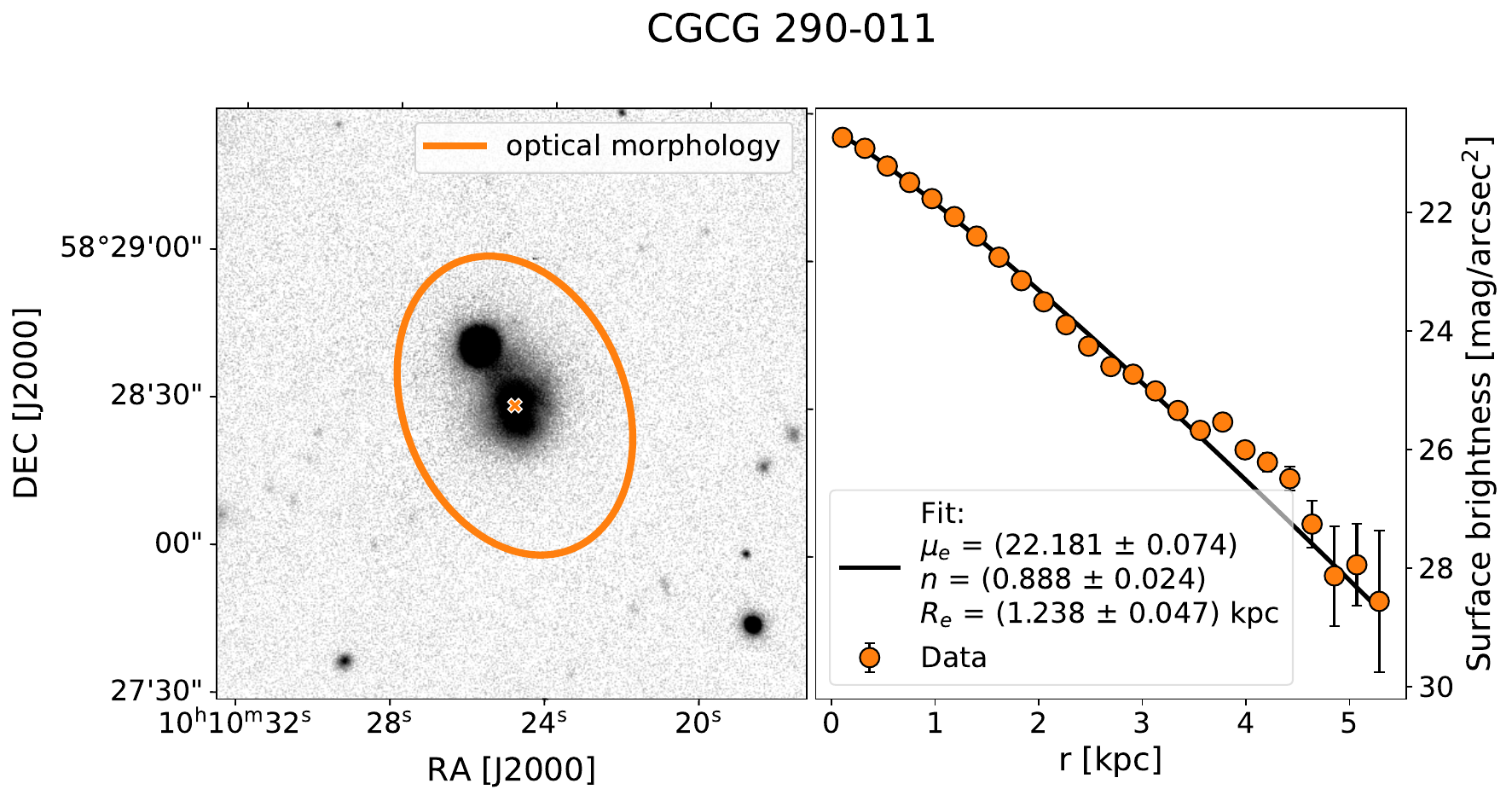}
        \includegraphics[width=0.89\linewidth]{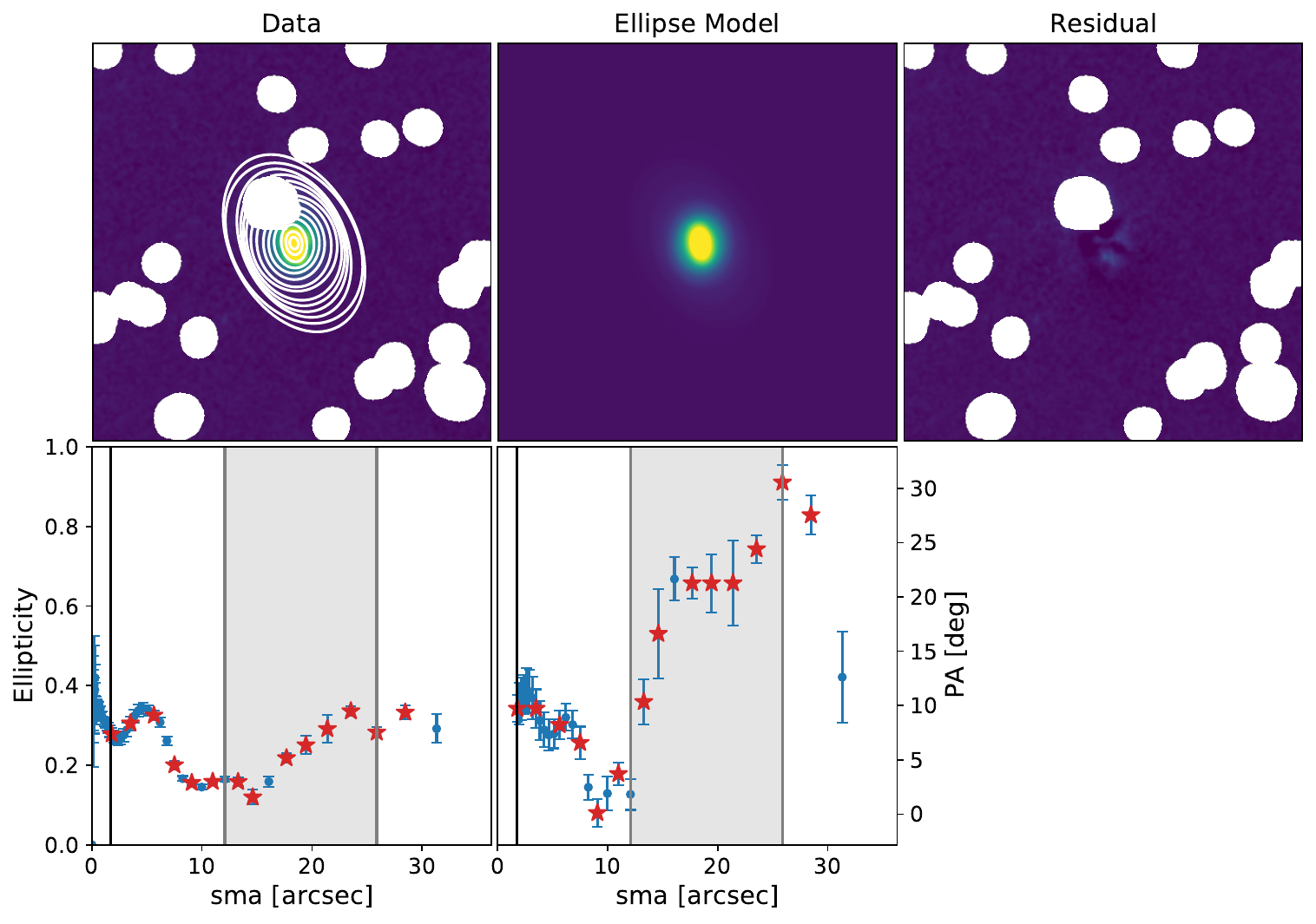}
        \caption{\pairthrleft. First row is the same as in Fig. \ref{fig:pavel-optical}. Bottom two rows are the same as in Fig. \ref{fig:isophot_fit-Pavel}.}
        \label{fig:HI-op-pair_3-left}
    \end{figure*}

\begin{figure*}
        \centering
        \includegraphics[width = 0.7 \textwidth]{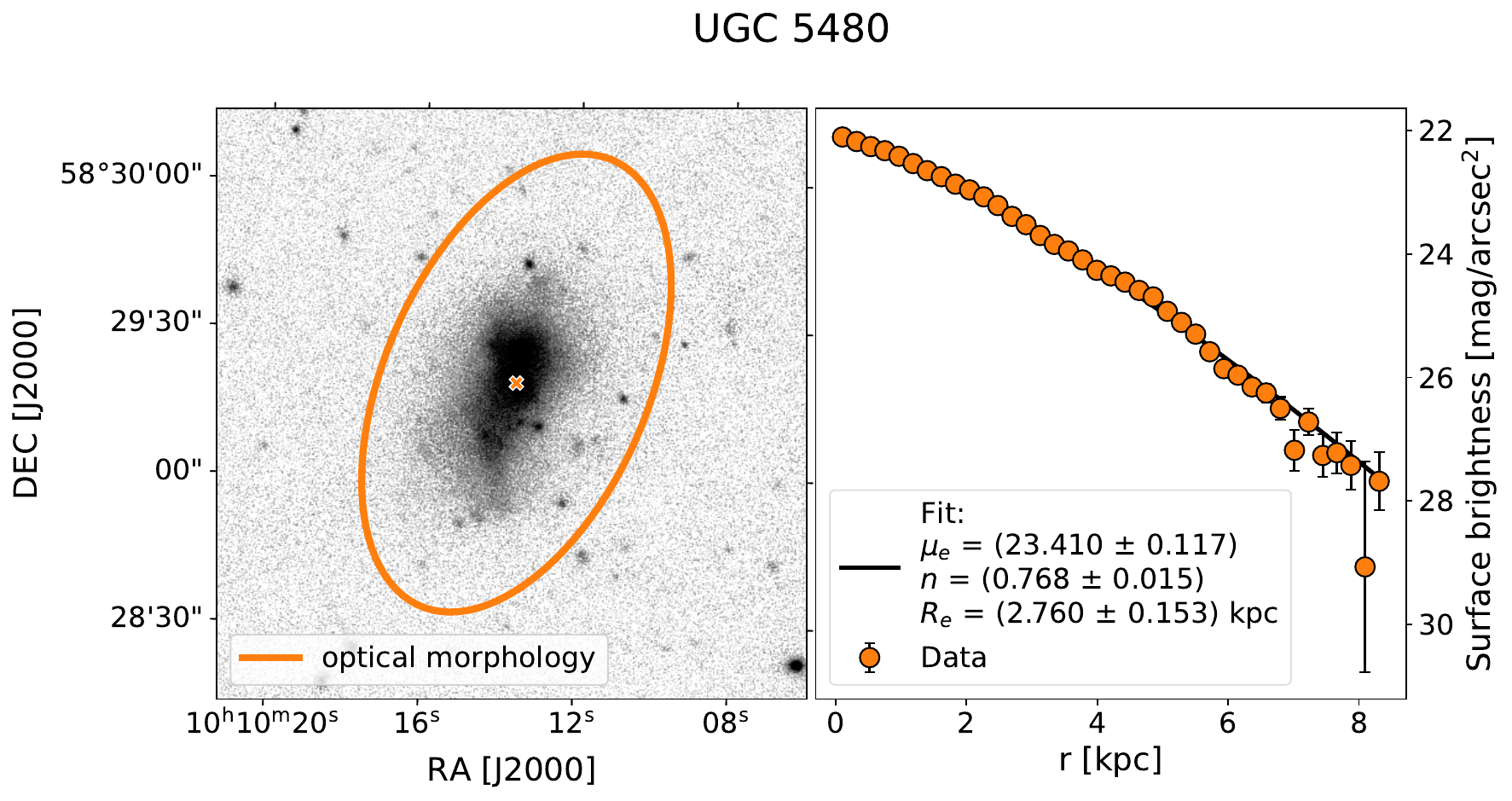}
        \includegraphics[width=0.89\linewidth]{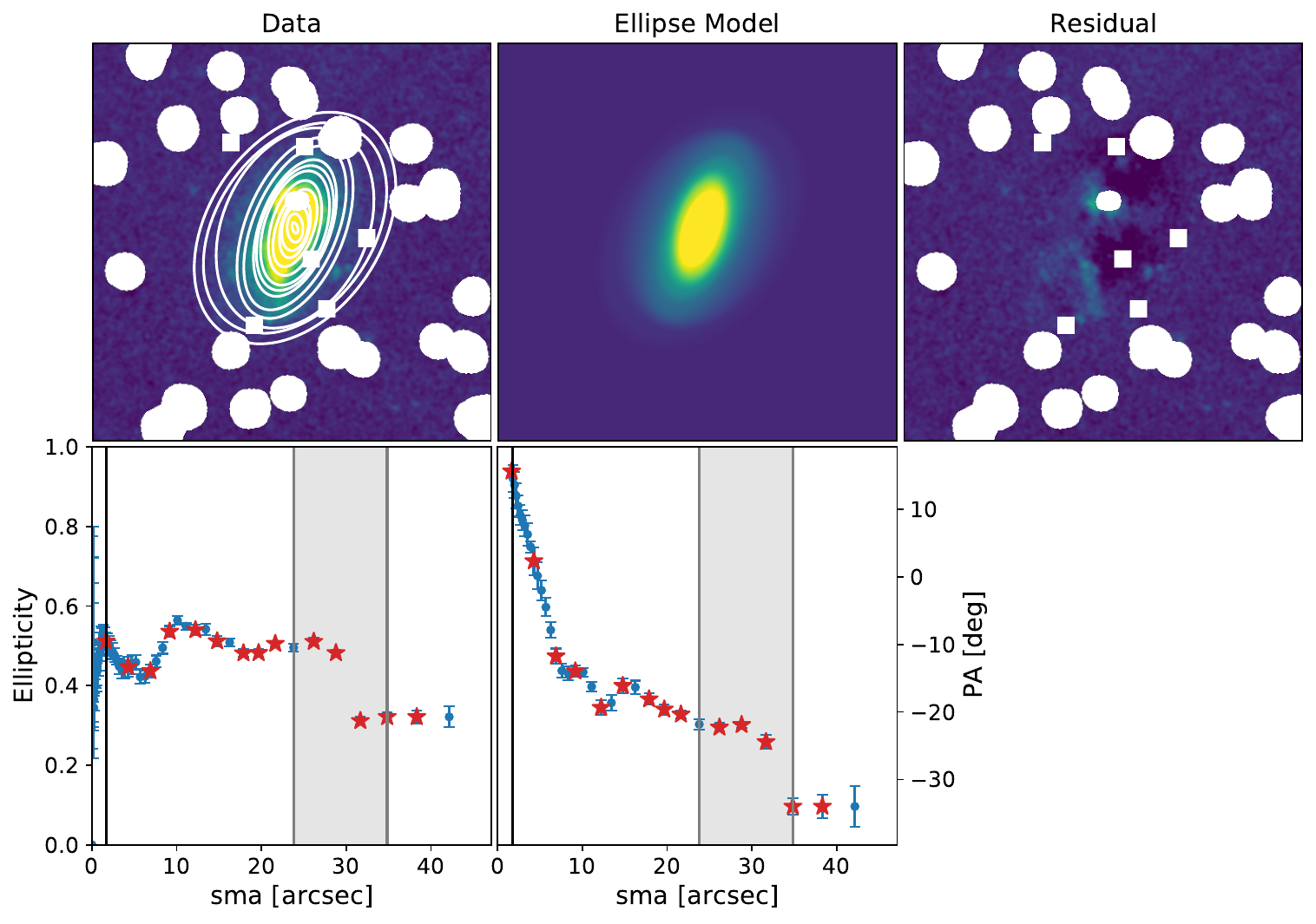}
        \caption{\pairthrright. First row is the same as in Fig. \ref{fig:pavel-optical}. Bottom two rows are the same as in Fig. \ref{fig:isophot_fit-Pavel}.}
        \label{fig:HI-op-pair_3-right}
    \end{figure*}

    \textbf{\pairthrleft\ and \pairthrright.} These galaxies make the third pair in our sample and are shown in the lowermost panel on Fig. \ref{fig:kinematics_pairs} with \pairthrleft\ being the eastward and \pairthrright\ the westward galaxy. The two galaxies are not directly connected in the 3D cube, due to a separation along the spectral axis. \pairthrright\ is the larger of the two galaxies and shows a clear velocity gradient in the velocity map. The stellar counterpart of \pairthrleft, shown in Fig. \ref{fig:HI-op-pair_3-left}, is close to a nearby star which might have slightly influenced the isophotal fitting and the extracted surface brightness profile. 
    The stellar counterpart of \pairthrright, shown in Fig. \ref{fig:HI-op-pair_3-right}, shows a drop in ellipticity in the outskirts, but an almost constant ellipticity in the rest of the galaxy.

\begin{figure*}
        \centering
        \includegraphics[width = 0.67 \textwidth]{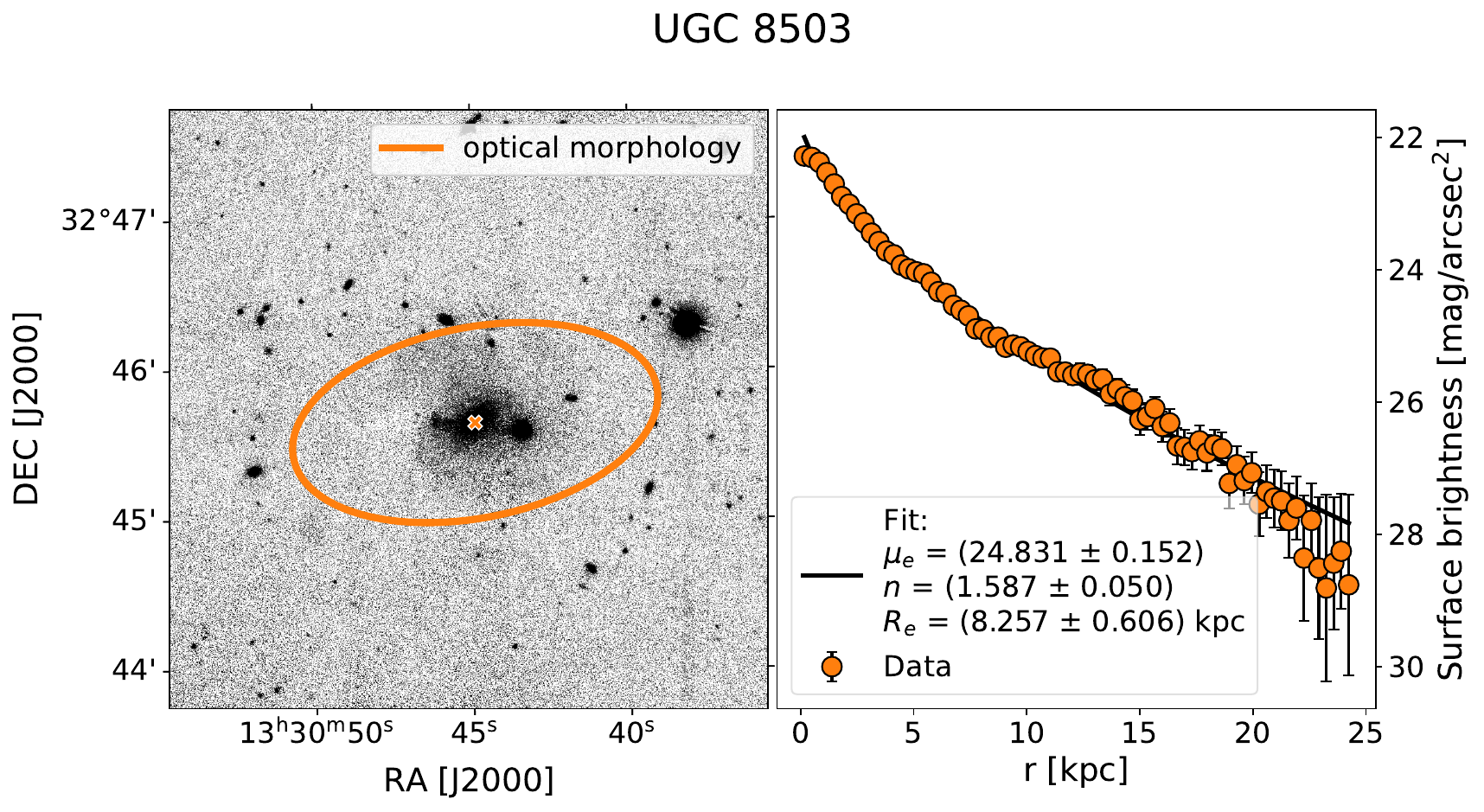}
        \includegraphics[width = 0.67 \textwidth]{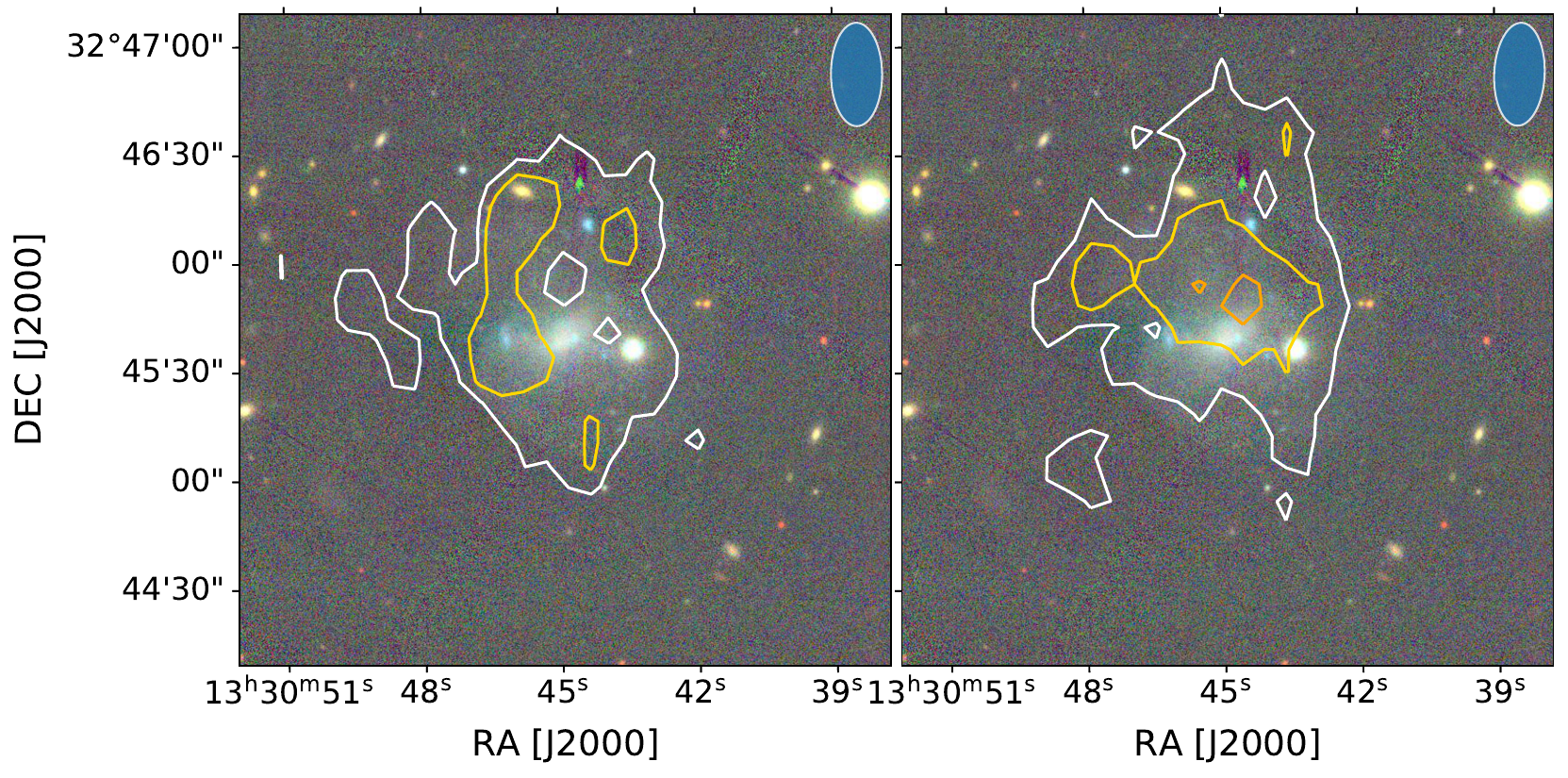}

        \includegraphics[width=0.86\linewidth]{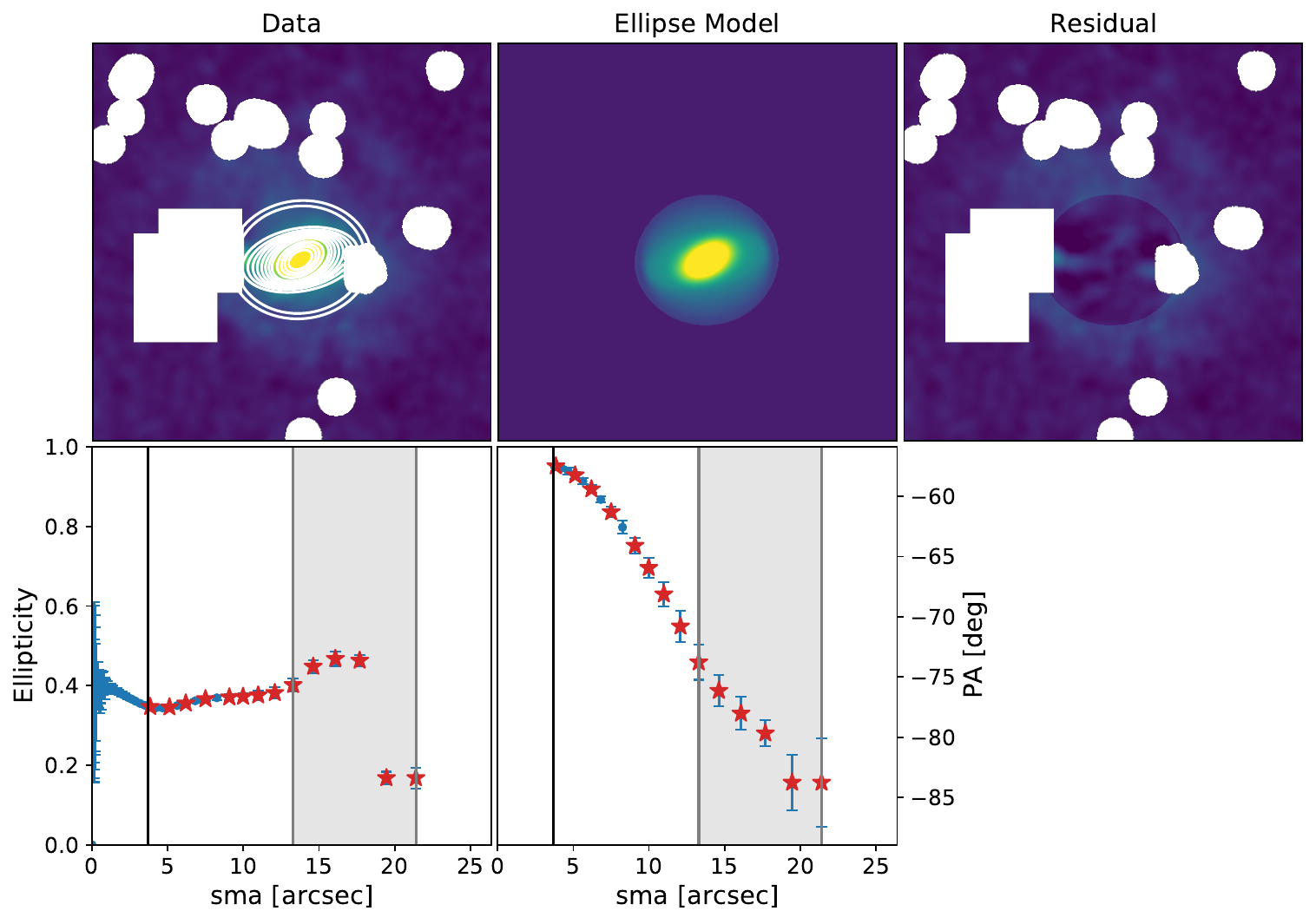}
        \caption{\bSNRone. First row is the same as in Fig. \ref{fig:pavel-optical}. The second row shows moment 0 of the two Apertif detections in two different fields (see Sect. \ref{sec:results}). The \hi\ contours correspond to column densities starting from $2.2 \times 10^{20}$ cm$^{-2}$ (in white) and growing by a factor of 2 in intensity towards contours in redder colors. Bottom two rows are the same as in Fig. \ref{fig:isophot_fit-Pavel}. }
        \label{fig:HI-op-bSNR_1}
    \end{figure*}

    \textbf{\bSNRone.} The galaxy is shown in Fig. \ref{fig:HI-op-bSNR_1}. This galaxy is positioned at the edges of two Apertif observed fields. As the fields are not being mosaicked together at the present moment, the galaxy was detected two times in the new source finding and was assigned two names. One of them reported in Table \ref{tab:general_properties}, and the second being AHC133045.2+324548. Due to being at the edge of the primary beam, the \hi\ data quality is bad and unreliable for this galaxy. Additionally, in the $i-$band of PS1, there is a drop in background counts eastward from the galaxy which was masked for the optical analysis. The stellar counterpart shows a slight twist in position angle, but an almost constant ellipticity with radius. There is an indication of a faint emission beyond the fitted ellipses which would probably result in a decrease in ellipticity at larger radii. 

\begin{figure*}
    \centering
        \includegraphics[width = 0.98 \textwidth]{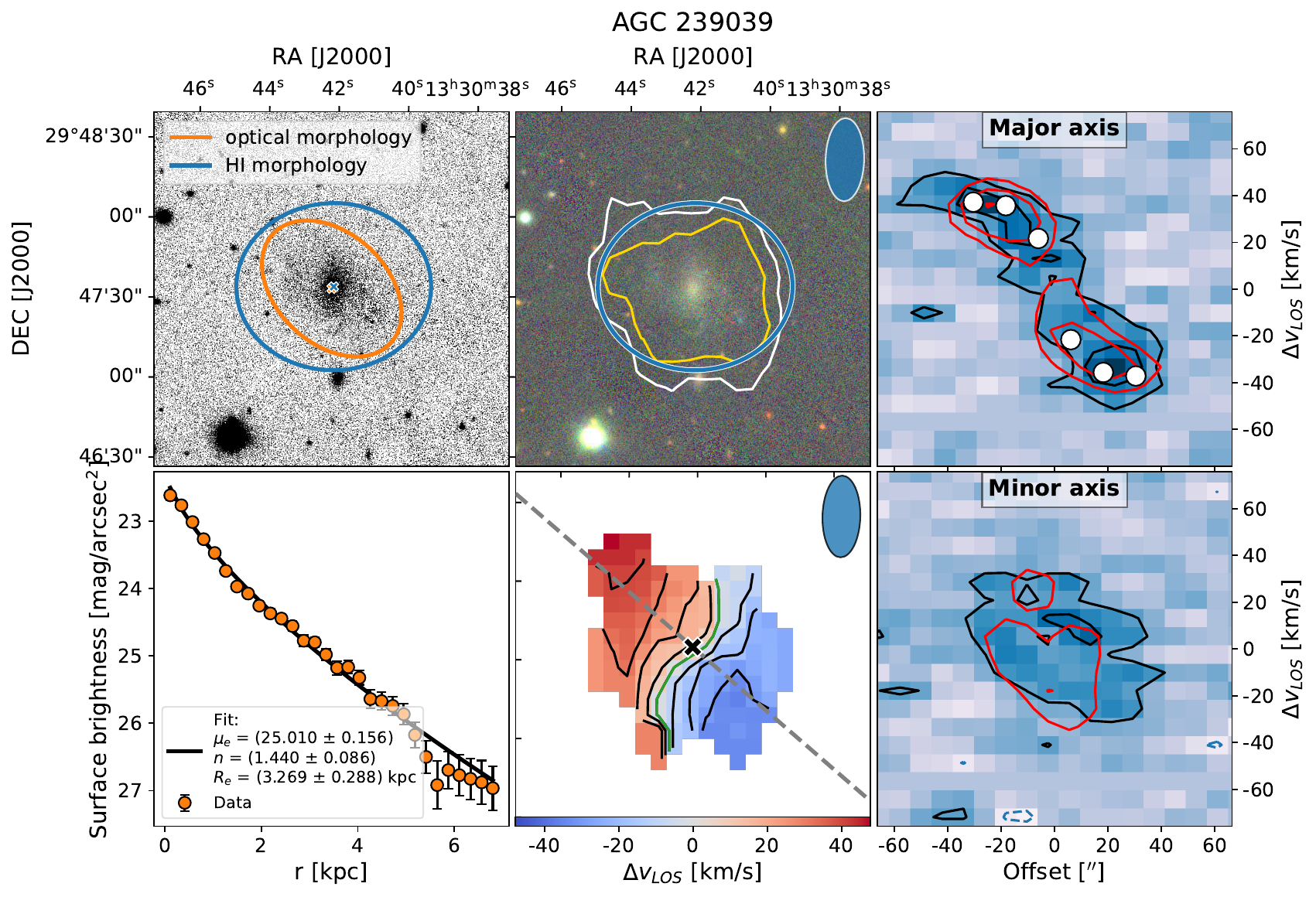}
        \includegraphics[width=0.87\linewidth]{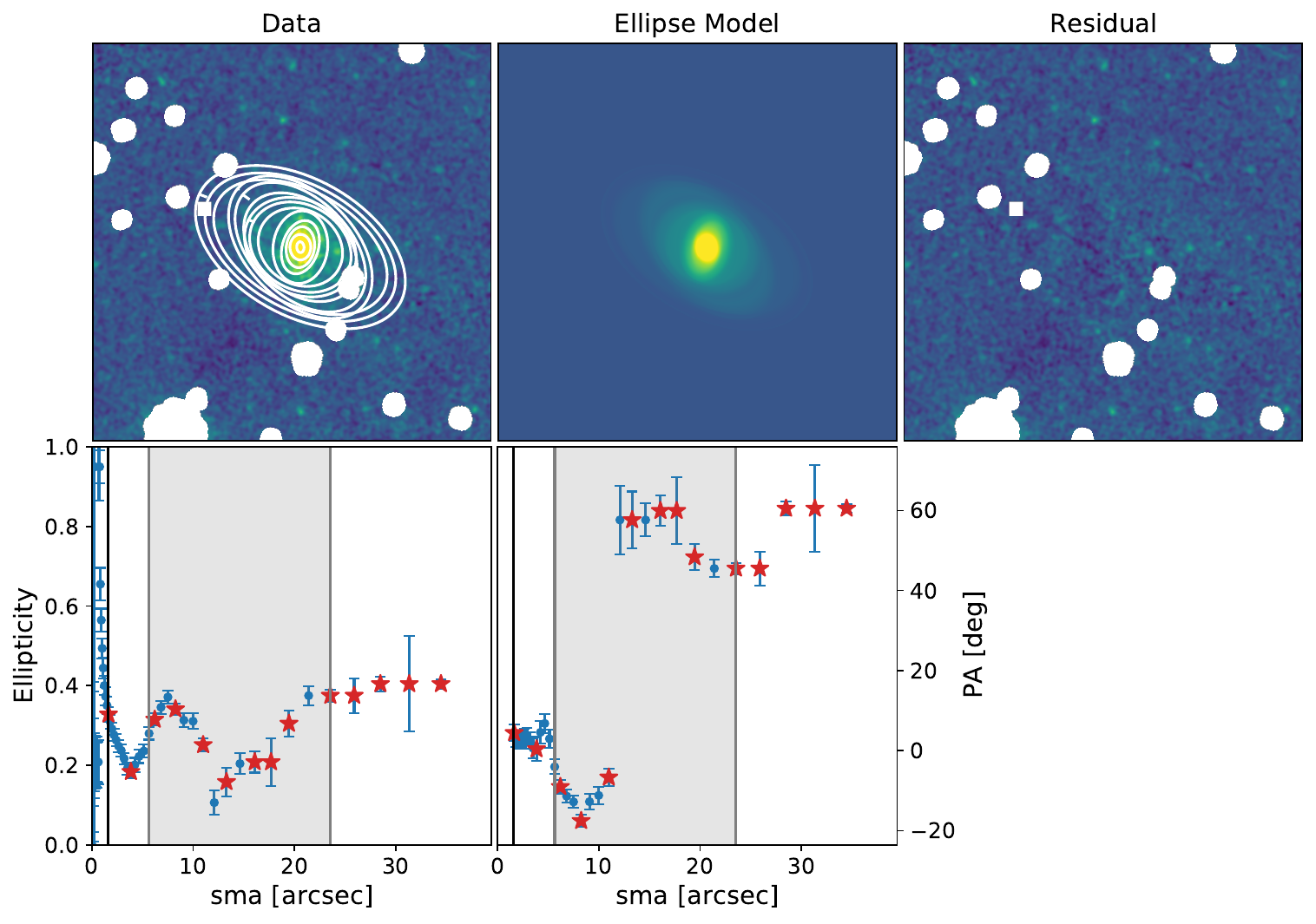}
        \caption{\svel. First two rows are the same as Figs. \ref{fig:pavel-kinematics} (middle and right column) and \ref{fig:pavel-optical} (left column) combined. The \hi\ contours in the middle upper panel correspond to column densities starting from $1.9 \times 10^{20}$ cm$^{-2}$ (in white) and growing by a factor of 2 in intensity towards contours in redder colors. Bottom two rows are the same as in Fig. \ref{fig:isophot_fit-Pavel}.}
        \label{fig:HI-op-svel}
    \end{figure*}

    \textbf{\svel.} The galaxy is shown in Fig. \ref{fig:HI-op-svel}. The kinematics shows a twist along the minor axis (as seen from its velocity field and from the velocity gradient in the PV slice along the minor axis) and an asymmetric PV slice along the major axis. The twist along the minor axis could be a sign of non-negligible radial motions. The asymmetry in the PV slice is caused by higher velocity dispersion in the approaching side of the galaxy indicating higher disturbance, but with no effect on the underlying rotation. There is also a large offset between the morphological and kinematic position angles of the \hi\ disk. In this case, CANNUBI was unable to constrain the inclination, showing consistency with inclinations down to 0\dg. Stellar morphology shows a twist in position angle as a function of radius.
    The inner region is almost perpendicular to the northeast-southwest orientation of the outskirts. However, the kinematic position angle from \hi\ is also oriented in the northeast-southwest direction indicating the orientation of the disk, which is why we adopt optical geometric parameters for the kinematic model of this galaxy.

\begin{figure*}
        \centering
        \includegraphics[width = 0.98 \textwidth]{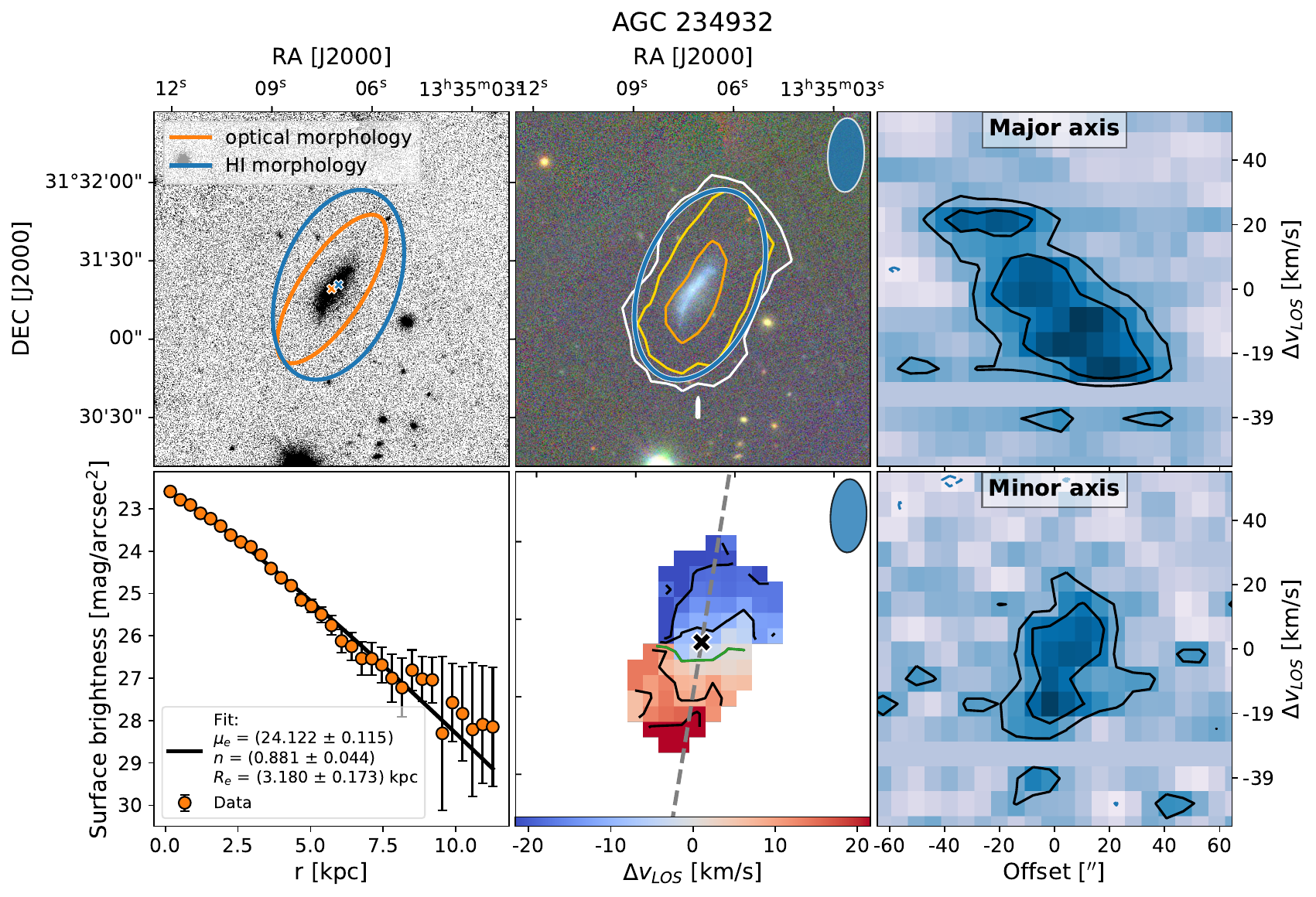}
        \includegraphics[width=0.87\linewidth]{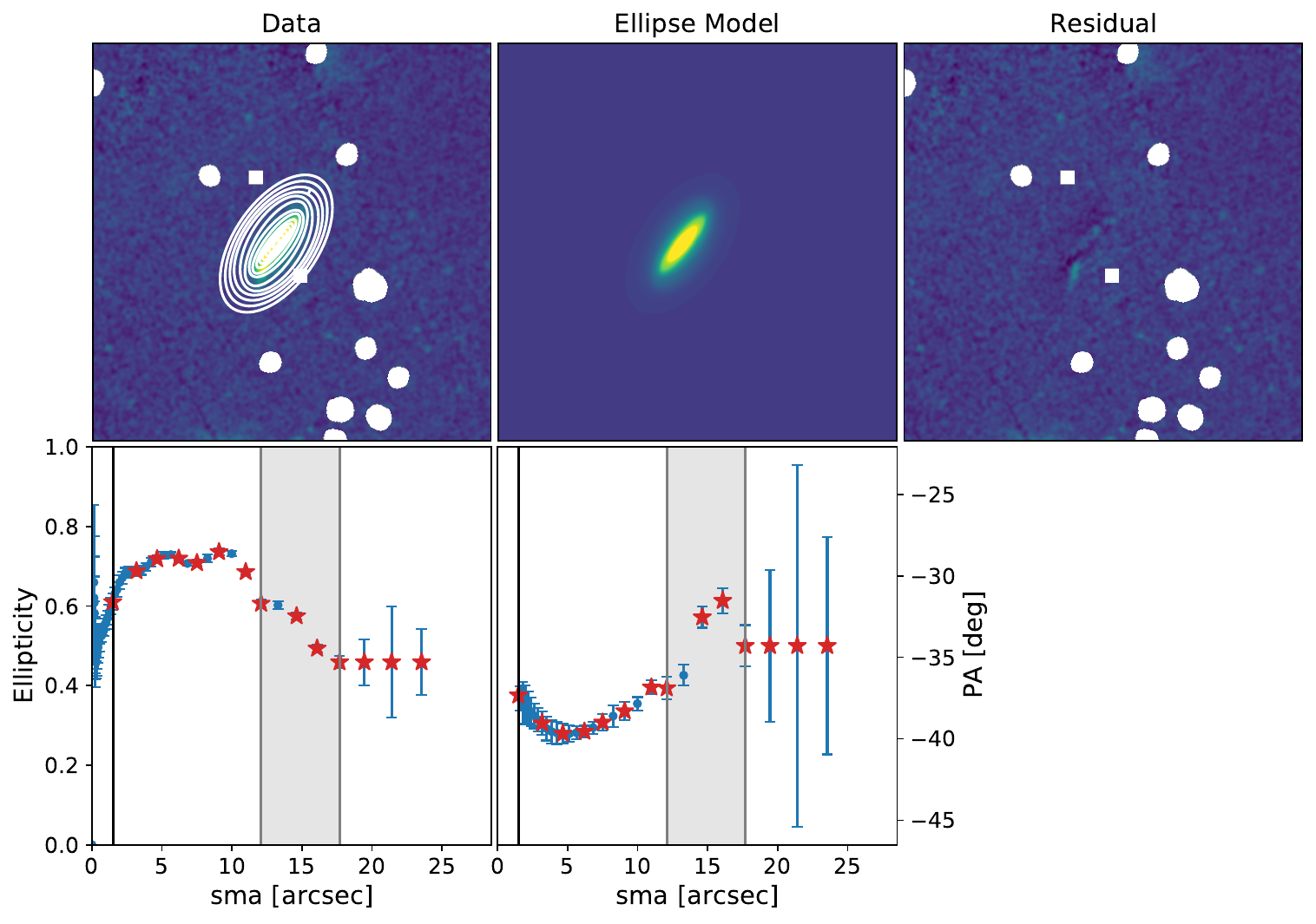}        
        \caption{\lsmerge. First two rows are the same as Figs. \ref{fig:pavel-kinematics} (middle and right column) and \ref{fig:pavel-optical} (left column) combined. The \hi\ contours in the middle upper panel correspond to column densities starting from $1.5 \times 10^{20}$ cm$^{-2}$ (in white) and growing by a factor of 2 in intensity towards contours in redder colors. Bottom two rows are the same as in Fig. \ref{fig:isophot_fit-Pavel}.}             
        \label{fig:HI-op-lsmerge}
    \end{figure*}

    \textbf{\lsmerge.} As shown in Fig. \ref{fig:HI-op-lsmerge}, the galaxy shows signs of disturbance in the disk seen from the asymmetry in PV slices in the form of a seemingly separate peak in emission in the receding part. Interestingly, the stellar disk displays a peculiar curvature in the same part of the disk. These features indicate that the galaxy might be a merger in a late stage. Even with these features, the galaxy seems morphologically regular and geometry was successfully constrained in both the optical and \hi. In the inner region of the optical counterpart, the ellipticity goes up to 0.7, which would be equal to the inclination of 90\dg\ for $q_0=0.3$. There also seems to be a broad \hi\ emission region in the PV slice along the major axis, which could be another signature of edge-on galaxies, but we note that the galaxy is not spatially well enough resolved to distinguish this from the effect of the beam.

\begin{figure*}
        \centering
        \includegraphics[width = 0.98 \textwidth]{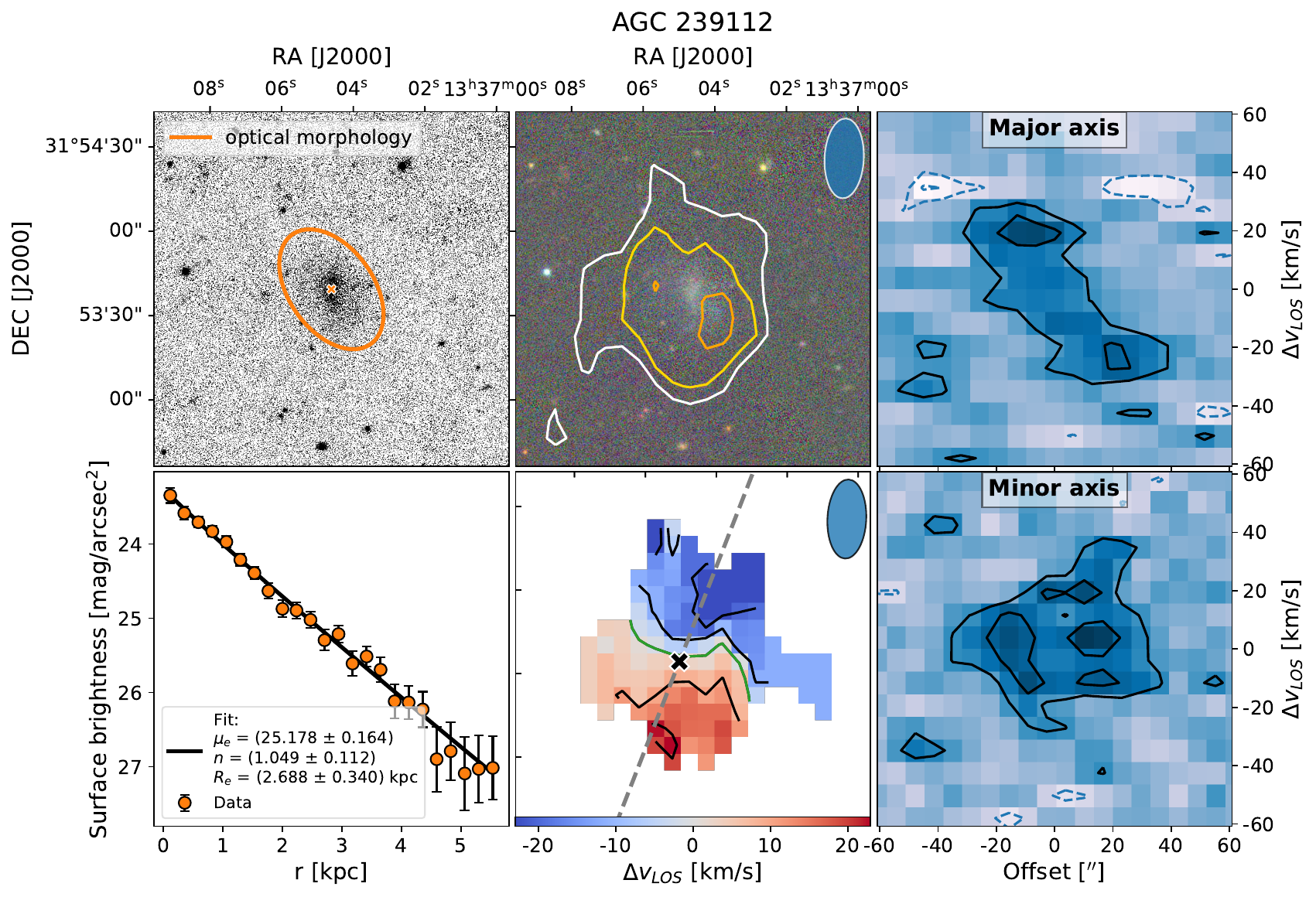}
        \includegraphics[width=0.87\linewidth]{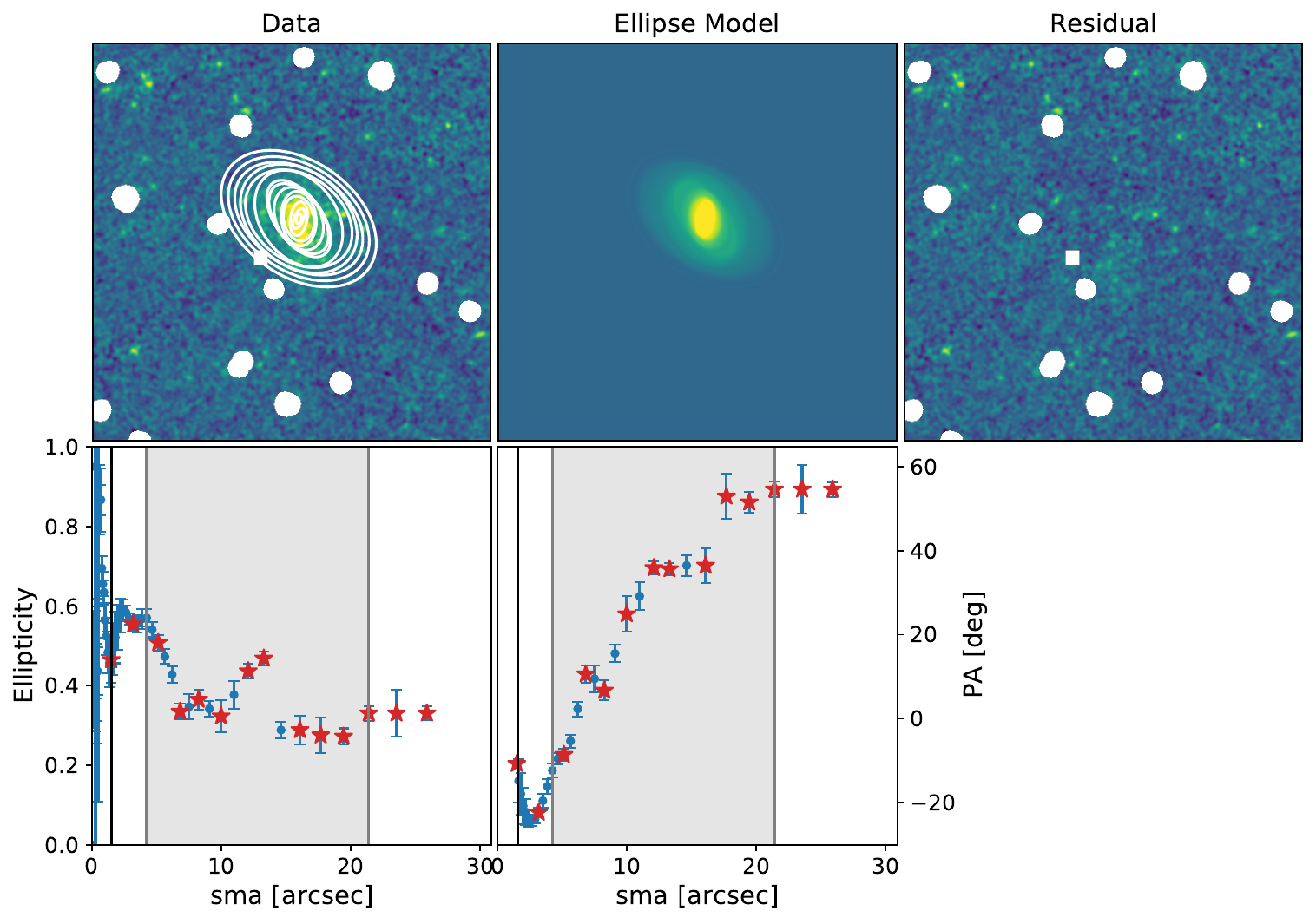}
        \caption{\fon. First two rows are the same as Figs. \ref{fig:pavel-kinematics} (middle and right column) and \ref{fig:pavel-optical} (left column) combined. The \hi\ contours in the middle upper panel correspond to column densities starting from $1.4 \times 10^{20}$ cm$^{-2}$ (in white) and growing by a factor of 2 in intensity towards contours in redder colors. Bottom two rows are the same as in Fig. \ref{fig:isophot_fit-Pavel}.}        \label{fig:HI-op-fon}
    \end{figure*}

    \textbf{\fon.} The galaxy is shown in Fig. \ref{fig:HI-op-fon}. The \hi\ total intensity map shows a very irregular morphology so the geometry of the \hi\ disk could not be constrained. The twist along the minor axis seen in the velocity field could be a sign of non-negligible radial motions. The stellar counterpart shows a twist in the position angle with radius, but this time towards higher misalignment with the kinematics of the \hi. 

\begin{figure*}
        \centering
        \includegraphics[width = 0.98 \textwidth]{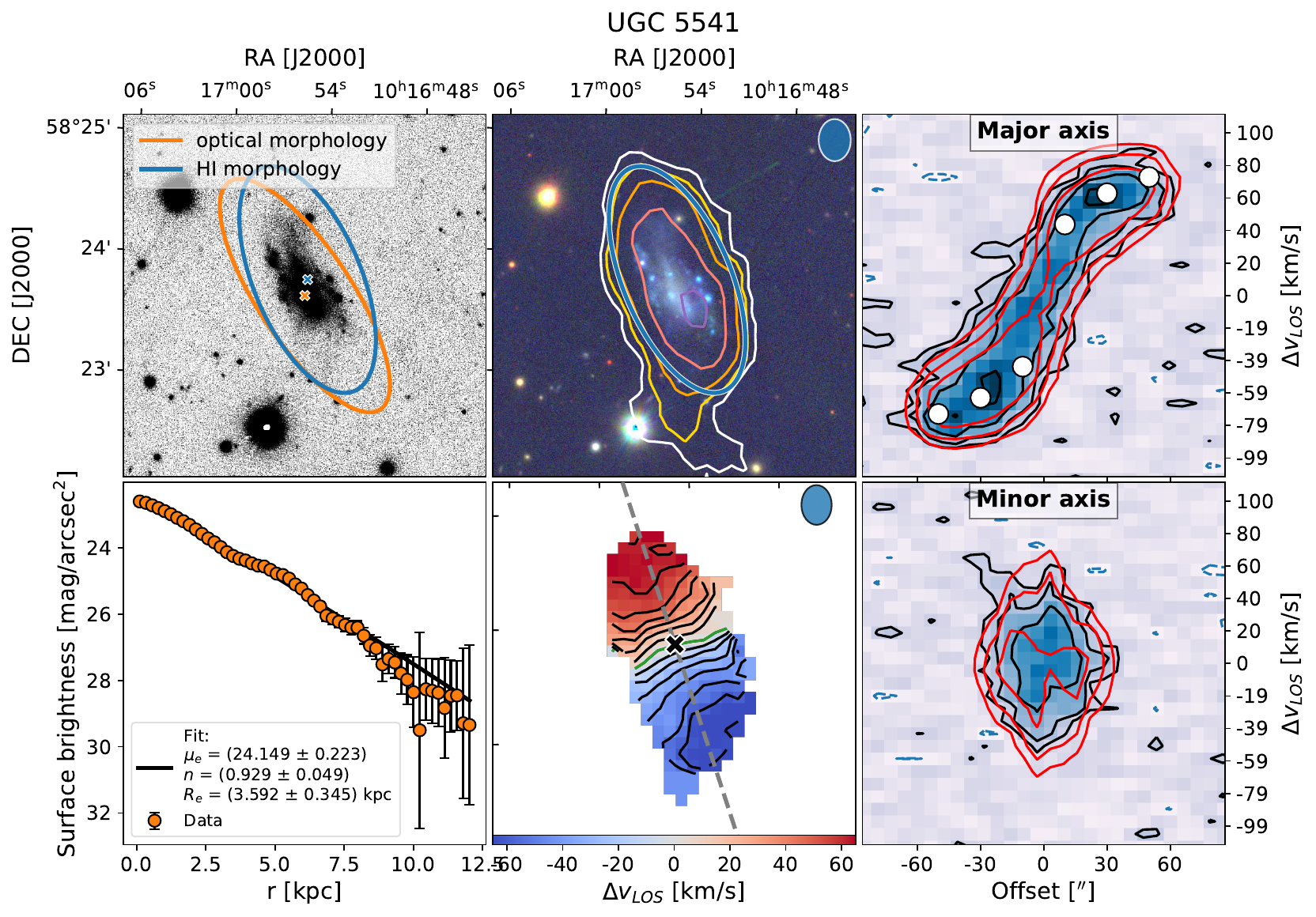}
        \includegraphics[width=0.87\linewidth]{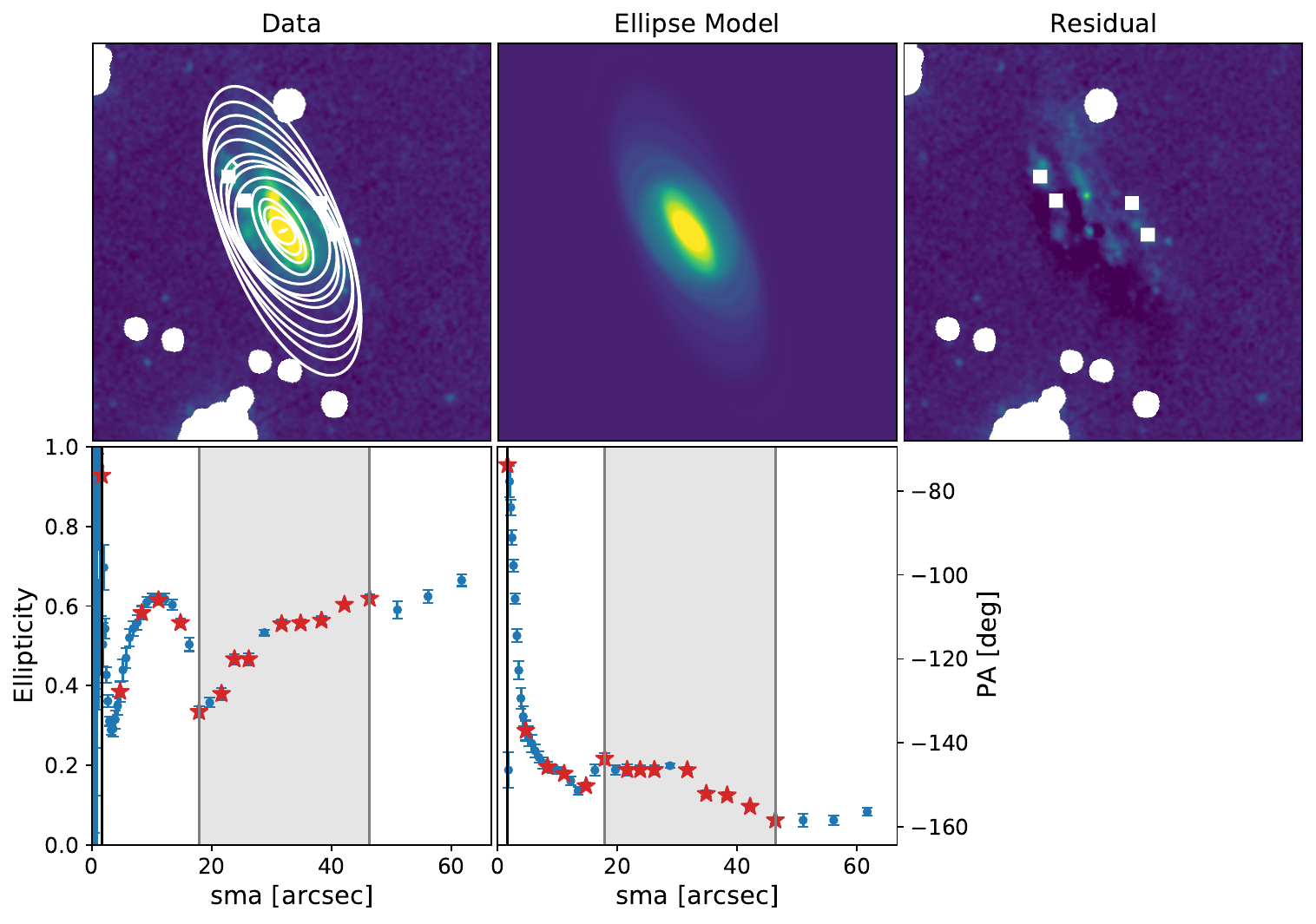}
        \caption{\inter. First two rows are the same as Figs. \ref{fig:pavel-kinematics} (middle and right column) and \ref{fig:pavel-optical} (left column) combined. The \hi\ contours in the middle upper panel correspond to column densities starting from $1.6 \times 10^{20}$ cm$^{-2}$ (in white) and growing by a factor of 2 in intensity towards contours in redder colors. Bottom two rows are the same as in Fig. \ref{fig:isophot_fit-Pavel}.}
        \label{fig:HI-op-inter}
    \end{figure*}

    \textbf{\inter.} The galaxy is shown in Fig. \ref{fig:HI-op-inter}. The \hi\ geometry was well constrained and galaxy shows a well ordered rotation. However, the galaxy has an \hi\ tail extending towards the south indicating a possible interaction. Going 250 kpc eastward, there is the NGC 3182 galaxy with stellar mass of $10^{10.3}$\msun\ \citepads{2023AJ....165..109P}, which could potentially be interacting with \inter. The optical counterpart shows high disturbance with many bright starforming regions in the outskirts, another possible signature of an interaction. Some of these bright regions were masked for the isophotal fitting, but the obtained position angle was still slightly influenced by the region in the northeast direction. The surface brightness profiles could not be robustly fitted with a Sérsic function for this galaxy due to its irregular shape.

\begin{figure*}
        \centering
        \includegraphics[width = 0.96 \textwidth]{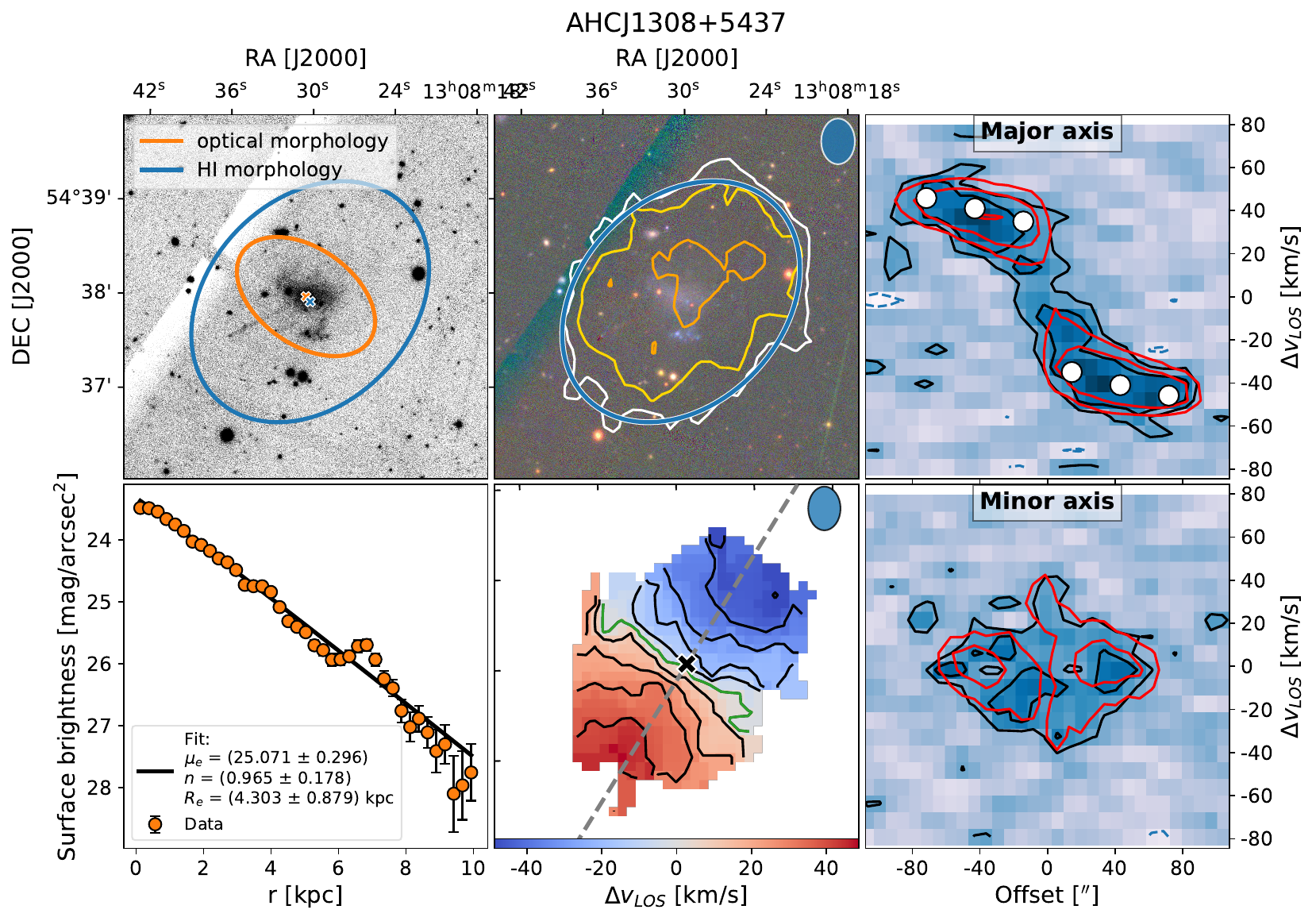}
        \includegraphics[width=0.87\linewidth]{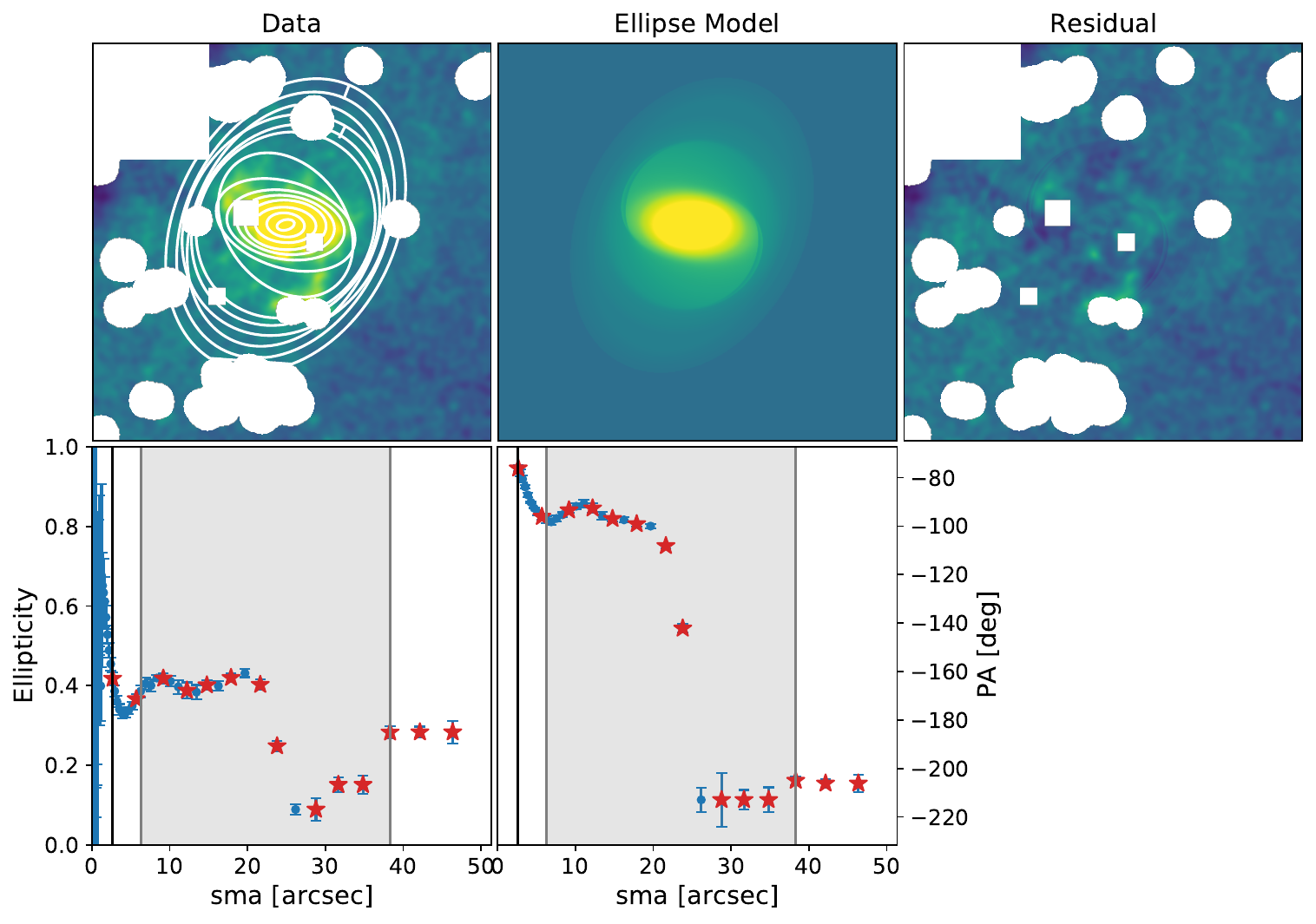}
        \caption{\bSNRtwo. First two rows are the same as Figs. \ref{fig:pavel-kinematics} (middle and right column) and \ref{fig:pavel-optical} (left column) combined. The \hi\ contours in the middle upper panel correspond to column densities starting from $1.4 \times 10^{20}$ cm$^{-2}$ (in white) and growing by a factor of 2 in intensity towards contours in redder colors. Bottom two rows are the same as in Fig. \ref{fig:isophot_fit-Pavel}.}
        \label{fig:HI-op-bSNR_2}
    \end{figure*}

    \textbf{\bSNRtwo.} The galaxy is shown in Fig. \ref{fig:HI-op-bSNR_2}. Due to low S/N, the cube was smoothed 1.5 times for the estimation of \hi\ geometry and the kinematic modeling. The galaxy shows regular rotation. The PS1 data in the $i$-band for this galaxy has a region with very low background counts in the northeast corner which was masked for the optical analysis. The stellar counterpart shows a large starforming region almost completely outside of the inner disk which was partially masked for the isophotal fitting, but unmasked for the extraction of the surface brightness profile. The obtained geometry of the inner disk was well constrained, but the derived surface brightness profile cannot be well fitted with the Sérsic function.

\begin{figure*}
        \centering
        \includegraphics[width = 0.98 \textwidth]{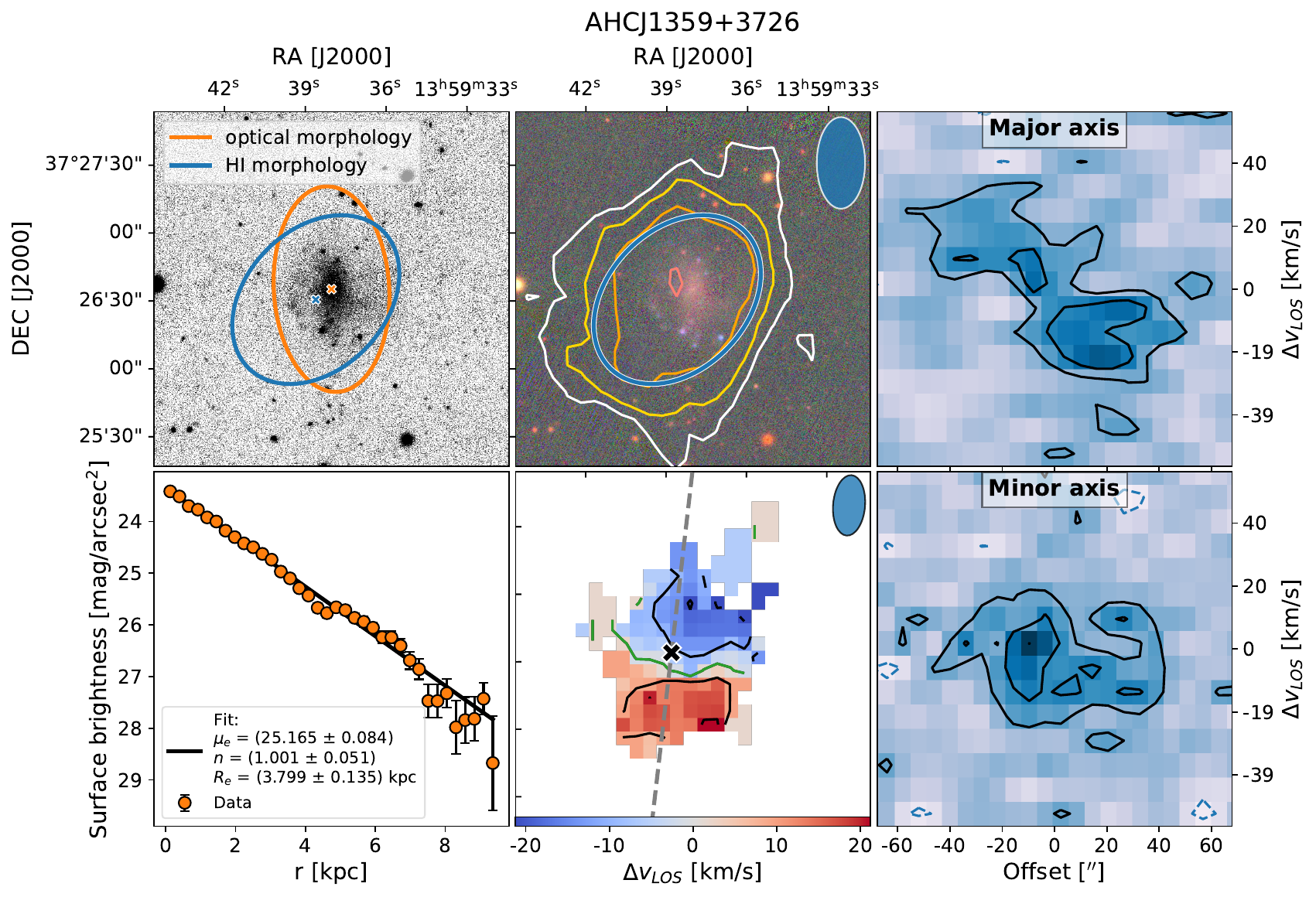}
        \includegraphics[width=0.87\linewidth]{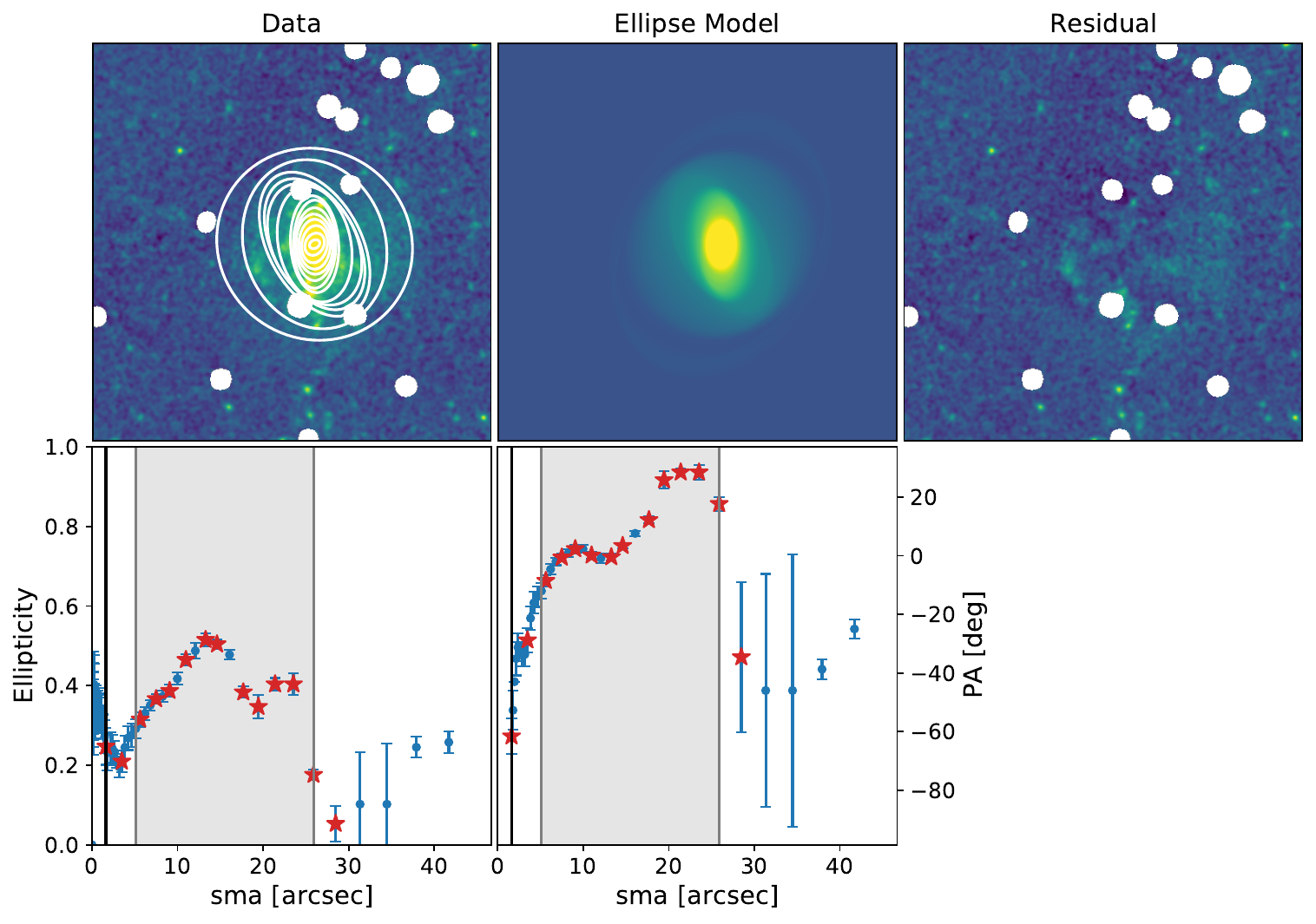}
        \caption{\tstd. First two rows are the same as Figs. \ref{fig:pavel-kinematics} (middle and right column) and \ref{fig:pavel-optical} (left column) combined. The \hi\ contours in the middle upper panel correspond to column densities starting from $8.7 \times 10^{19}$ cm$^{-2}$ (in white) and growing by a factor of 2 in intensity towards contours in redder colors. Bottom two rows are the same as in Fig. \ref{fig:isophot_fit-Pavel}.}
        \label{fig:HI-op-tstd}
    \end{figure*}

    \textbf{\tstd.} The galaxy is shown in Fig. \ref{fig:HI-op-tstd}. Again, due to low S/N, the cube was smoothed 1.5 times for the estimation of \hi\ geometry. The \hi\ kinematics are very irregular as seen in the PV slices. Therefore, we could not produce reliable kinematic model for this galaxy. Optical counterpart shows significant variations of geometric parameters with radius. Specifically, the ellipticity grows from the central part and falls off at the outer part. The obtained optical geometry is in accordance with the kinematic position angle from \hi, but not with \hi\ geometry.

\begin{figure*}
        \centering
        \includegraphics[width = 0.96 \textwidth]{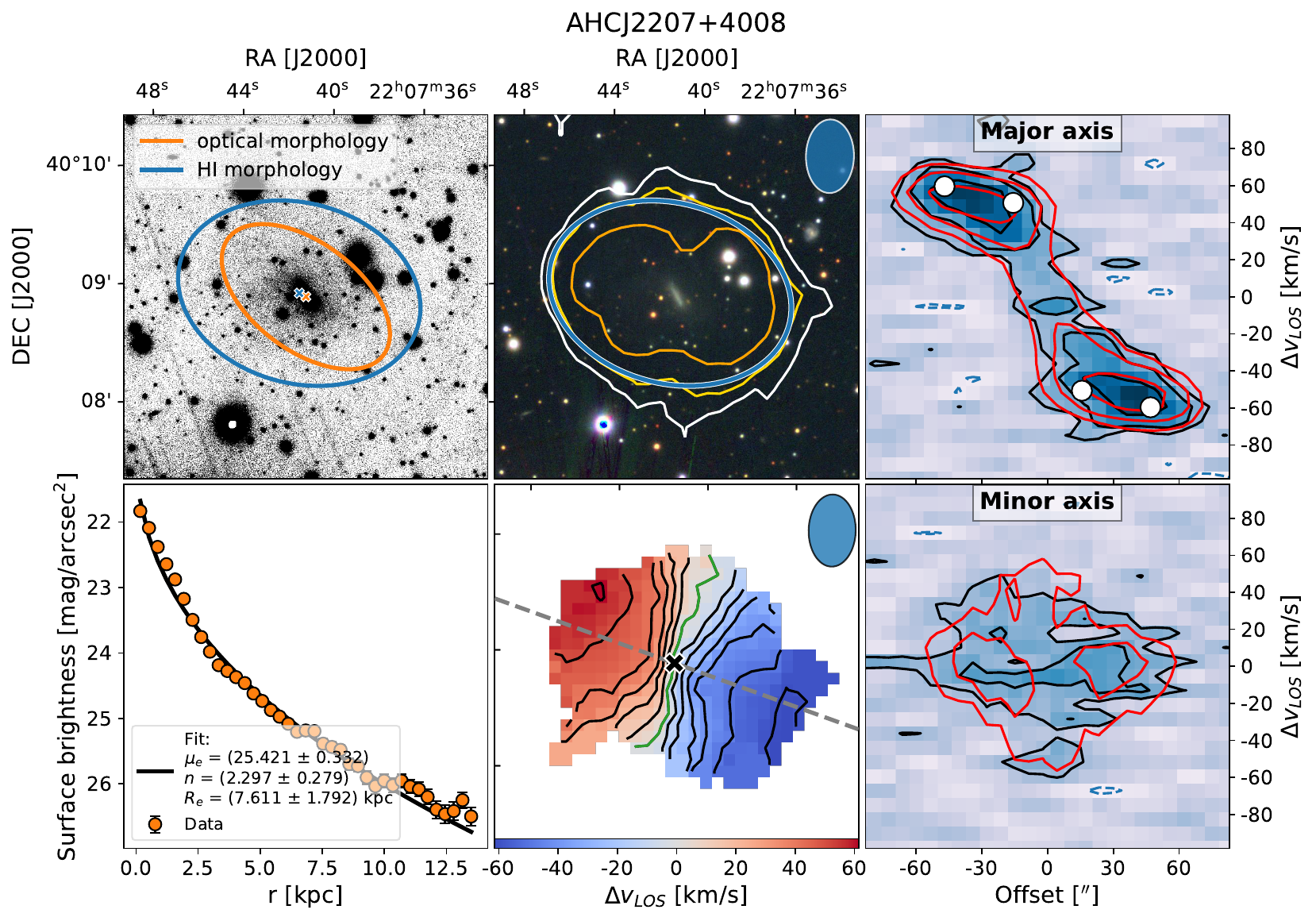}
        \includegraphics[width=0.87\linewidth]{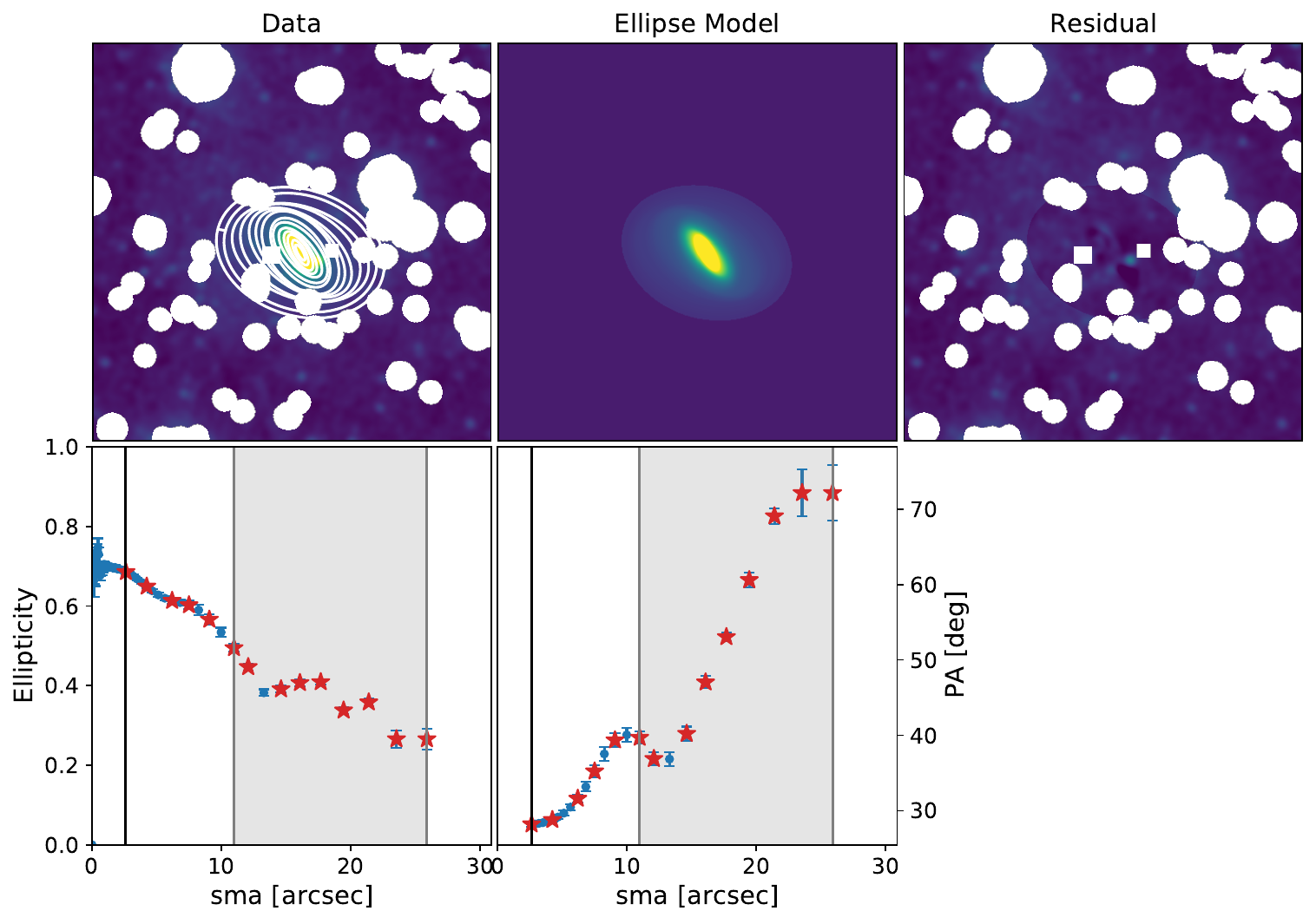}
        \caption{\nce. First two rows are the same as Figs. \ref{fig:pavel-kinematics} (middle and right column) and \ref{fig:pavel-optical} (left column) combined. The \hi\ contours in the middle upper panel correspond to column densities starting from $7.9 \times 10^{19}$ cm$^{-2}$ (in white) and growing by a factor of 2 in intensity towards contours in redder colors. Bottom two rows are the same as in Fig. \ref{fig:isophot_fit-Pavel}.} 
        \label{fig:HI-op-nce}
    \end{figure*}

    \textbf{\nce.} The galaxy is shown in Fig. \ref{fig:HI-op-nce}. The \hi\ geometry and kinematics of the galaxy are well constrained. Its stellar counterpart shows a variation in position angle and ellipticity with radius. Interestingly, the evolution of position angle with radius is moving towards the position angle seen in \hi. 

\begin{figure*}
        \centering
        \includegraphics[width = 0.98 \textwidth]{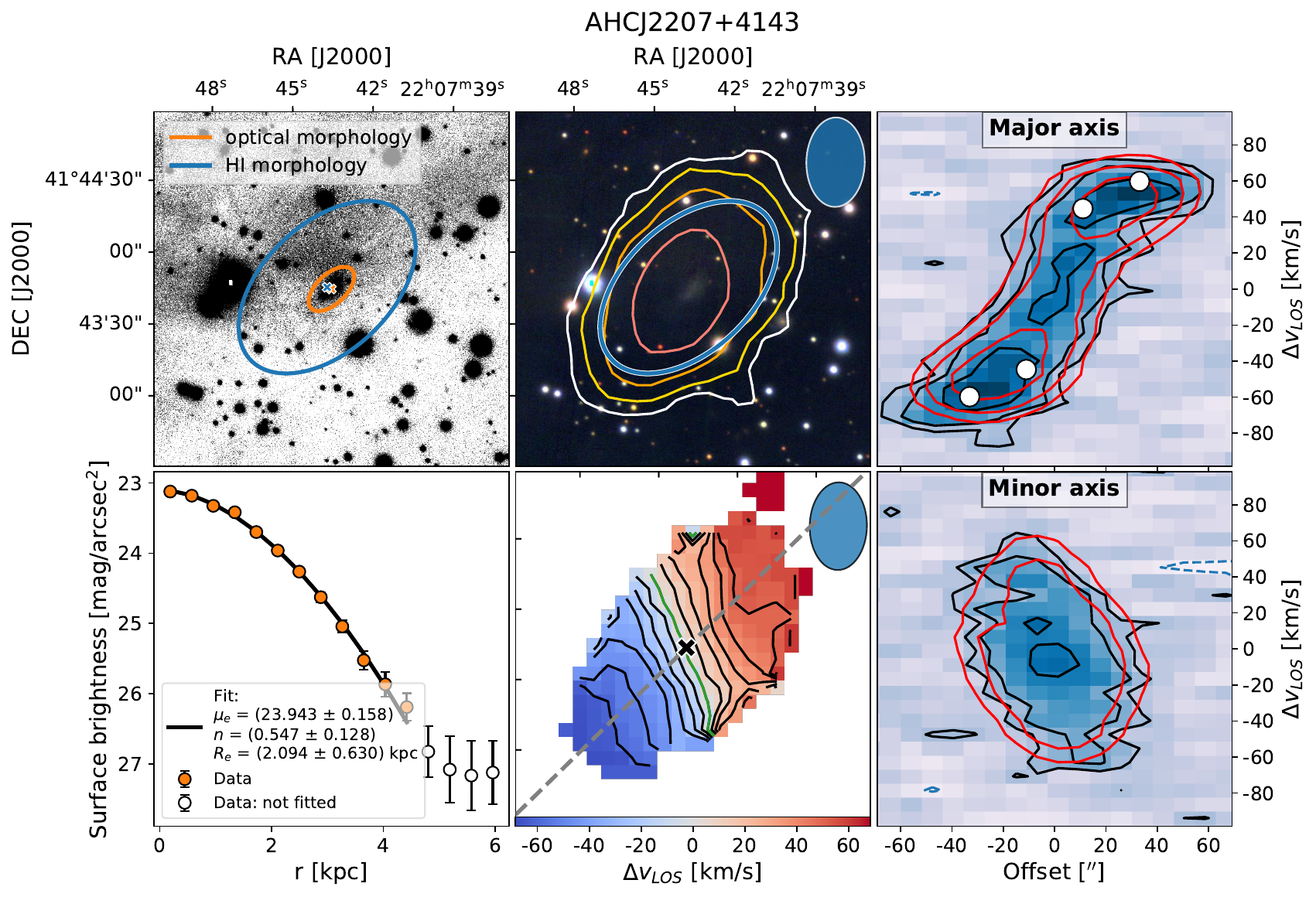}
        \includegraphics[width=0.87\linewidth]{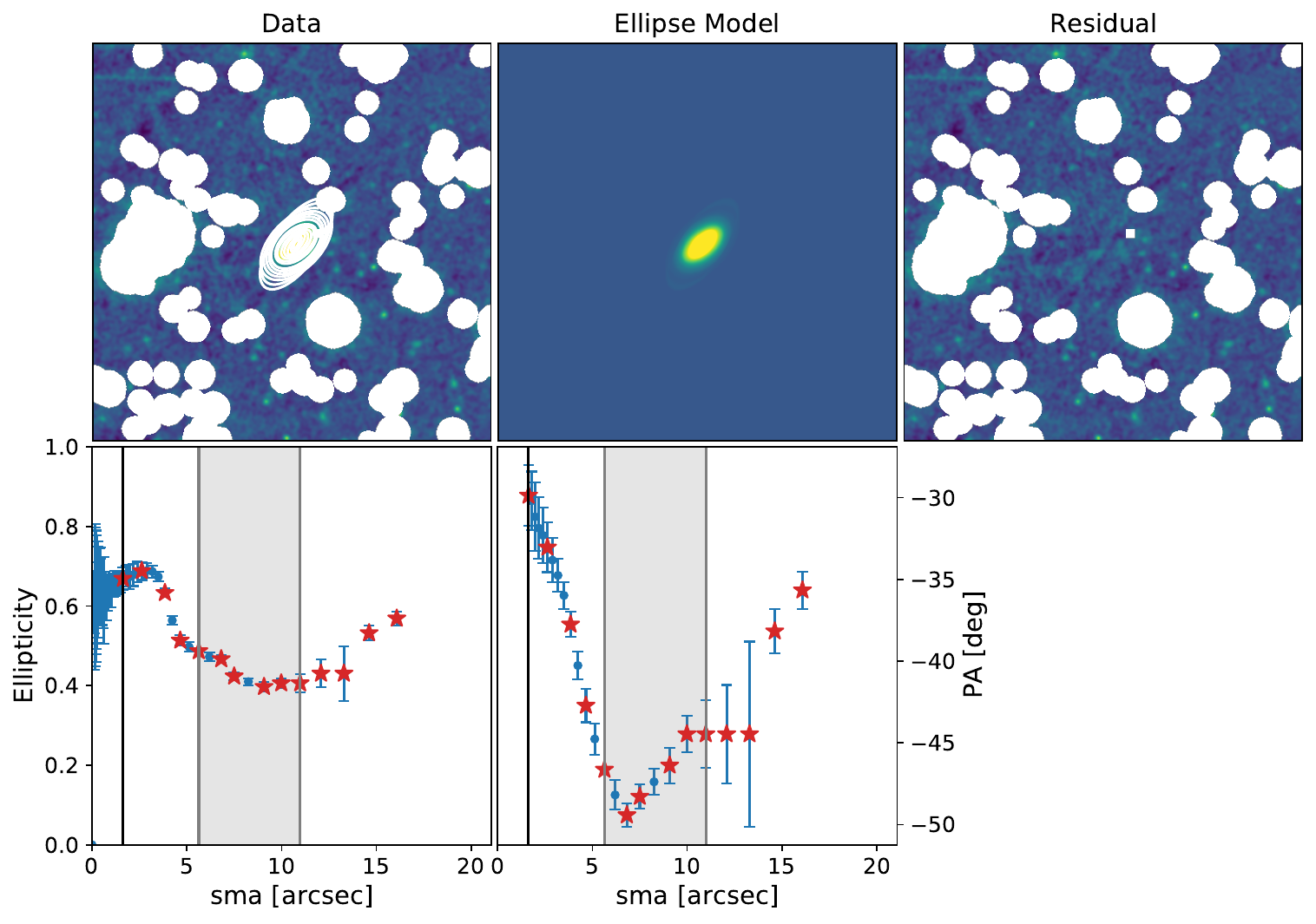}
        \caption{\beaut. First two rows are the same as Figs. \ref{fig:pavel-kinematics} (middle and right column) and \ref{fig:pavel-optical} (left column) combined. The \hi\ contours in the middle upper panel correspond to column densities starting from $8.0 \times 10^{19}$ cm$^{-2}$ (in white) and growing by a factor of 2 in intensity towards contours in redder colors. Bottom two rows are the same as in Fig. \ref{fig:isophot_fit-Pavel}.} 
        \label{fig:HI-op-beaut}
    \end{figure*}

    \textbf{\beaut.} The galaxy is shown in Fig. \ref{fig:HI-op-beaut}. Its kinematics shows a very well ordered disk, but a significant non-orthogonality of major axis with its systemic velocity contour seen from the velocity field map. This also causes the velocity gradient in the PV along the minor axis and is a likely consequence of radial motions of gas. Unfortunately, the galaxy is positioned close to Galactic cirrus, making the optical analysis less reliable. Otherwise, the galaxy seems to be regular.

\begin{figure*}
        \centering
        \includegraphics[width = 0.98 \textwidth]{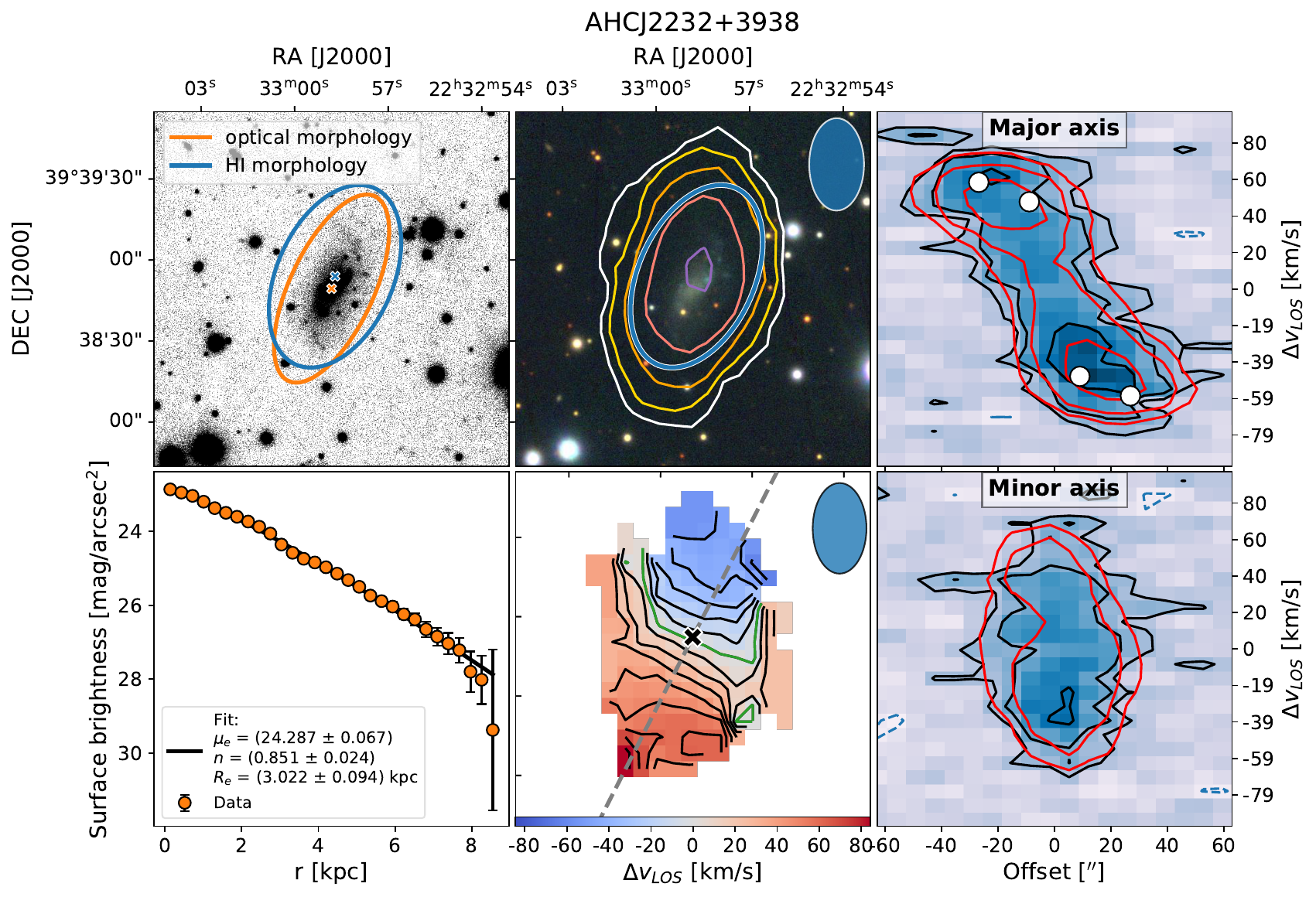}
        \includegraphics[width=0.87\linewidth]{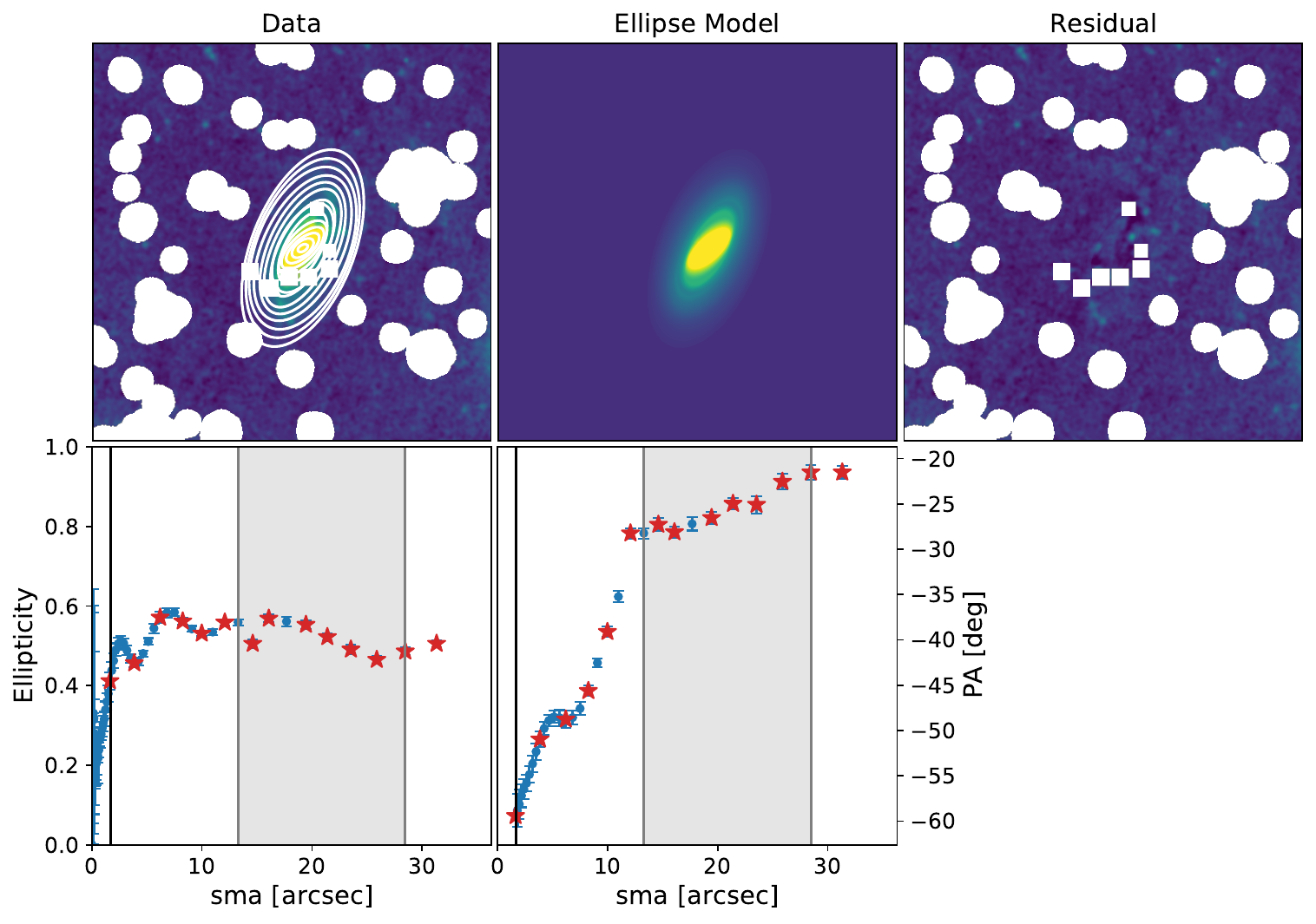}
        \caption{\stick. First two rows are the same as Figs. \ref{fig:pavel-kinematics} (middle and right column) and \ref{fig:pavel-optical} (left column) combined. The \hi\ contours in the middle upper panel correspond to column densities starting from $7.7 \times 10^{19}$ cm$^{-2}$ (in white) and growing by a factor of 2 in intensity towards contours in redder colors. Bottom two rows are the same as in Fig. \ref{fig:isophot_fit-Pavel}.} 
        \label{fig:HI-op-stick}
    \end{figure*}

    \textbf{\stick.} The galaxy is shown in Fig. \ref{fig:HI-op-stick}. The \hi\ geometry was well constrained and the \hi\ kinematics shows a well ordered disk. The optical counterpart shows little variation of geometric parameters with radius and is well constrained.

\begin{figure*}
        \centering
        \includegraphics[width = 0.98 \textwidth]{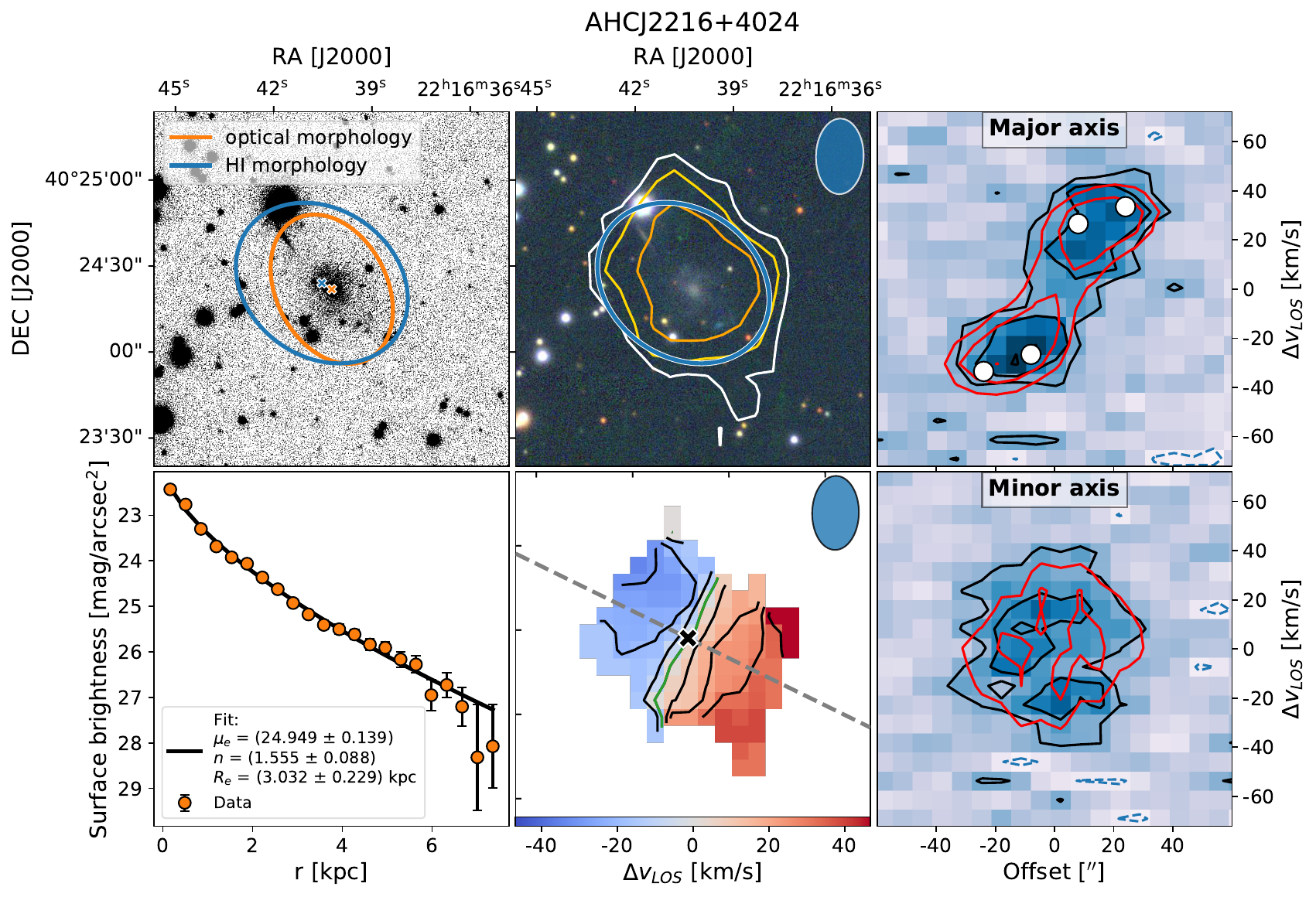}
        \includegraphics[width=0.87\linewidth]{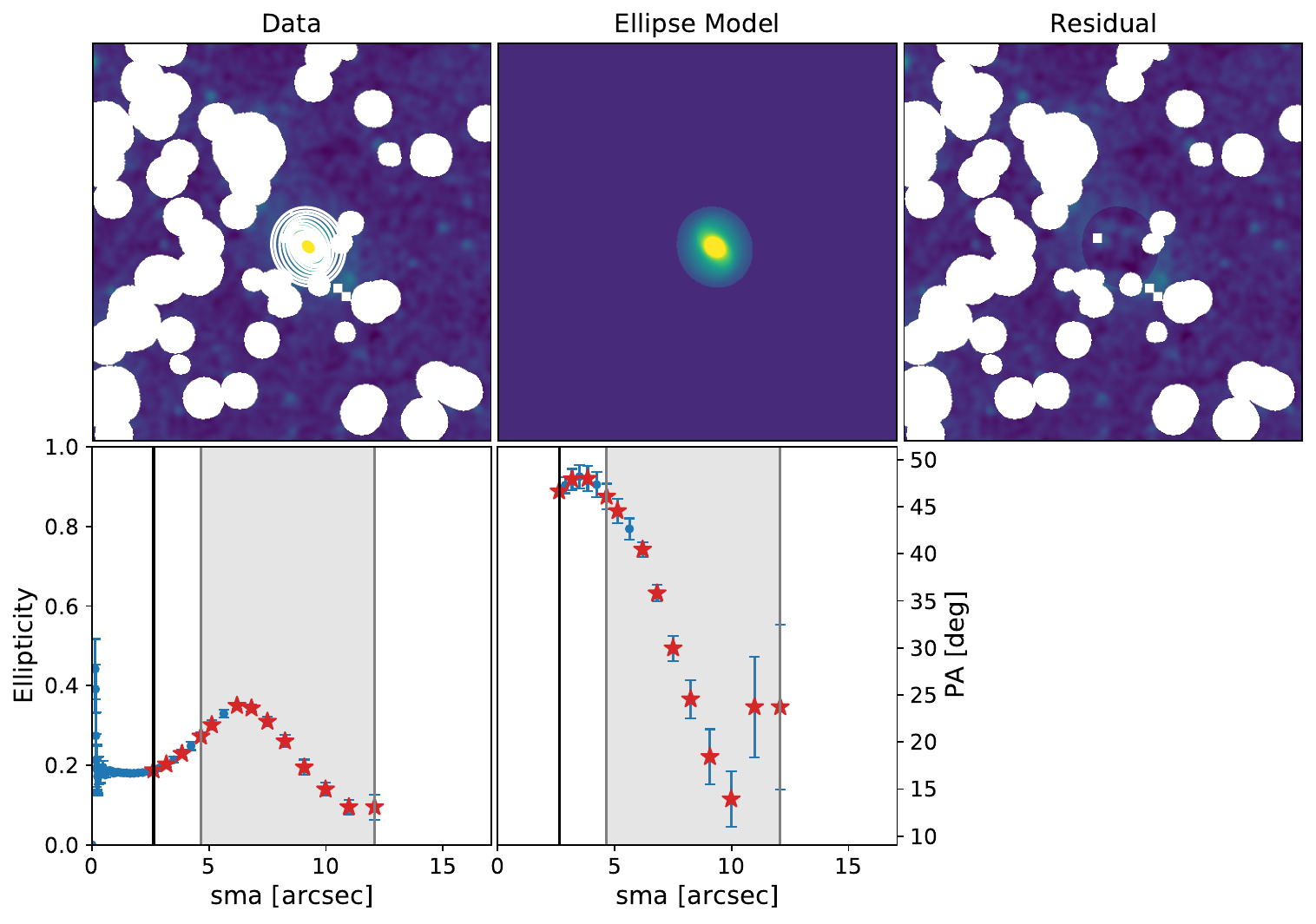}
        \caption{\reg. First two rows are the same as Figs. \ref{fig:pavel-kinematics} (middle and right column) and \ref{fig:pavel-optical} (left column) combined. The \hi\ contours in the middle upper panel correspond to column densities starting from $9.5 \times 10^{19}$ cm$^{-2}$ (in white) and growing by a factor of 2 in intensity towards contours in redder colors. Bottom two rows are the same as in Fig. \ref{fig:isophot_fit-Pavel}.} 
        \label{fig:HI-op-reg}
    \end{figure*}

    \textbf{\reg.} The galaxy is shown in Fig. \ref{fig:HI-op-reg}. The \hi\ geometry was well constrained and the kinematics seems regular. The optical counterpart shows some variation in position angle with radius, but it is not unexpected due to low inclination. 

\begin{figure*}
        \centering
        \includegraphics[width = 0.98 \textwidth]{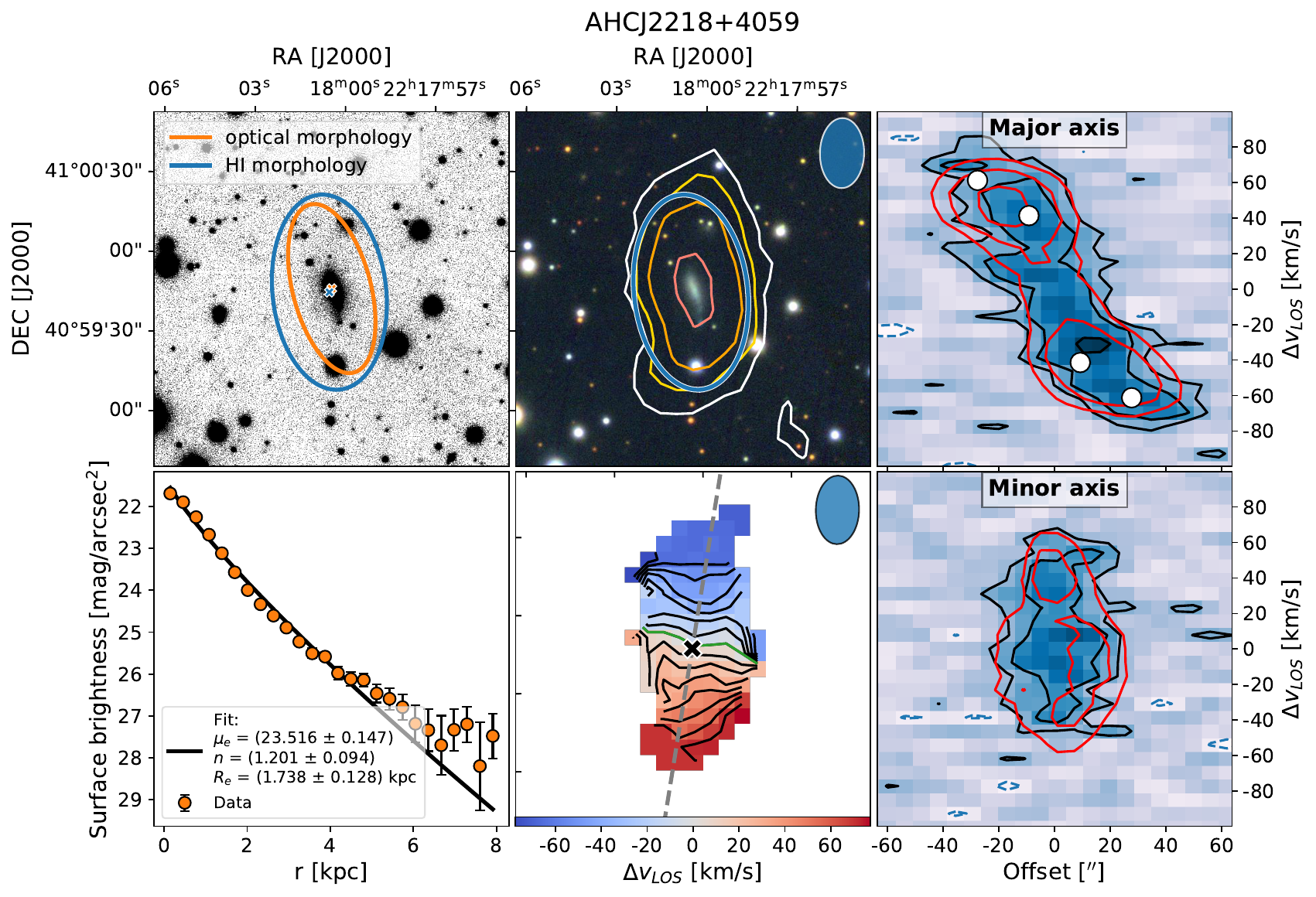}
        \includegraphics[width=0.87\linewidth]{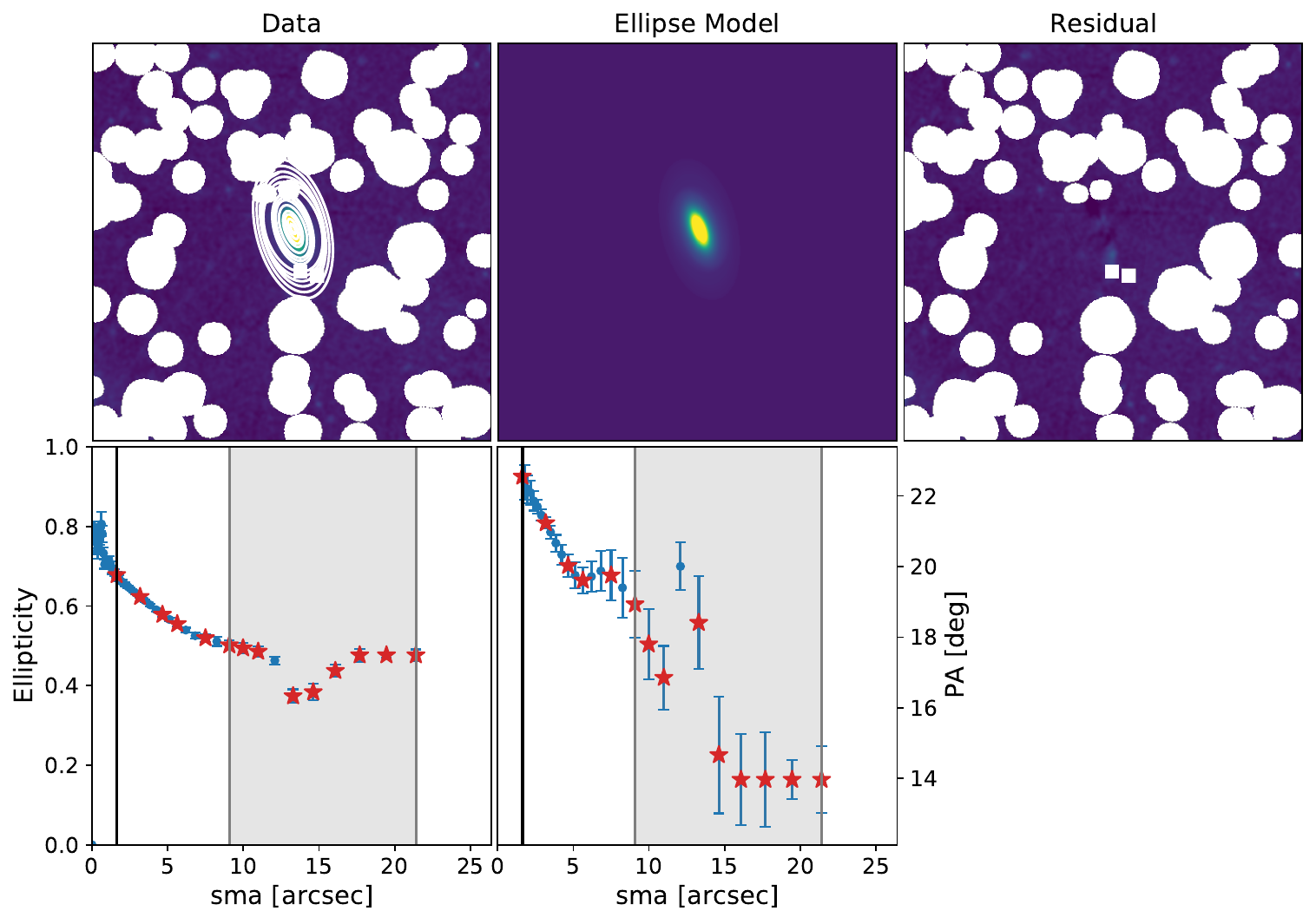}
        \caption{\paver. First two rows are the same as Figs. \ref{fig:pavel-kinematics} (middle and right column) and \ref{fig:pavel-optical} (left column) combined. The \hi\ contours in the middle upper panel correspond to column densities starting from $1.2 \times 10^{20}$ cm$^{-2}$ (in white) and growing by a factor of 2 in intensity towards contours in redder colors. Bottom two rows are the same as in Fig. \ref{fig:isophot_fit-Pavel}.} 
        \label{fig:HI-op-PA_180}
    \end{figure*}

    \textbf{\paver.} The galaxy is shown in Fig. \ref{fig:HI-op-PA_180}. The geometry of the \hi\ disk was well constrained. However, the galaxy shows a warp in the position angle seen as a twist in the velocity field, which is also apparent in the \hi\ morphology. Its optical counterpart also has a well constrained morphology and little variation of geometric parameters with radius. We note that in the very inner part, the ellipticity is around 0.7, corresponding to the inclination of 90\dg\ for $q_0=0.3$. As before, there also seems to be a broad \hi\ emission region in the PV slice along the major axis, which could be another signature of edge-on galaxies. Again, we note that the galaxies in this sample are not spatially well enough resolved to distinguish this broadening from the effect of the beam.

\begin{figure*}
        \centering
        \includegraphics[width = 0.98 \textwidth]{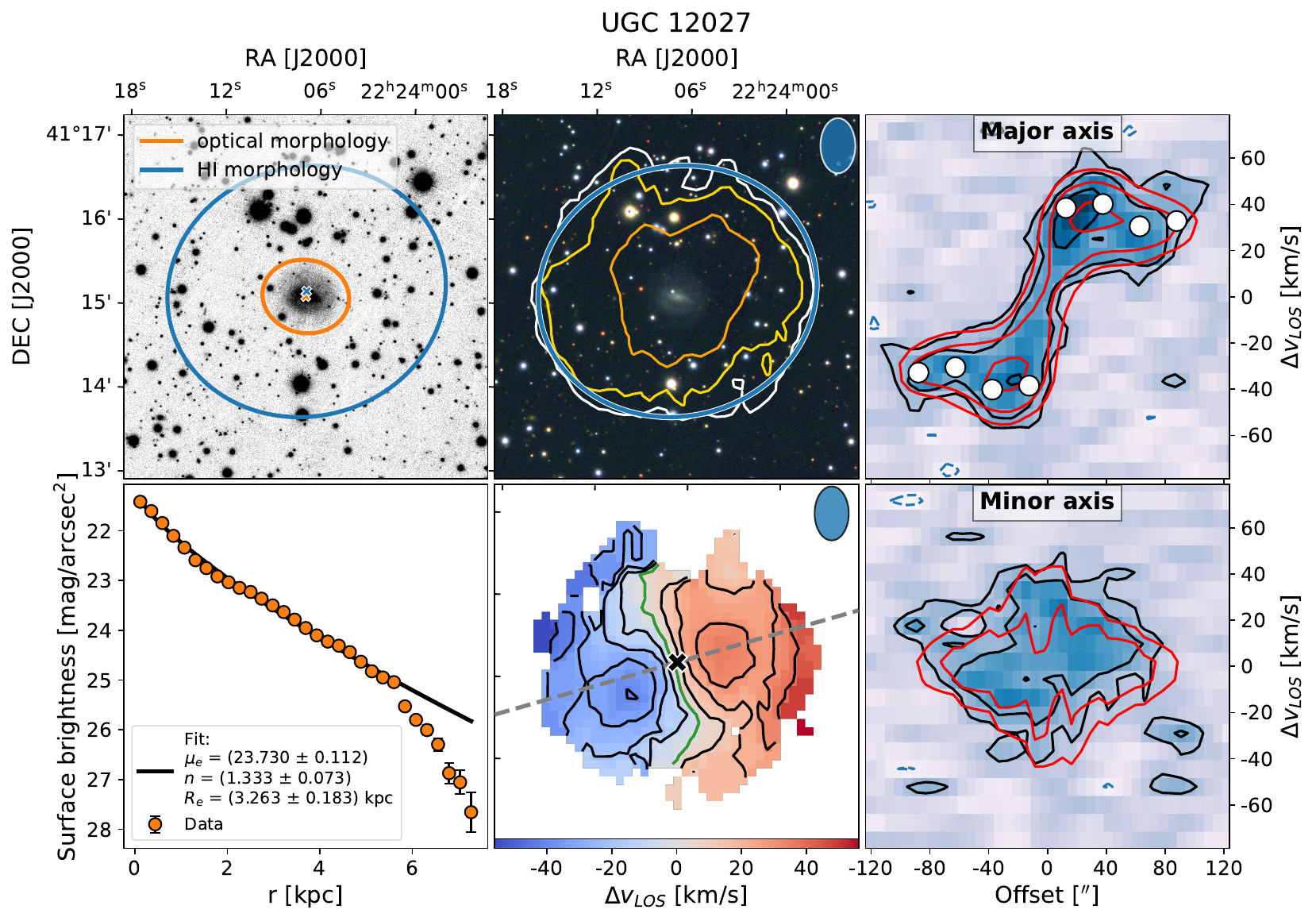}
        \includegraphics[width=0.87\linewidth]{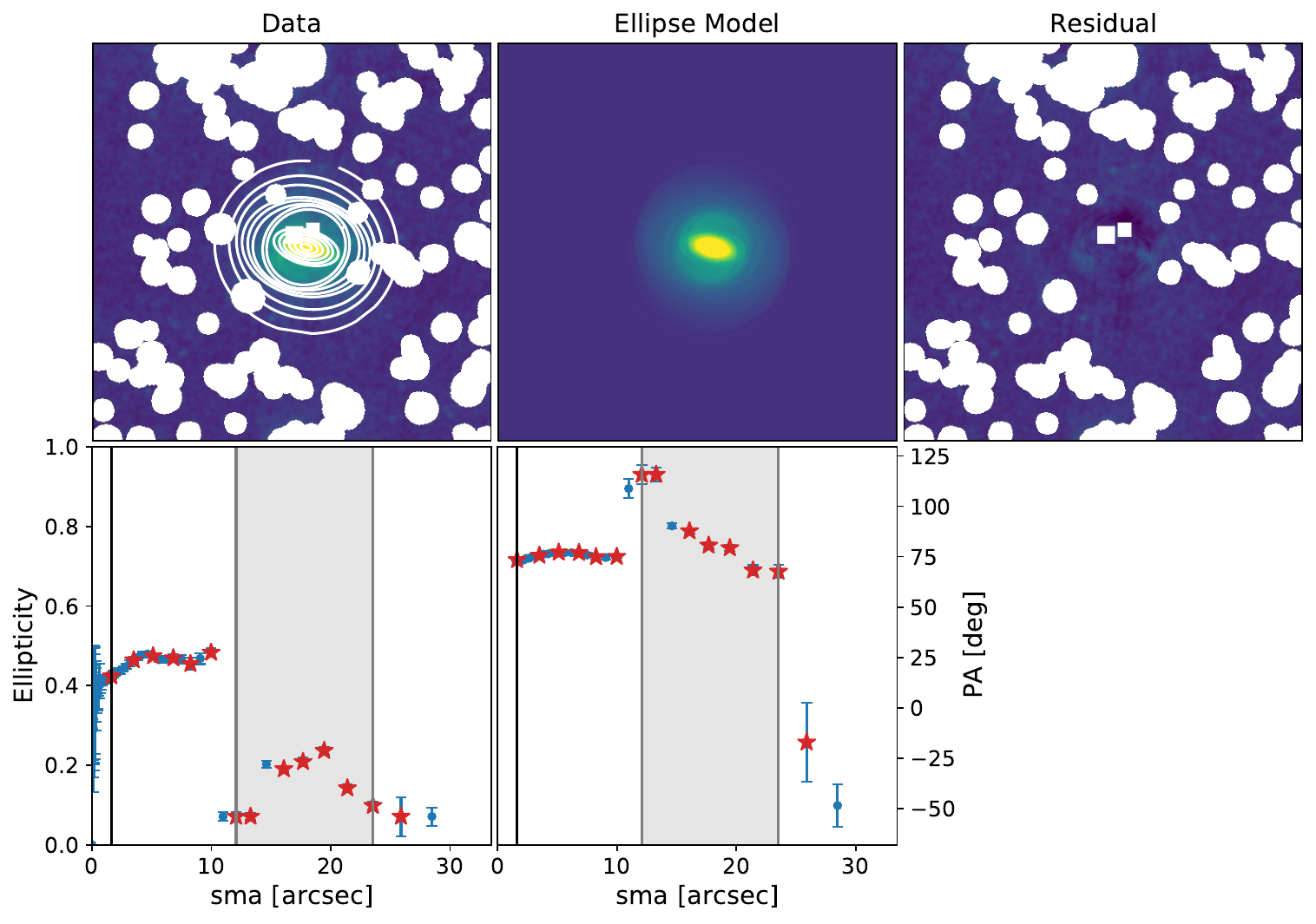}
        \caption{\mess. First two rows are the same as Figs. \ref{fig:pavel-kinematics} (middle and right column) and \ref{fig:pavel-optical} (left column) combined. The \hi\ contours in the middle upper panel correspond to column densities starting from $5.4 \times 10^{19}$ cm$^{-2}$ (in white) and growing by a factor of 2 in intensity towards contours in redder colors. Bottom two rows are the same as in Fig. \ref{fig:isophot_fit-Pavel}.} 
        \label{fig:HI-op-mess}
    \end{figure*}

    \textbf{\mess.} The galaxy is shown in Fig. \ref{fig:HI-op-mess}. Due to low S/N, the \hi\ cube of this galaxy was smoothed 1.5 times for the estimation of \hi\ geometry and the kinematic modeling. The galaxy shows a drop in rotational velocity at larger radii which is also apparent in the velocity map. This might be a sign of a warp in inclination and possibly the position angle. The change in inclination (or ellipticity) also seems to be present in its optical counterpart, with higher inclination in the inner part and a drop in inclination in the outskirts.

\begin{figure*}
        \centering
        \includegraphics[width = 0.98 \textwidth]{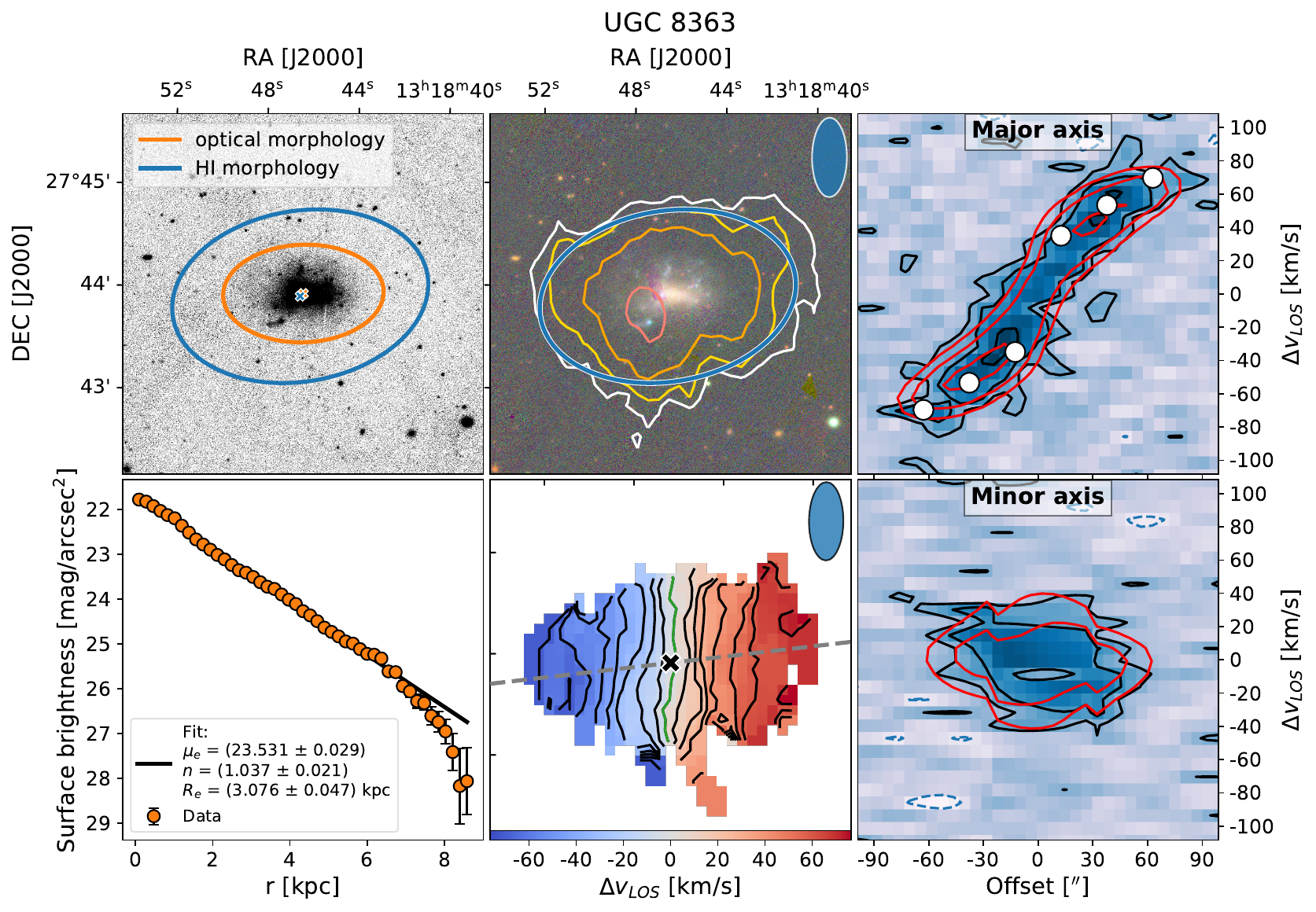}
        \includegraphics[width=0.87\linewidth]{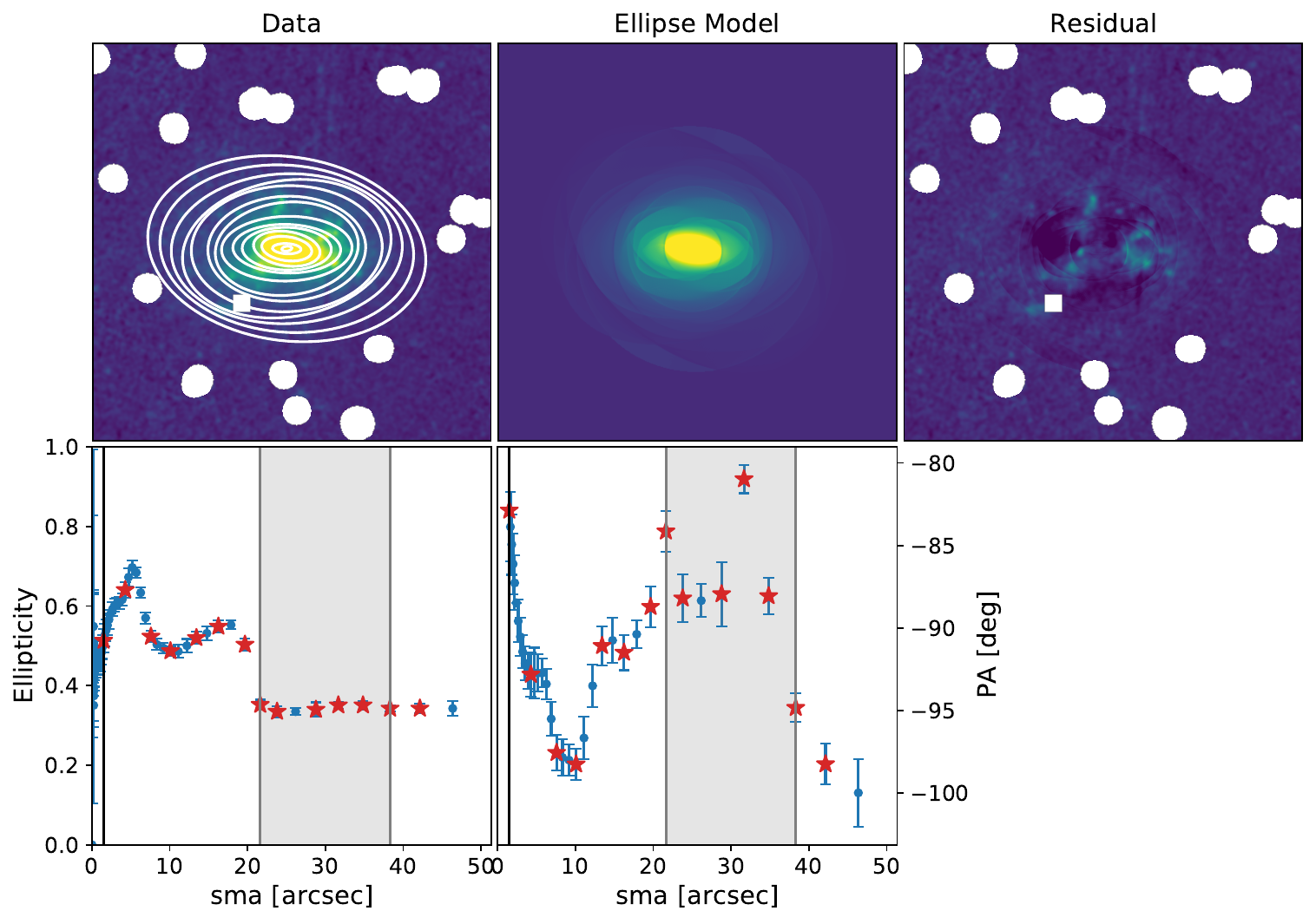}
        \caption{\wtm. First two rows are the same as Figs. \ref{fig:pavel-kinematics} (middle and right column) and \ref{fig:pavel-optical} (left column) combined. The \hi\ contours in the middle upper panel correspond to column densities starting from $1.4 \times 10^{20}$ cm$^{-2}$ (in white) and growing by a factor of 2 in intensity towards contours in redder colors. Bottom two rows are the same as in Fig. \ref{fig:isophot_fit-Pavel}.} 
        \label{fig:HI-op-WTM}
    \end{figure*}

    \textbf{\wtm.} The galaxy is shown in Fig. \ref{fig:HI-op-WTM}. The \hi\ geometry was well constrained and the kinematics shows a well ordered rotating disk. The optical counterpart shows a change in ellipticity from a higher constant value in the inner part to a slightly lower constant value in the outer parts. The change is minimal in our region of interest for the estimation of geometric parameters and is in accordance with the \hi\ geometry.

\begin{figure*}
        \centering
        \includegraphics[width = 0.98 \textwidth]{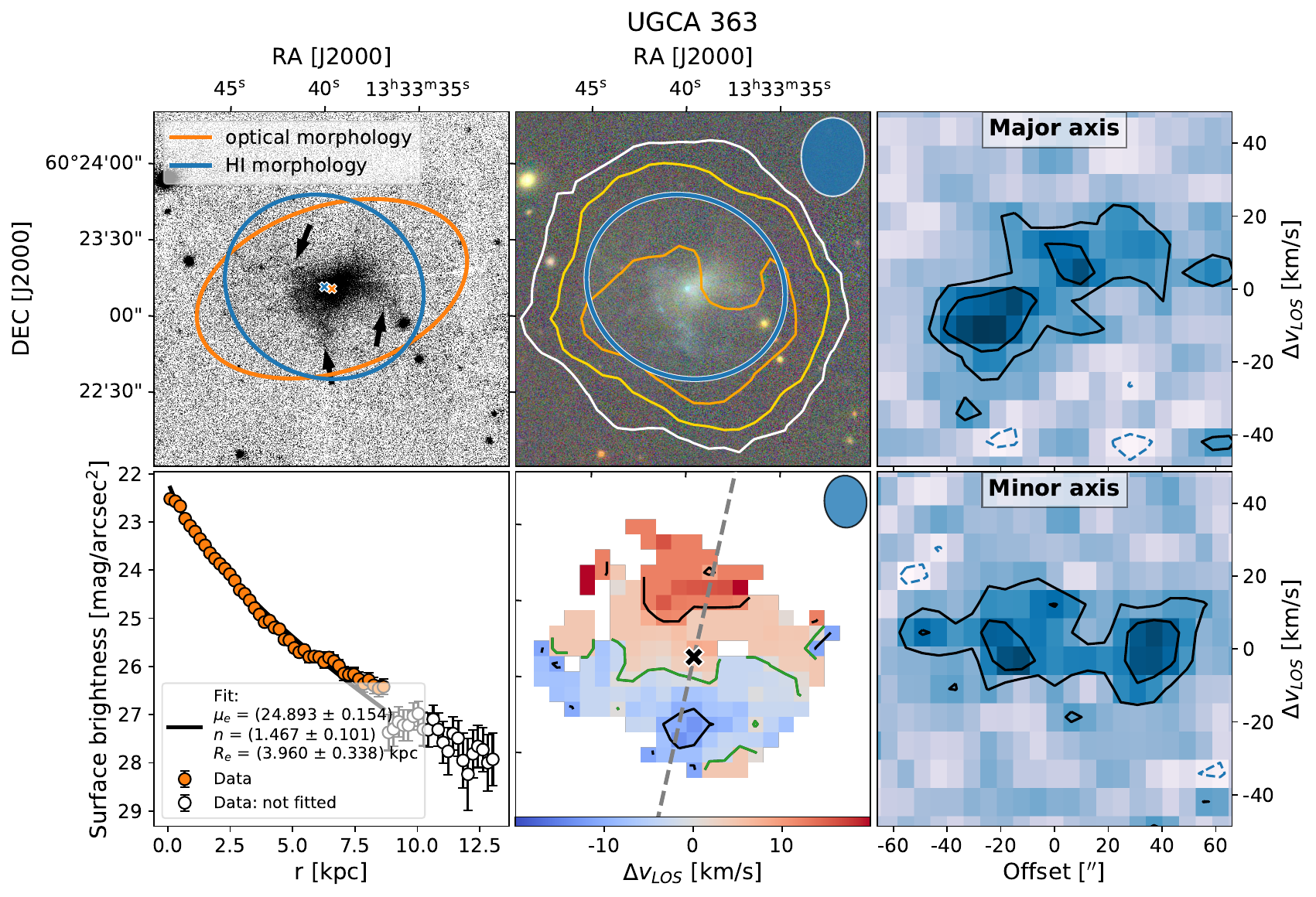}
        \includegraphics[width=0.87\linewidth]{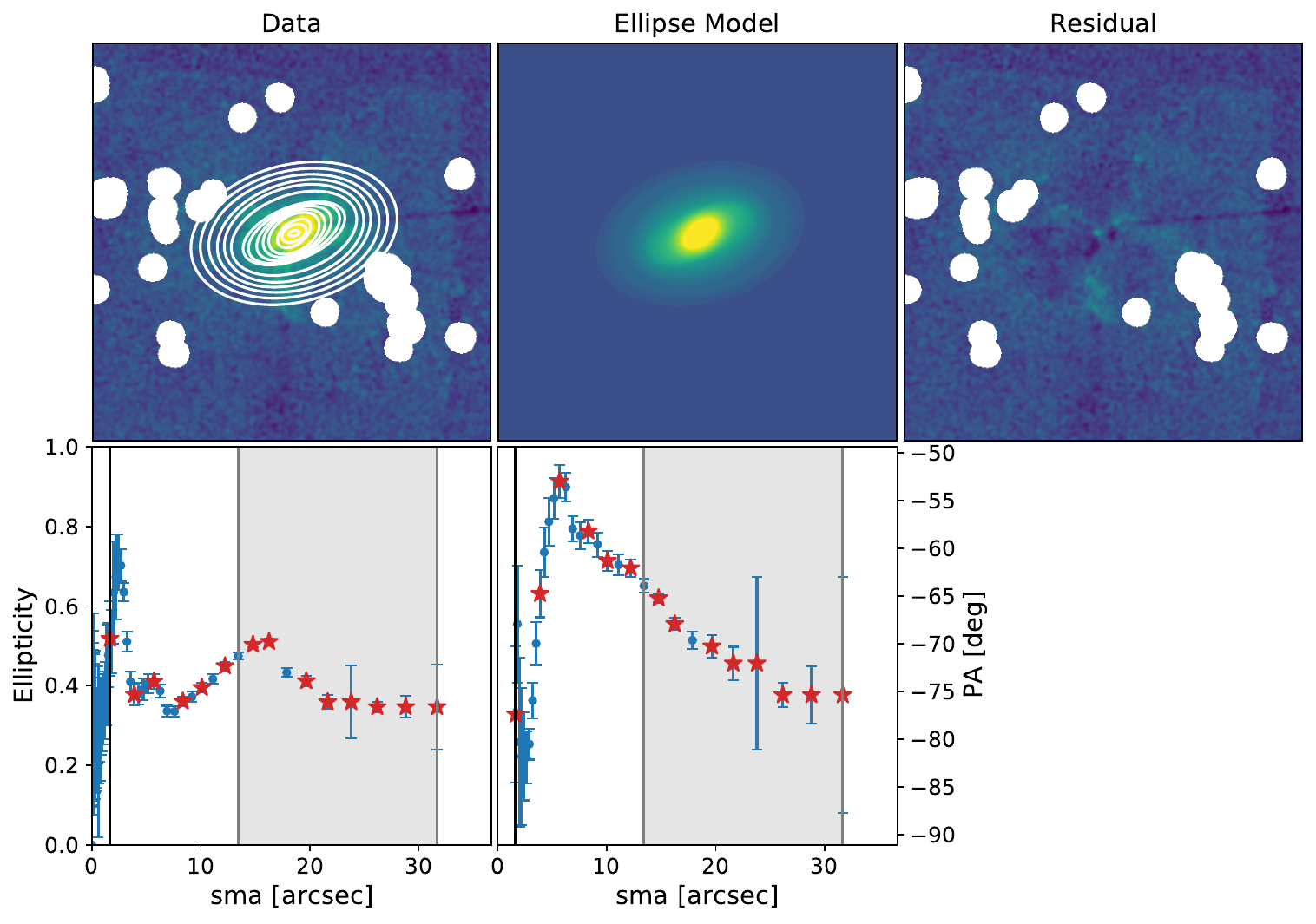}
        \caption{\dwtm. First two rows are the same as Figs. \ref{fig:pavel-kinematics} (middle and right column) and \ref{fig:pavel-optical} (left column) combined. Black arrows in the upper left panel denote extended stellar structures in the outskirt of the galaxy (see the text). The \hi\ contours in the middle upper panel correspond to column densities starting from $1.3 \times 10^{20}$ cm$^{-2}$ (in white) and growing by a factor of 2 in intensity towards contours in redder colors. Bottom two rows are the same as in Fig. \ref{fig:isophot_fit-Pavel}.}
        \label{fig:HI-op-dwtm}
    \end{figure*}

    \textbf{\dwtm.} The galaxy is shown in Fig. \ref{fig:HI-op-dwtm}. Due to low S/N, the \hi\ cube was smoothed 1.5 times for the estimation of \hi\ geometry. However, kinematic modeling was not possible for this galaxy due to its high disturbance and the apparent complete misalignment between the morphological and kinematic position angle in \hi\ as well as the misalignment between the optical position angle and the kinematic one. 
    The stellar counterpart shows signs of three stellar arms towards the northeast, south and southwest directions. 
    The obtained position angle is in accordance with the orientation of the inner disk. 

\begin{figure*}
        \centering
        \includegraphics[width = 0.98 \textwidth]{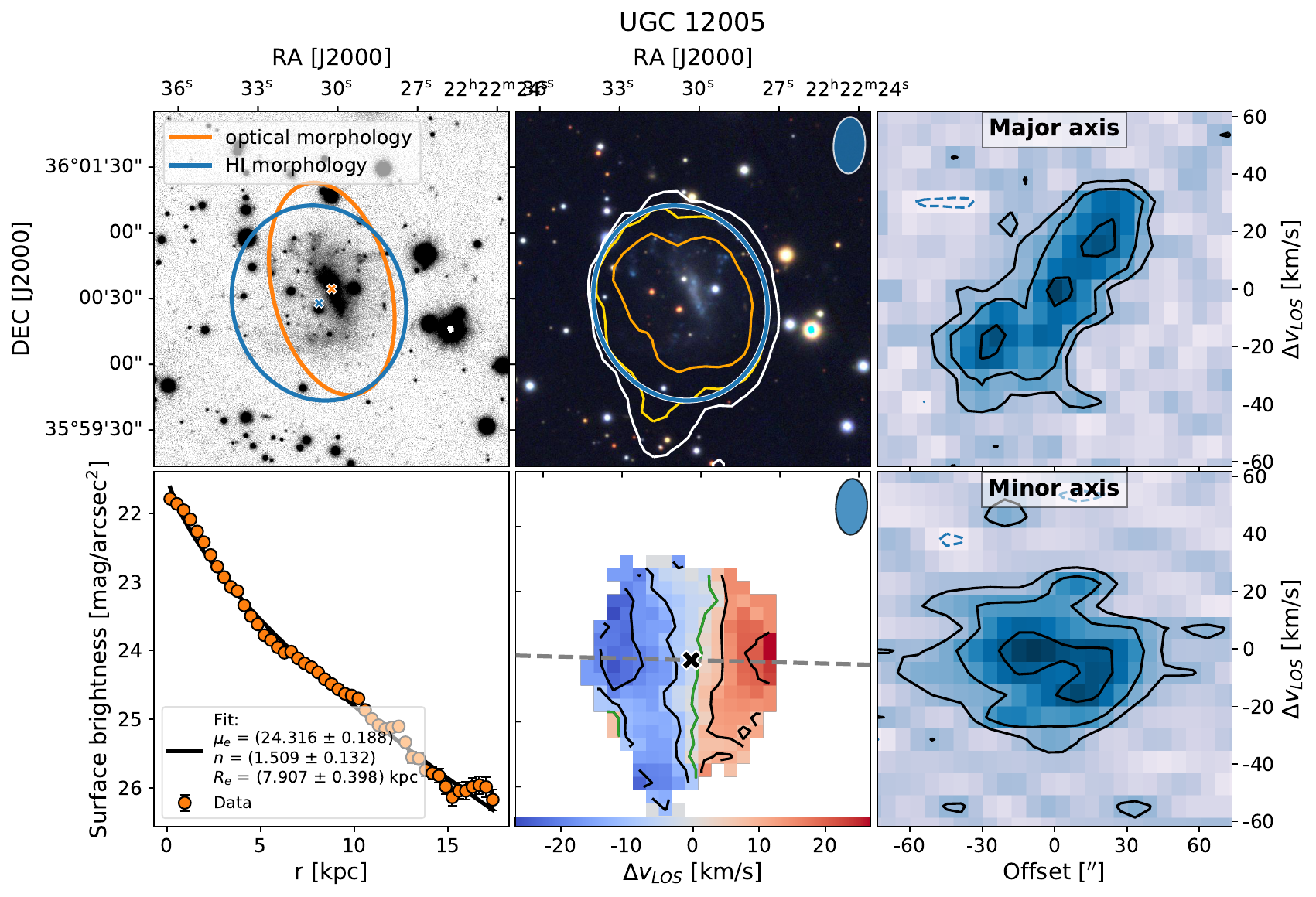}
        \includegraphics[width=0.87\linewidth]{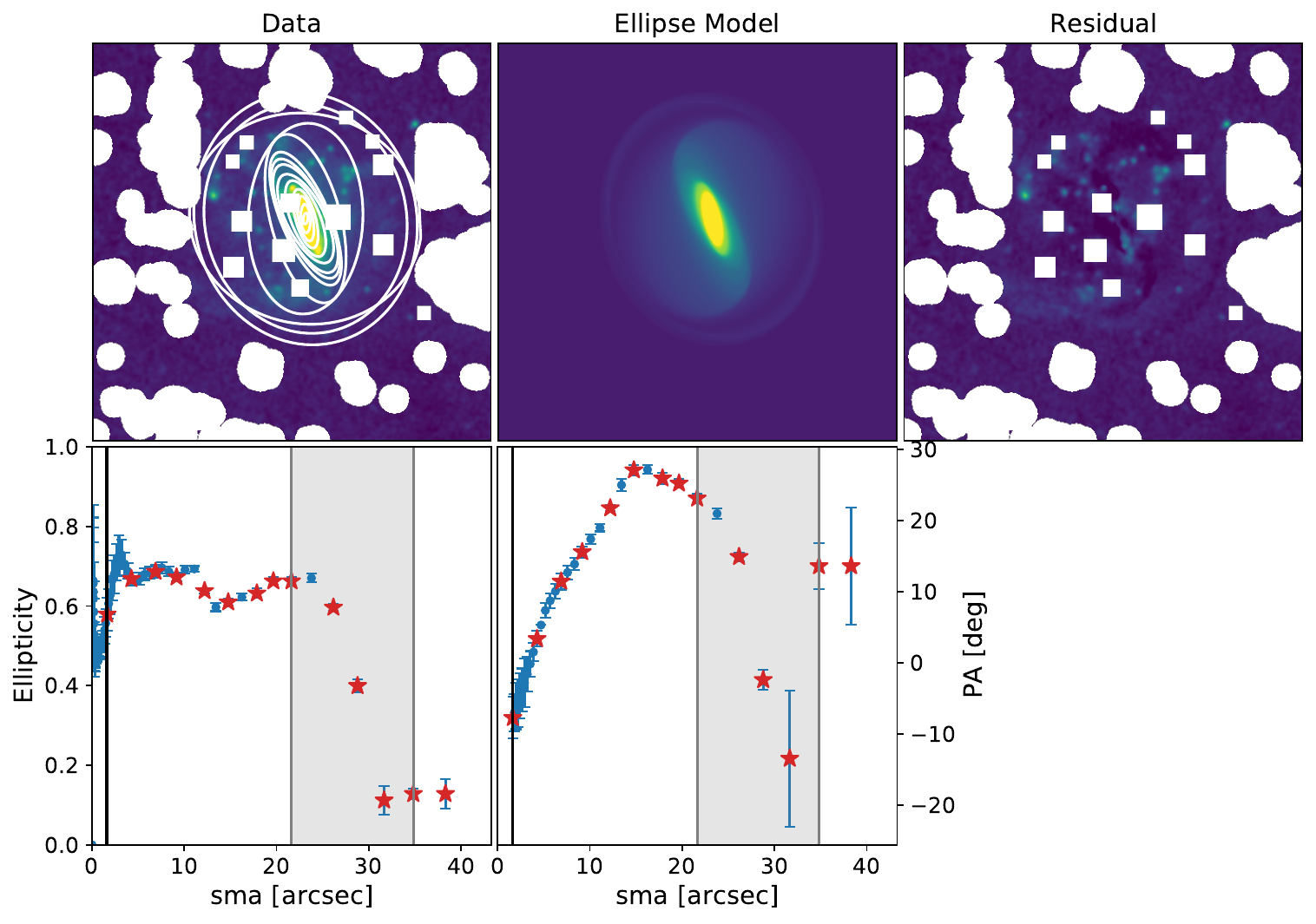}
        \caption{\misal. First two rows are the same as Figs. \ref{fig:pavel-kinematics} (middle and right column) and \ref{fig:pavel-optical} (left column) combined. The \hi\ contours in the middle upper panel correspond to column densities starting from $1.6 \times 10^{20}$ cm$^{-2}$ (in white) and growing by a factor of 2 in intensity towards contours in redder colors. Bottom two rows are the same as in Fig. \ref{fig:isophot_fit-Pavel}.}
        \label{fig:HI-op-misal}
    \end{figure*}

    \textbf{\misal.} The galaxy is shown in Fig. \ref{fig:HI-op-misal}. The geometry of the \hi\ disk was well constrained with CANNUBI. Interestingly, the obtained morphological position angle is almost perpendicular to the kinematic one (70\dg\ difference). Because of this and the fact that the obtained optical morphology is also misaligned with the kinematic position angle, the kinematic modeling was not possible for this galaxy. Additionally, the PV along the major axis shows the galaxy is kinematically lopsided meaning its rotational velocity might not be a good tracer of its underlying potential. The stellar counterpart looks disturbed with a high number of starforming regions in the outskirts. These regions cause the drop of ellipticity in the outermost part of the disk. In the inner part, the galaxy has ellipticity of around 0.7, corresponding to inclination of 90\dg\ for $q_0=0.3$. As before, there is an indication of a broad \hi\ emission region in the PV slice along the major axis, which could be another signature of edge-on galaxies, but is indistinguishable from the effect of the beam due to having too few resolution elements. 

\begin{figure*}
        \centering
        \includegraphics[width = 0.98 \textwidth]{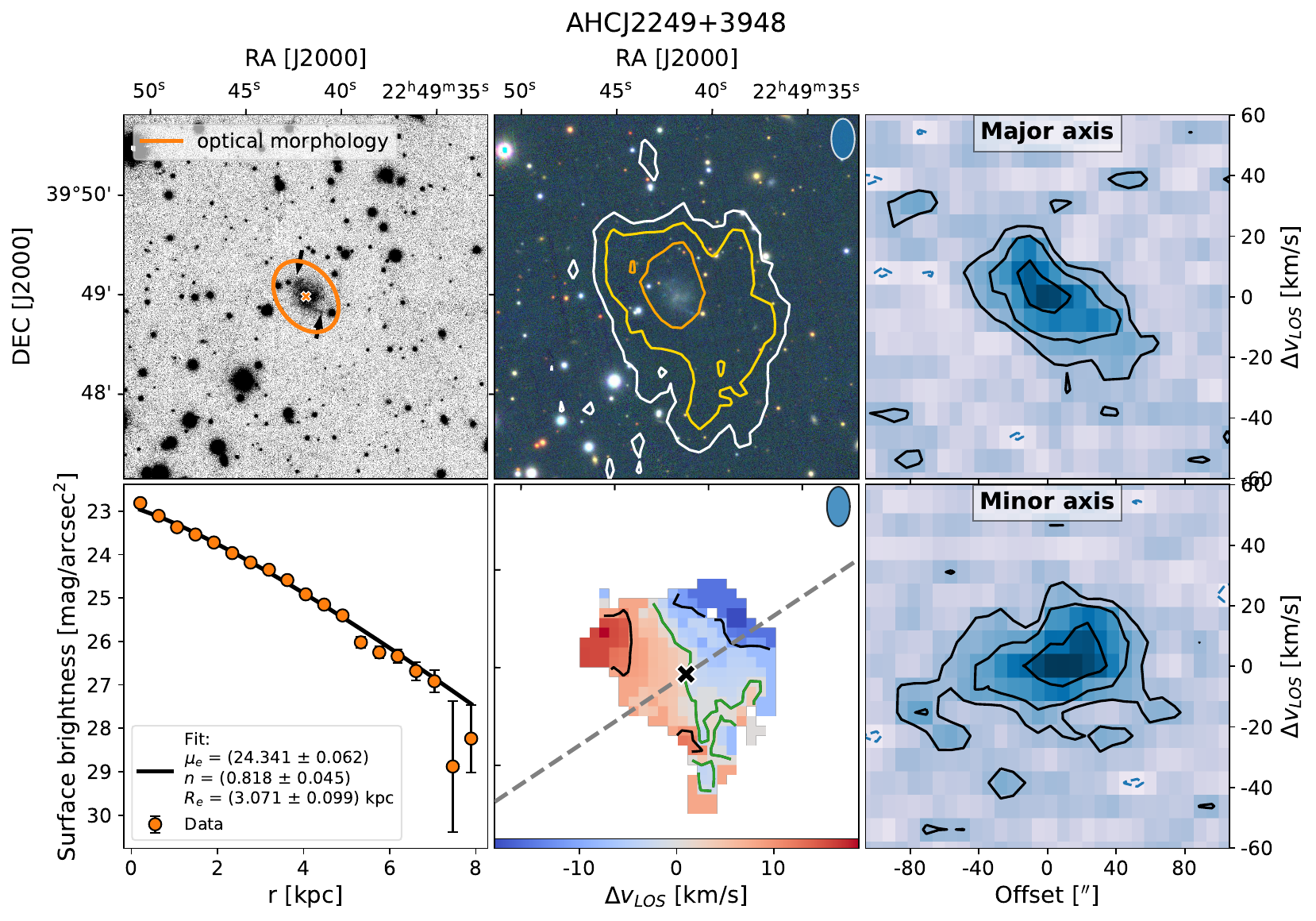}
        \centering
        \includegraphics[width=0.87\linewidth]{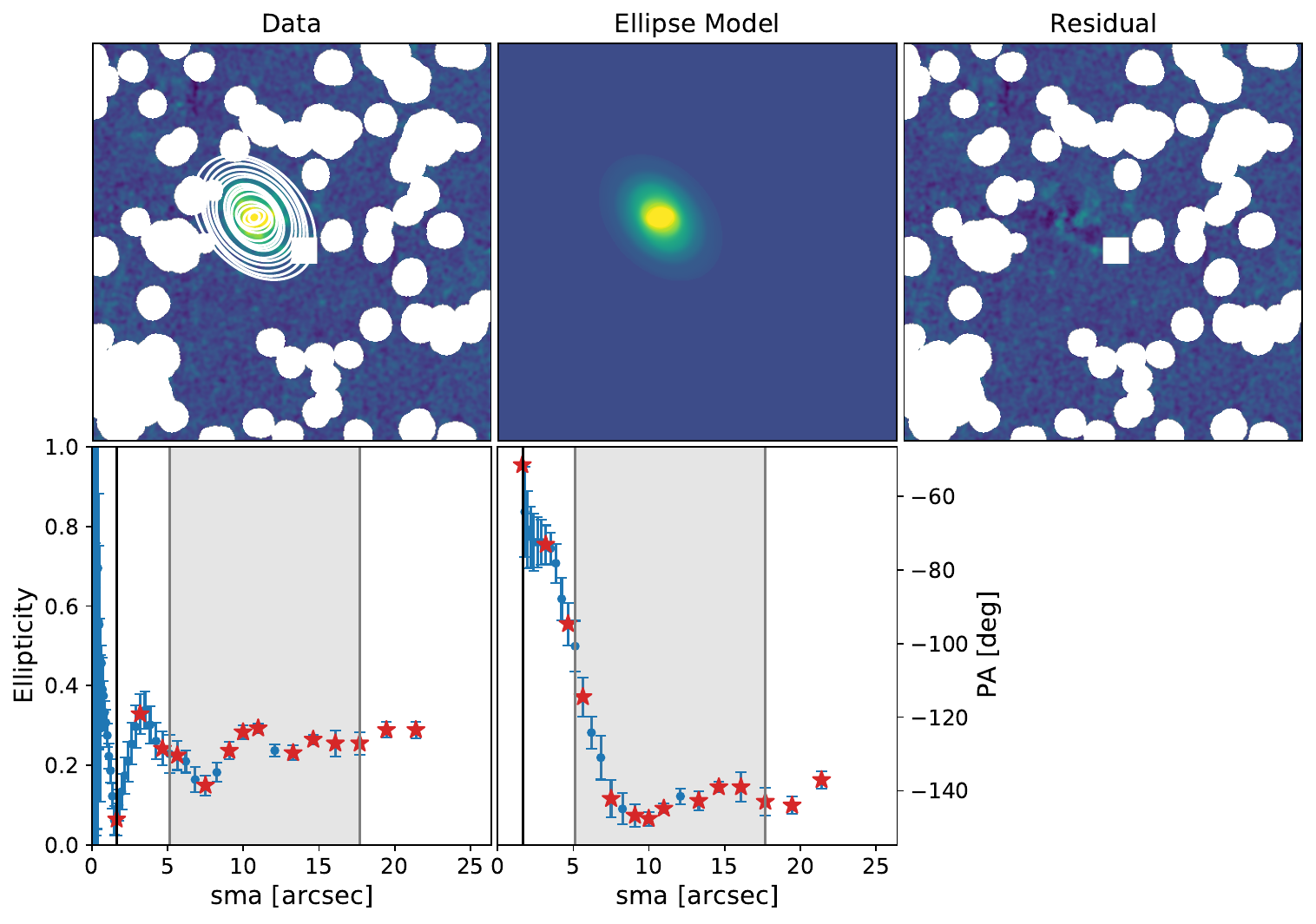}
        \caption{\weir. First two rows are the same as Figs. \ref{fig:pavel-kinematics} (middle and right column) and \ref{fig:pavel-optical} (left column) combined. Black arrows in the upper left panel denote extended stellar structures in the outskirt of the galaxy (see the text). The \hi\ contours in the middle upper panel correspond to column densities starting from $1.5 \times 10^{20}$ cm$^{-2}$ (in white) and growing by a factor of 2 in intensity towards contours in redder colors. Bottom two rows are the same as in Fig. \ref{fig:isophot_fit-Pavel}.}
        \label{fig:HI-op-weir}
    \end{figure*}

    \textbf{\weir.} The galaxy is shown in Fig. \ref{fig:HI-op-weir}. The \hi\ geometry of the disk could not be constrained for this galaxy due to its peculiar morphology with a significant extension of \hi\ towards the south. Additionally, velocity field map shows kinematic disturbances along the disk which could be a sign of interaction with another galaxy. However, there is no recorded galaxy with a redshift measurement in NED within 600 kpc from \weir\ that could confirm this assumption. The stellar counterpart shows two stellar arms in the northeast and southwest direction causing the variation of position angle with radius. 
    The \hi\ kinematics and optical morphology in the outer parts are completely misaligned.

\end{appendix}

\end{document}